\documentclass[fleqn,usenatbib]{mnras}

\usepackage[T1]{fontenc}
\usepackage{longtable}
\usepackage{xcolor}

\DeclareRobustCommand{\VAN}[3]{#2}
\let\VANthebibliography\thebibliography
\def\thebibliography{\DeclareRobustCommand{\VAN}[3]{##3}\VANthebibliography}

\usepackage{graphicx}	
\usepackage{amsmath}	
\usepackage{amssymb}	
\usepackage{lscape}     

\newcommand{\teff}{$T_{\rm eff}$}
\renewcommand{\vec}[1]{\mathbf{#1}}
\newcommand{\vectheta}{\boldsymbol{\theta}}

\title[RR Lyrae in the GALAH survey]{The GALAH survey: tracing the Milky Way's formation and evolution through RR Lyrae stars}

\author[V. D'Orazi et al.]
{Valentina D'Orazi,$^{1,2,3,4}$\thanks{E-mail: vdorazi@roma2.infn.it}
Nicholas Storm,$^{5}$
Andrew R. Casey,$^{3,4}$
Vittorio F. Braga,$^{6}$
Alice Zocchi,$^{7}$
\newauthor
Giuseppe Bono,$^{1,6}$
Michele Fabrizio,$^{8,6}$
Christopher Sneden,$^{9}$
Davide Massari,$^{10}$ 
Riano E. Giribaldi, $^{11}$
\newauthor
Maria Bergemann,$^{5}$
Simon W. Campbell,$^{3,4}$
Luca Casagrande,$^{12,4}$
Richard de Grijs,$^{13,14,15}$
\newauthor
Gayandhi De Silva,$^{13,4}$
Maria Lugaro,$^{16,17,18,3}$
Daniel B. Zucker,$^{13,4}$
Angela Bragaglia,$^{10}$
Diane Feuillet,$^{19}$
\newauthor
Giuliana Fiorentino,$^{6}$
Brian Chaboyer,$^{20}$
Massimo Dall'Ora,$^{21}$
Massimo Marengo,$^{22}$
\newauthor
Clara E. Mart\'inez-V\'azquez,$^{23}$
Noriyuki Matsunaga,$^{24,25}$
Matteo Monelli,$^{26,27,6}$
Joseph P. Mullen,$^{28}$
\newauthor
David Nataf,$^{29}$
Maria Tantalo,$^{26,27,6}$
Frederic Thevenin,$^{30}$
Fabio R. Vitello,$^{31}$
Rolf-Peter Kudritzki,$^{32,33}$
\newauthor
Joss Bland-Hawthorn,$^{34,4}$
Sven Buder,$^{12,4}$
Ken Freeman,$^{12,4}$
Janez Kos,$^{35}$
Geraint F. Lewis,$^{34}$
\newauthor
Karin Lind,$^{36}$
Sarah Martell,$^{37,4}$
Sanjib Sharma,$^{34,4}$
Dennis Stello,$^{37,34,4}$
Toma\v{z} Zwitter$^{35}$.
\\
\\
Affiliations at the end of the manuscript
}
\date{Accepted 2024 April 25. Received 2024 April 25; in original form 2024 March 19}
\pubyear{2024}
\begin{document}
\label{firstpage}
\pagerange{\pageref{firstpage}--\pageref{lastpage}}
\maketitle

\begin{abstract}
Stellar mergers and accretion events have been crucial in shaping the evolution of the Milky Way (MW). These events have been dynamically identified and chemically characterised using red giants and main-sequence stars. RR Lyrae (RRL) 
variables can play a crucial role in tracing the early formation of the MW since they are ubiquitous, old (t$\ge$10 Gyr) low-mass stars and accurate distance indicators. We exploited Data Release 3 of the GALAH survey to identify 78 field RRLs suitable for chemical analysis. Using synthetic spectra calculations, we determined atmospheric parameters and abundances of Fe, Mg, Ca, Y, and Ba. Most of our stars exhibit halo-like chemical compositions, with an iron peak around [Fe/H]$\approx -$1.40, and enhanced Ca and Mg content. Notably, we discovered a metal-rich tail, with [Fe/H] values ranging from $-$1 to approximately solar metallicity. This sub-group includes almost  1/4 of the sample, it is characterised by thin disc kinematics and displays sub-solar $\alpha$-element abundances, marginally consistent with the majority of the MW stars. Surprisingly, they differ distinctly from typical MW disc stars in terms of the s-process elements  Y and Ba. We took advantage of similar data available in the literature and built a total sample of 535 field RRLs for which we estimated kinematical and dynamical properties. We found that metal-rich RRLs (1/3 of the sample) likely represent an old component of the MW thin disc. We also detected RRLs with retrograde orbits and provided preliminary associations with the Gaia-Sausage-Enceladus, Helmi, Sequoia, Sagittarius, and Thamnos stellar streams.
\end{abstract}
%

\section{Introduction}
RR Lyrae stars (RRLs) play a crucial role in our understanding of the Universe. 
In the foundational review of the field, \cite{preston1964} articulated 
that these stars act as essential tools in unravelling the evolutionary processes that shaped our Galaxy.
Through their unique characteristics and patterns, RRLs provide insights into the age, structure, and formation history of our Milky Way (MW, e.g., \citealt{dekany2013}; \citealt{belokurov2018}). A wealth of studies have leveraged the insights provided by RRLs to delve into the chemical and dynamical transformations characterising the MW evolutionary journey (see e.g., \citealt{for2011}; \citealt{hansen2016}; \citealt{crestani2021a,crestani2021b}; \citealt{fabrizio2021}, \citealt{ablimit2022}; \citealt{li2022} and references therein). 

RRLs are low-mass, old ($\approx$ 10 Gyr) stars that reside on the horizontal branch, undergoing core helium burning, and are characterised by their pulsations within the instability strip, with typical periods shorter than one day (\citealt{bono2011}; \citealt{marconi2012}).
They are radial variables oscillating in the fundamental mode (RRab), in the first overtone (RRc), and mixed-mode (RRd, two modes are simultaneously excited), which offer valuable insights into the physics of stellar evolution and pulsation (e.g., \citealt{catelan2007}, \citealt{sos2011}).

Dating back to the works of \citet{baade1958} and \citet{sandage1962}, RRL stars were identified as reliable standard candles because their absolute visual magnitude is inversely related to their metallicity. The precision of using RRL stars as distance indicators significantly improved with the discovery that they adhere to well-defined Period-Luminosity (PL) relations for wavelengths exceeding the $R$-band as detailed by \citep{longmore1989, bono2011, braga2021}. This means that they are invaluable tools for distance determination within our Galaxy and beyond (\citealt{neeley2019}; \citealt{latham2023} and references therein). With this capability, distances can be easily determined with a remarkable level of precision better than 5\% (see \citealt{beaton2018} for a review).
Owing to their distinct pulsation patterns and relatively high luminosity, this class of variable stars stands out in time-domain surveys, serving as indispensable tools 
for probing diverse structures in our Galaxy \citep{clementini2024,bhardwaj2024}. As a result, RRLs serve as pivotal markers for uncovering the formation of Galactic substructures and potential merger events, as exemplified by 
the sample of nine RRLs in \cite{feuillet2022} and the deduced accreted origin from their (low) aluminium content.

While there has been considerable research on the metallicities of RR Lyrae stars (recent works including \citealt{liu2013}; \citealt{fabrizio2019}; \citealt{gilligan2021})
the field remains limited in terms of high-resolution spectroscopic studies that delve into their detailed chemical abundances. \cite{crestani2021b} conducted the most extensive high-resolution spectroscopic analysis of RR Lyrae stars to date. Their sample covered 162 stars,  with an additional 46 obtained from online archival repositories. The dataset encompassed a diverse range of spectral sources, including UVES,  HARPS, and FEROS at ESO, HRS at SALT and $echelle$ at DuPont.
Using the python wrapper \texttt{pymoogi} (\citealt{adamow2017})\footnote{This code is developed and maintained by M. Adamow and can be downloaded at \url{https://github.com/madamow/pymoogi}}of the 2019 version of {\sc moog} \citep{sneden1973} and adopting the LTE assumption. These authors derived abundances for Fe, Mg, and Ca through equivalent width (EW) methods. Notably, this work represents a significant milestone as it marks the detection of subsolar [$\alpha$/Fe]\footnote{We adopt the standard spectroscopic notation such that
\begin{math}
[\mathrm{Fe}/\mathrm{H}] = \log\left(\frac{N_{\mathrm{Fe}}}{N_{\mathrm{H}}}\right)_{\bigstar} -  \log\left(\frac{N_{\mathrm{Fe}}}{N_{\mathrm{H}}}\right)_{\odot} \mathrm{and} ~
[\mathrm{X}/\mathrm{Fe}] = [\mathrm{X}/\mathrm{H}] - [\mathrm{Fe}/\mathrm{H}]
\end{math}
for a given species \(\mathrm{X}\). Units are in dex.} 
ratios in old stellar tracers based on a large sample of stars, as previously introduced by \cite{prudil2020}. 
Crestani and collaborators found a resemblance in the trend of [$\alpha$/Fe] with metallicity between $\alpha$-poor, metal-rich RR Lyrae stars and red giants (RGs) in the Sagittarius dwarf galaxy, as well as between $\alpha$-enhanced, metal-poor RR Lyrae stars and RGs in ultra-faint dwarf (UDF) galaxies. Instead, \cite{prudil2020}, utilising kinematic evidence, asserted that the metal-rich, $\alpha$-poor RRL population genuinely belongs to the Galactic disc. Yet, this introduces a dilemma. The age of the thin disc is less than $\approx$ 8 Gyr \citep{kilic2017}, while RRLs are understood to exceed 10 Gyr in age. On the other hand, their chemical composition does not align with the predominant makeup of the Galactic thick disc (which is $\alpha$-rich), suggesting that they must represent an exceptionally $\alpha$-poor segment of the thick disc \citep{prudil2020}. 


We used the DR3 of the GALAH survey (GALactic Archeology with Hermes, \citealt{desilva2015}; \citealt{buder2021}) and identified a sample of 
78 RRL stars, suitable for chemical analysis. We provide atmospheric parameters and detailed abundance values for iron (Fe), $\alpha$-elements (Mg, Si, Ca), and $n$-capture elements (Y, Ba). This represents the most extensive high-resolution, single-source database of detailed abundance determinations for RRLs in the current literature. 
The organisation of the paper is structured in the following manner: Section~\ref{sec:sample} introduces the sample and discusses the pulsation properties of the RRL stars examined in this study. Section~\ref{sec:abu_analysis} details the methodology of the abundance analysis. The findings related to chemistry, kinematics and dynamics are presented and explored in Sections~\ref{sec:chemistry} and \ref{sec:kinematics}. In Section \ref{sec:discussion} we discuss the scientific implications of our findings; finally, Section~\ref{sec:conclusions} concludes the paper by summarising the key results.

\section{RRL Sample and pulsation properties}\label{sec:sample}

We began our study with a catalogue of RR Lyrae stars, assembled over the past few years \citep{fabrizio2019,fabrizio2021,braga2021}. The foundation of the catalogue is built upon Gaia DR3 \citep{gaia2023} and its comprehensive catalogue of RRL stars, as detailed in \cite{clementini2023}, but it includes RRLs from almost all the known photometric surveys (Catalina, \citealt{drake2009,drake2017}; ASAS, \citealt{pojmanski1997}; ASAS-SN, \citealt{jayasinghe2019} to name a few, see \citealt{fabrizio2021,braga2021} for the details). The catalogue contains 286,135 RRLs, all of which have a valid Gaia DR3 \texttt{source id}, despite not all of them being in the \texttt{gaiadr3.vari\_rrlyrae} table on the ESA archive\footnote{\url{https://gea.esac.esa.int/archive/}}. As already discussed in \citet{braga2021}, for 10,413 of them, we have associated spectra, but only for a minority (192), we have high-resolution spectra. From the original catalogue of 268,135 RRLs, we performed a cross-match with the GALAH DR3 main catalogue and found 331 objects in common. However, we must set specific criteria for our study: 1) exposure times for stacked spectra cannot exceed 60 consecutive minutes (3$\times$20 minutes) because the pulsation cycle of RRLs is shorter than 1 day and we avoid stack spectra collected at very different phases (meaning significantly different atmospheric parameters) and 2) a Signal-to-Noise Ratio (SNR) per pixel of at least 30 (in the red channel, i.e., $\lambda \approx 6500 $\AA$ $) is essential for abundance analysis. 
Applying these criteria, our study focuses on a sub-sample of 78 RRL.
Table ~\ref{tab:photometric} displays the pulsation properties of our RRL sample. Note that we provide, for each RRL, both the epoch of maximum light ($T_{\rm max}$) and the epoch of the mean magnitude on the rising branch $T_{\rm ris}$. While the former one is the most used in spectroscopic studies and more familiar to the astronomical community, the latter ensures a more precise phasing and is more suited to apply to radial velocity templates \citep{braga2021}.

We prioritised using light curves from the $V$-band provided by ASAS-SN~V and Catalina whenever possible. However, in cases where these were not available for some RR Lyrae stars, we used Gaia's $G$-band light curves instead. It is important to mention that the amplitude ratio between the $G$ and $V$ bands is nearly 1, according to \cite{clementini2019}. For the RR Lyrae stars whose luminosity amplitudes were used to determine the barycentric radial velocity (RV) with RV templates, we adjusted the amplitudes from the $G$-band to the $V$-band by following the procedure outlined by \citet{clementini2019}.

\subsection{Validation of the RRL sample}\label{sec:validation}
\begin{figure}
\begin{center}
\includegraphics[width=0.50\textwidth]{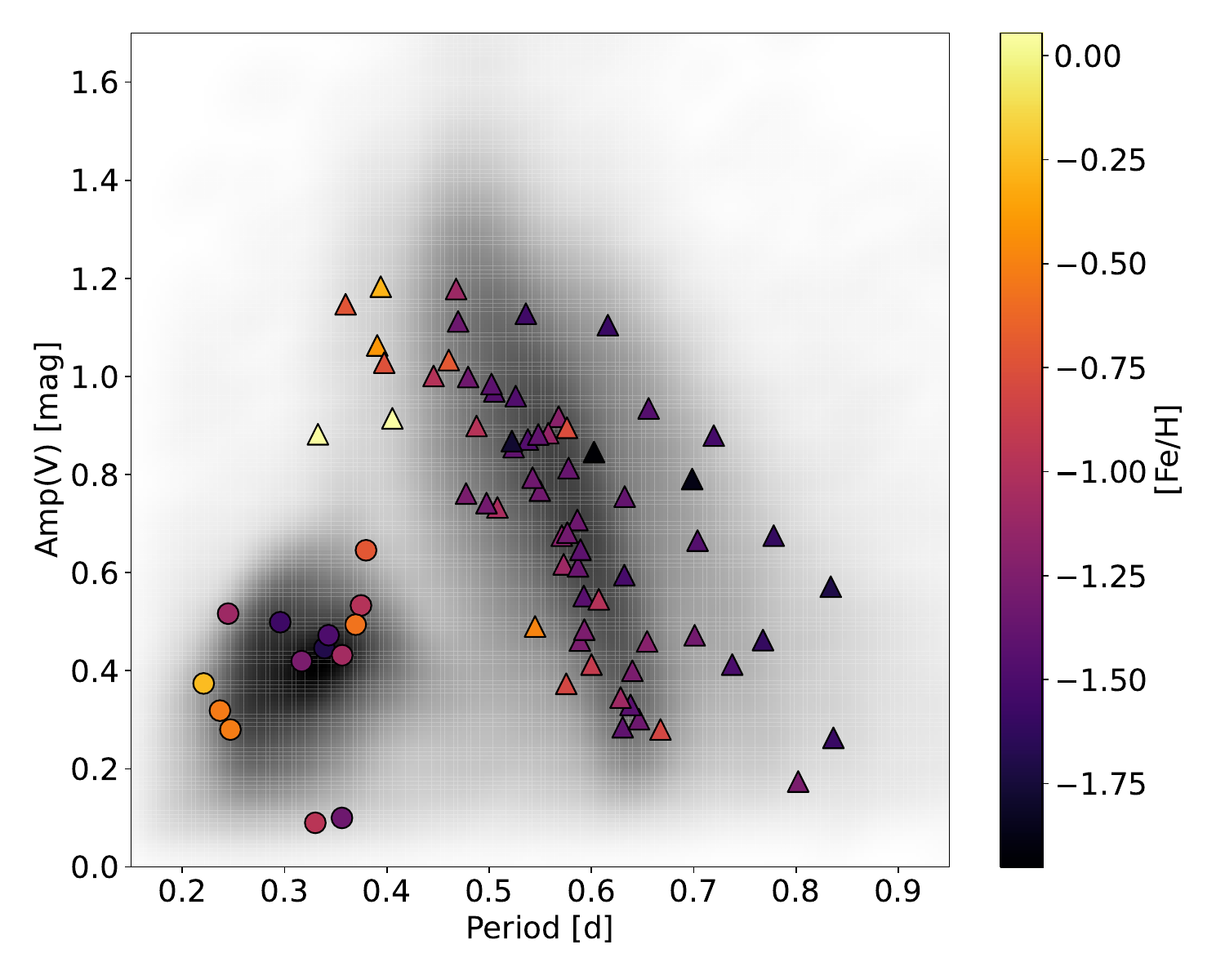}
\caption{$V$-band luminosity amplitude versus pulsation period (Bailey Daigram). RRLs in the 
GALAH's sample were plotted as triangles (RRab) and circles (RRc). They are colour-coded according to iron abundance
(see scale on the right-hand side). The grey-shaded area shows the distribution of field RRLs included in our photometric catalogue.}
\label{Bailey_V.pdf}
\end{center}
\end{figure}

To determine the characteristics of the RRLs in the GALAH sample, Figure~\ref{Bailey_V.pdf} presents the distribution of the selected RRLs (marked with triangles and circles for RRab and RRc, respectively) in the plane of $V$-band luminosity amplitude versus pulsation period (known as the Bailey diagram). The symbols within the diagram are colour-coded to reflect the iron abundance, as indicated by the scale on the right side of the figure. This serves as a reliable diagnostic for pulsation because it is based on two observables that are unaffected by the uncertainties associated with individual distances and reddening corrections. Data in this panel show that both 
RRab and RRc variables are located in the regions (grey-shaded areas) typical of Galactic field RRLs. Note that more metal-rich RRLs are located, as expected, in the short period tail of RRab variables, while the more metal-poor RRLs are in the long period tail \citep{fiorentino2015, crestani2021a, fabrizio2021}. 
The current sample of RRc variables is limited and the trend with metal abundance is less clear.

\begin{figure}
\begin{center}
\includegraphics[width=0.40\textwidth]{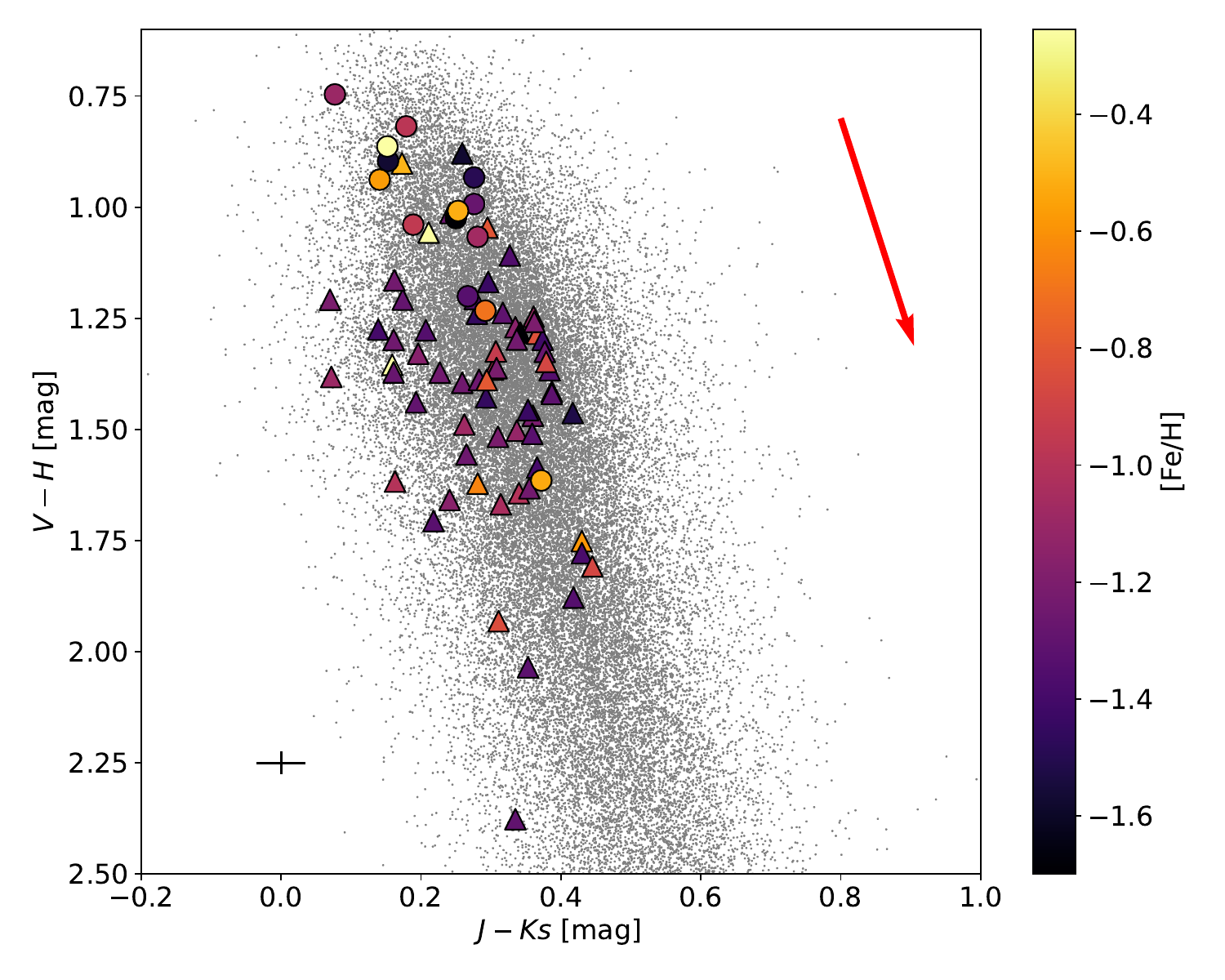} 
\includegraphics[width=0.40\textwidth]{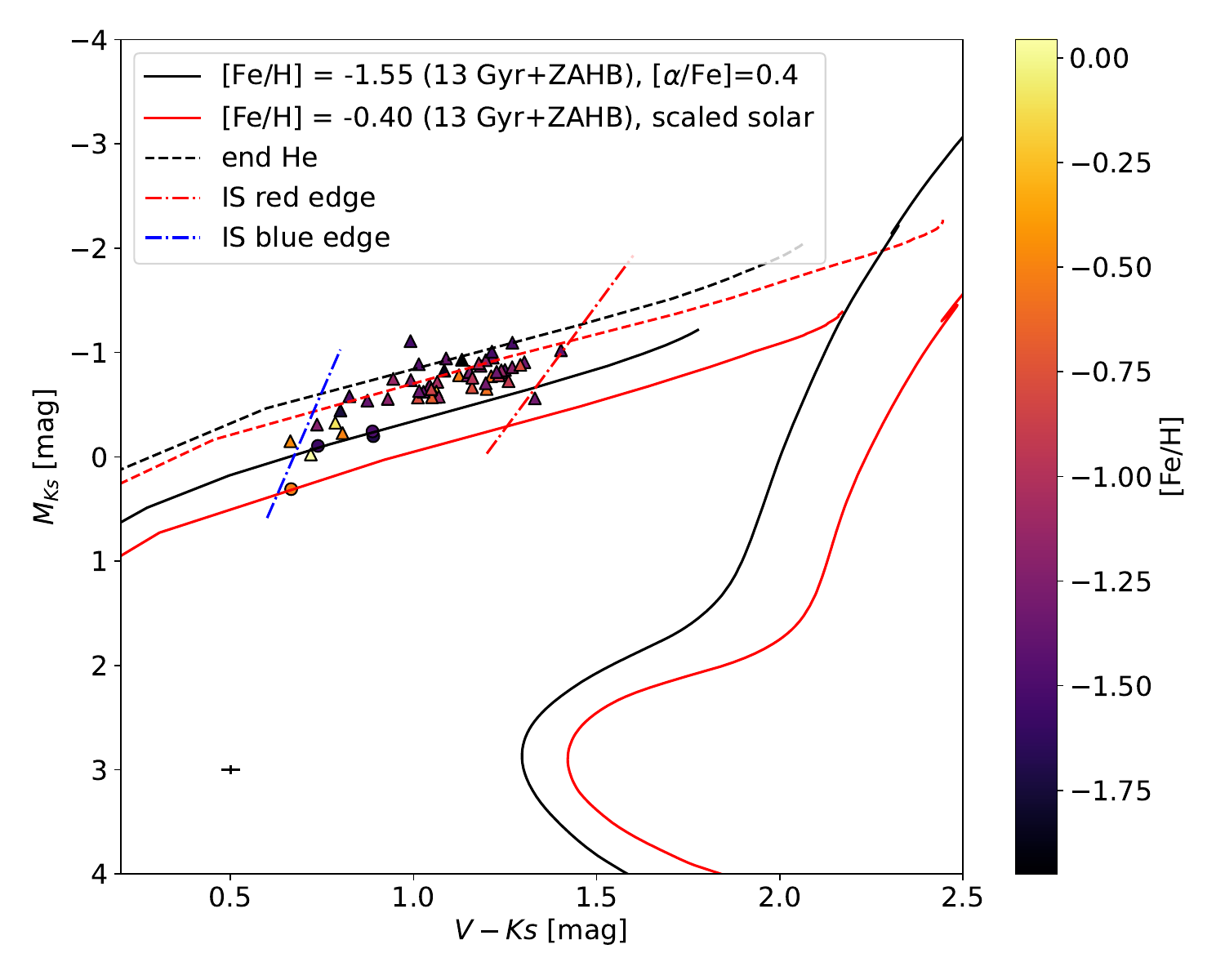}
\caption{Top: $V-H$ vs $J-K$ colour-colour diagram of the RRLs in the sample (circles).
The symbols are colour-coded according to the iron content (see the scale on the right).  
The grey dots show the distribution of field RRLs included in our photometric catalogue. 
The red arrow displays the reddening vector. The black cross indicates the average uncertainty on the RRLs of our sample.
Bottom: NIR ($V-K$ vs $K$) Colour-Magnitude Diagram of the RRLs in the GALAH's sample.
The RRL were plotted in this plane by using individual distance determinations and 
individual reddening estimates (see text for more details).
The black and the red lines display cluster isochrones for t=13 Gyr and different 
$\alpha$-enhanced chemical compositions (see labelled values). The horizontal 
red and black solid lines display the Zero-Age-Horizontal-Branch for the two selected 
chemical compositions, while the dashed lines indicate the end of core helium burning phases.
Symbols are the same as in the top panel. The almost vertical dashed-dotted lines display the hot (blue) and the cool (red) edge of the RRL instability strip (IS) were derived using the pulsation models in \citet{marconi2015}, including metallicities from Z=0.0006 to 0.02. The black cross shows the average uncertainties for our RRL sample.}
\label{ccd_cmd}
\end{center}
\end{figure}

To enhance our understanding of the pulsation and evolutionary characteristics of our RRLs, the upper section of Figure~\ref{ccd_cmd} displays the chosen RRLs plotted in the colour-colour diagram, specifically in the optical$-$NIR ($G-H$) versus NIR ($J-K$) plane. The average $J$, $H$, $K$ magnitudes were sourced from the 2MASS catalogue. 
It is important to note that the 2MASS data rely on a limited set of observations, meaning the averages we utilize are derived from these direct measurements rather than being calculated by fitting the individual light curves. We selected this colour-colour plane as chosen for its sensitivity to variations in reddening estimates while remaining unaffected by the uncertainties common to individual distance measurements. The distribution of RRL colours in the GALAH dataset shows a strong concordance with that of field RRLs (grey points), as evidenced by their similar colour range and their collective alignment along the reddening vector, indicated by the red arrow. In addition, the RRLs in the GALAH sample were plotted in an optical-NIR CMD to constrain their evolutionary properties. This diagnostic tool is influenced by the uncertainties associated with both reddening and distance calculations. To mitigate the impact of individual reddening variations, their distances were determined using the mid-infrared (MIR) Period-Luminosity-Metallicity (PLZ) relationships for fundamental and first overtone RRLs as given by \cite{neeley2017}. The individual metallicities are the same adopted in the current investigation, while mean W1 magnitudes for these stars were obtained from the NEOWISE catalogue \citep{wright2010}, with their periods and mode classifications sourced from the Gaia database.
Dereddening the apparent magnitudes involved using reddening estimates derived from the map provided by \cite{schlafly11} and applying the reddening law established by \cite{cardelli1989}. To enhance the diagnostic's sensitivity to temperature variations, the RRLs within the GALAH dataset were plotted on a $V-K$ versus $K$ CMD, which can be viewed in the lower panel of Figure~\ref{ccd_cmd}.

To constrain their evolutionary status we adopted cluster isochrones for $t$=13 Gyr 
from the BASTI data set. We selected two $\alpha$-enhanced chemical compositions 
(see labelled values) that are representative of the metal-rich tail ([Fe/H]=$-$0.40, red) and of the peak ([Fe/H]=$-$1.55, black) in the metallicity distribution of the GALAH's sample. Moreover, we also adopted the Zero-Age-Horizontal-Branch (ZAHB, solid lines) and the end of core helium burning (dashed lines). 

The ZAHB and the end of core helium burning were selected because theoretical 
and empirical evidence indicate that the bulk of HB and RRL stars should be distributed 
within this well-defined magnitude interval. This evidence relies on a solid theoretical
prediction: the Asymptotic Giant Branch phase is $\approx$1/2 orders of magnitude faster than the HB phase. Data plotted in the bottom panel, display, within the errors, a very good agreement with the expected distribution along the HB. Moreover, they display the expected ranking in metal content. Indeed, more metal-rich RRLs are on average fainter than more metal-poor ones. Finally, RRc variables attain $G-K$ colours that 
are on average systematically bluer than RRab variables \citep{bono2011}.

\section{Abundance analysis}\label{sec:abu_analysis}
Our analysis exploits spectra obtained through the HERMES spectrograph  \citep{sheinis2015} at the 3.9-meter Anglo-Australian Telescope, located at the Siding Spring Observatory. These spectra are included in the Data Release 3 of the GALAH survey (see \citealt{desilva2015}; \citealt{buder2021}).
HERMES has a nominal spectral resolution R$\approx$ 28 000 in 4 spectral channels: blue (471.5 $-$ 490.0 nm), green (564.9 $-$ 587.3 nm), red (647.8 $-$ 673.7 nm) and infra-red (758.5 $-$ 788.7 nm). Details on data reduction (including wavelength solution and continuum normalisation) can be found in \cite{kos2017} and \cite{buder2021}.
We calculated the radial velocities (RVs) of our sample stars by using only channels blue and green of HERMES spectra because RRLs possess most of the absorption features in the bluest region of the visible bands (their fluxes typically peak at roughly 450 nm). We employed {\tt iSpec} \citep{blanco2019} to conduct cross-correlation with an F-type stellar template and calculate the RVs, as reported in Table~\ref{tab:kinematics}. The mean difference with GALAH DR3 value 
is $ \Delta {\rm (RV)}= -0.17 \pm 0.21 $ km s$^{-1}$, which reflects an essentially perfect agreement. 
Given the distinctive characteristics of RR Lyrae stars, we refrained from adopting the parameters and abundances published in the official GALAH DR3. This decision stems from the fact that the survey (industrial) approach is suboptimal for addressing the unique properties of these warm, variable, and mostly metal-poor stars. A primary issue is the survey's computation of microturbulence velocities ($\xi_{\rm t}$) that are notably low, ranging approximately between $\xi_{\rm t}  \approx$ 1.2 and 2 km s$^{-1}$, and more typical of cool red giant-branch stars. In contrast, RRL stars exhibit microturbulence velocities larger than 2.5 $-$ 4 km s$^{-1}$, as highlighted in studies such as \cite{for2011}, \cite{chadid2017}, \cite{crestani2021b}.

Furthermore, due to the restricted number of spectral lines available, also resulting from gaps between spectral channels, we optimised the effective temperatures (\teff) by fitting the wings of the H$\alpha$ profiles under non-local thermodynamic equilibrium (NLTE) conditions using $\chi^2$ minimisation. Emission lines and P Cygni profiles were present in many spectra, causing biased results when fitting effective temperatures. To ameliorate this we used \texttt{Korg} \citep{korg1,korg2} to compute synthetic spectra around the H$\alpha$ line for stars between 5000\,K and 8000\,K, with $\log{g}$ between 2 and 3, [Fe/H] between $-2.5$ and $+0.5$, and $[\alpha$/Fe] between 0 and $+0.4$. With this grid of continuum-normalised fluxes $f_\lambda$ we defined $\vec{X} = 1 - f_\lambda$ to be `line absorption' then used non-negative matrix factorisation \citep{nmf} to approximate the matrix $\vec{X}$ by two non-negative matrices $\vec{W}$ and $\vec{H}$ such that $\vec{X} \approx \vec{W}\vec{H}$. We chose to use 24 components for this approximation: $\vec{H}$ is a set of 24 non-negative eigenspectra and $\vec{W}$ is a set of 24 amplitudes per theoretical spectrum. With the eigenspectra $\vec{H}$ we then constructed a linear model to describe the stellar flux $y = 1 - \vectheta\vec{H}$, where $\vectheta$ is a set of unknown amplitudes for some observed spectrum. The linearity of this model allows us to add complexity (e.g., emission lines, P Cygni profiles) and ensure that inference remains stable and fast. We added emission and absorption line profiles with unknown parameters that enter multiplicatively to $y$, and fit these parameters simultaneously with $\vec\theta$, radial velocity $v_{\rm rad}$, and macroscopic broadening $v_{\rm broad}$. After fitting, we constructed a mask based on the fitted emission line centroids and standard deviations, and we used this mask to exclude pixels from the subsequent fitting of effective temperature from H$_\alpha$ lines.

To determine $T_{\rm eff}$ values, and for the abundance analysis outlined in this manuscript, we adopted the MARCS stellar grid (1D, plane-parallel models, \citealt{gustafsson2008}) and the Python wrapper {\tt TSFitPy} of the {\tt Turbospectrum v.20} \citep{plez2012} as described in \cite{gerber2023}. This tool enables the computation of synthetic stellar spectra with NLTE effects for multiple chemical species at once by fitting the normalised synthetic spectra from {\tt Turbospectrum} by $\chi^2$ minimisation using the Nelder-Mead algorithm \citep{gerber2023}. The tool was further optimised and had the addition of some extra features as described in \citet{storm2023}. For this particular project, \teff~fitting and $\xi_{\rm t}$ were added. The latter was done in the following fashion: $\xi_{\rm t}$ and [Fe/H] were fitted using a two-step optimisation process to minimise the $\chi^2$ statistic and break the degeneracy between them, as both influence the EW. For each tentative metallicity value, the corresponding optimal $\xi_{\rm t}$ was identified using the Nelder-Mead algorithm. The final fitted values for both [Fe/H] and $\xi_{\rm t}$ were then obtained by taking their means. 
\begin{figure*}
\begin{center}
\includegraphics[width=0.95\textwidth]{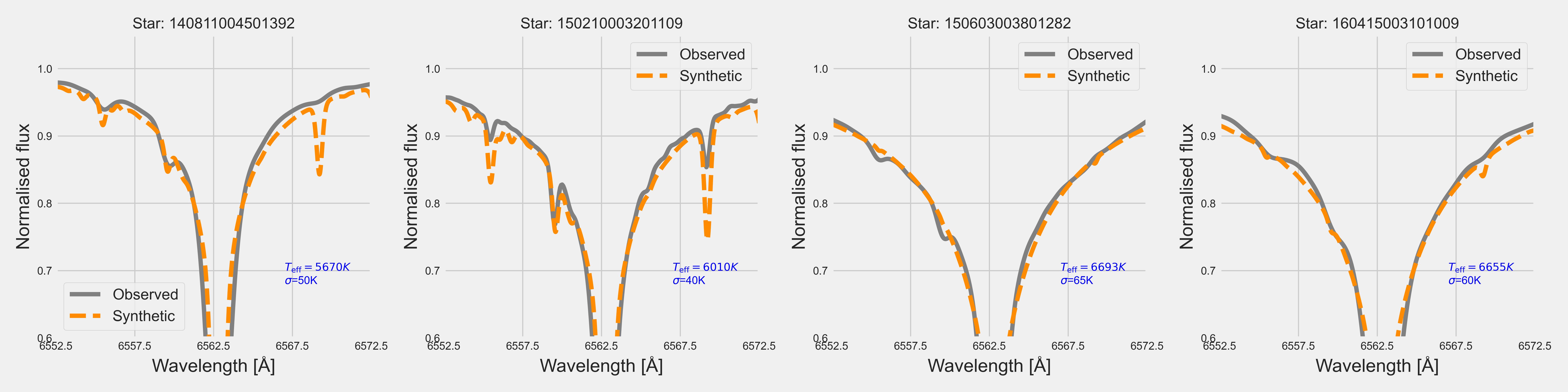}
\caption{Example of spectral fitting of H$\alpha$ profile in 1D NLTE by TSFitPy.}
\label{fig:halpha}
\end{center}
\end{figure*}
An example of the fitting procedure of H$_\alpha$ wings for our sample RRLs is provided in Figure \ref{fig:halpha}, whereby along with synthetic and observed spectrum we report the best-fit \teff and the statistical error. The error was calculated based on the $\chi^2$ interval, where the fitted \teff~was adjusted until the desired $\Delta \chi^2 = \chi^2 - \chi^2_{\textrm{min}} \approx \sigma^2$ was reached.

The optimisation of surface gravity ($\log g$) was achieved by leveraging the ionisation balance between Fe~{\sc i} and Fe~{\sc ii} lines, a practice commonly referenced in literature work. It is important to emphasise that the incorporation of NLTE is pivotal in this context (see e.g., \citealt{ruchti2013}). Our method involves fitting all Fe~{\sc i} and Fe~{\sc ii} lines within the spectra during each iteration, adjusting the gravity to achieve the least internal scatter. 
The error on our estimated $\log g$ is then calculated in a similar way as for \teff, that is  $\Delta \chi^2 \approx \sigma^2$. Due to the limited presence of ionised iron lines, the internal precision of the surface gravity we derived seldom is less than 0.25 dex. Nonetheless, this limitation does not significantly affect our inferred metallicities and elemental abundances.

We meticulously constructed our line list (reported in Table \ref{tab:line-list}) by choosing notably isolated, un-blended, and possibly strong lines that fall within our specified wavelength range and are backed by relatively accurate transition probabilities. We sourced atomic parameters from the Gaia-ESO survey \citep{randich2022}; for a more comprehensive description, readers are directed to \cite{heiter2021}. Our adopted solar abundances are: A(Fe)$_\odot$=7.50 dex, A(Ca)$_\odot$=6.37 dex, A(Si)$_\odot$=7.59 dex, A(Y)$_\odot$=2.21 dex, A(Ba)$_\odot$=2.18 dex (see \citealt{storm2023} and references therein).

The uncertainties in our abundance ratios ([X/Fe]) stem from two main sources: line-by-line scatter, which is represented as error bars in all plots in this paper, and errors associated with stellar parameters. 
To assess the impact of stellar parameters (i.e., the sensitivities), 
we altered the \teff, $\log g$, and microturbulence velocity by 100 K, 0.25 dex, and 0.2 km s$^{-1}$, respectively. After each modification, we reran our analysis to observe the resultant changes in abundance. This method does not account for the covariance between parameters, meaning it overlooks their interdependence. However, the impact of this covariance is relatively minor, with \teff~being the most influential parameter. We calculated errors by considering the internal errors of each parameter and combining them in quadrature. The typical values range between 0.10 and 0.15 dex.

We note that in our work all the published [X/Fe] ratios are calculated adopting a NLTE approach; references for each species are as follows: H \citep{mashonkina2008}, Mg \citep{bergemann2017}, Si \citep{bergemann2013, magg2022}, Ca and Fe \citep{semenova2020}, Y \citep{storm2023}, Ba \citep{gallagher2020}.

\section{CHEMISTRY}\label{sec:chemistry}
\begin{figure}
\begin{center}
\includegraphics[width=0.5\textwidth]{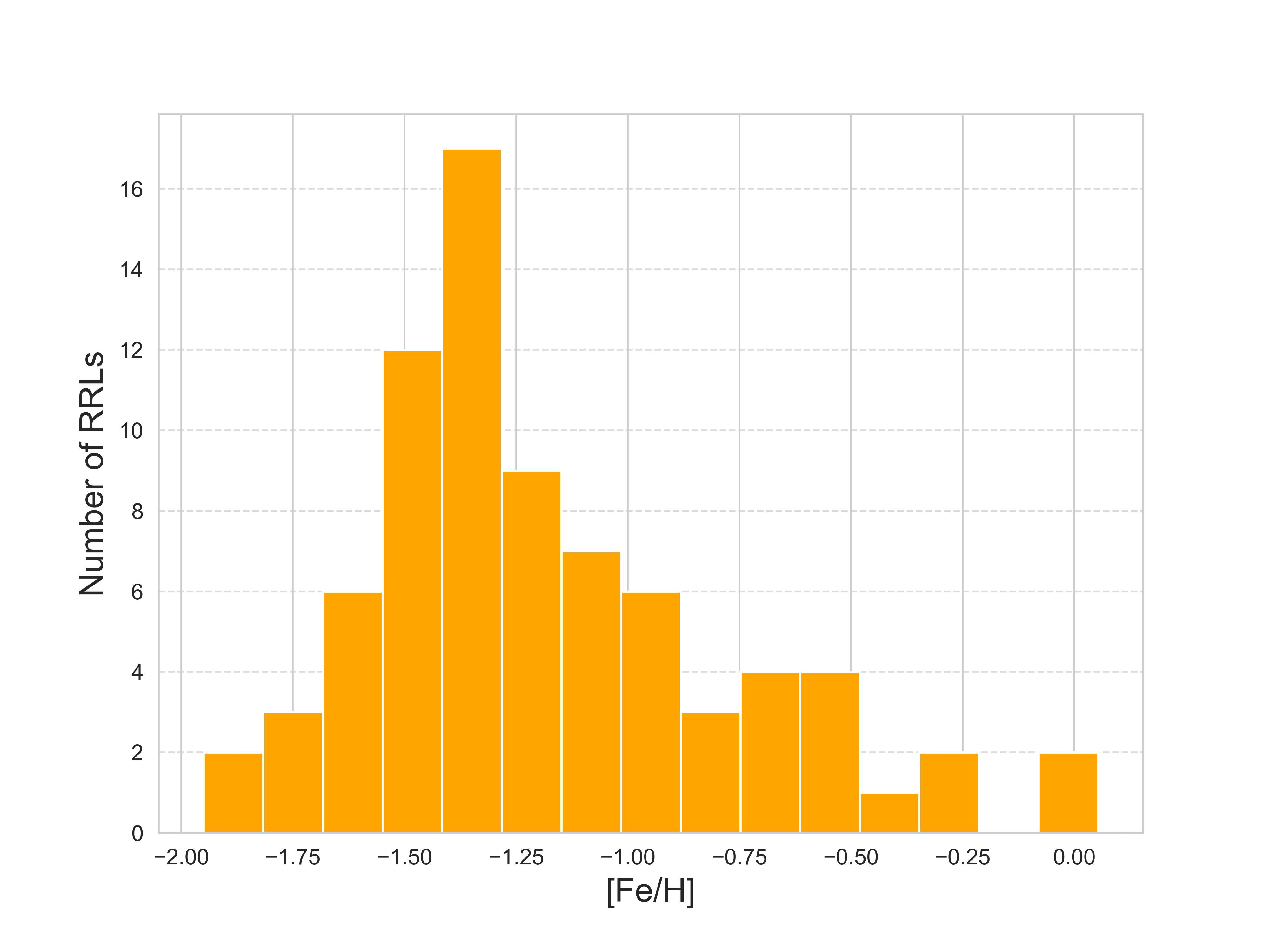}
\caption{Histogram of [Fe/H] values for the RRL analysed in the present work.}
\label{fig:mdf}
\end{center}
\end{figure}

\begin{figure}
\begin{center}
\includegraphics[width=0.5\textwidth]{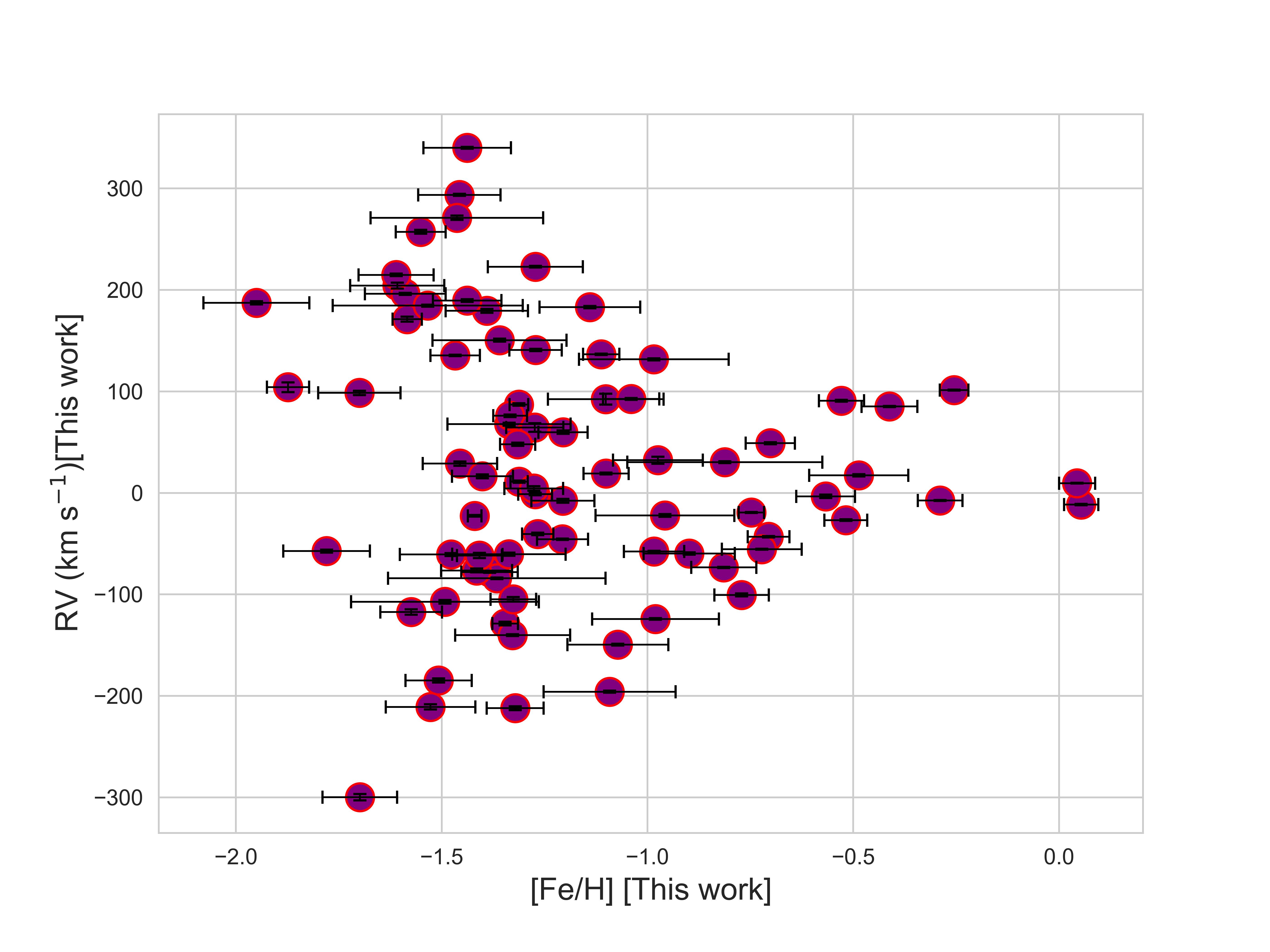}
\caption{RV values as a function of [Fe/H]}
\label{fig:rv_galah}
\end{center}
\end{figure}

The metallicity ([Fe/H]) for our sample of RRLs are presented in Figure \ref{fig:mdf}. The majority of RRLs align with the metallicity distribution function typical of halo stars. This is further supported kinematically, as depicted in Figure \ref{fig:rv_galah}. However, the undeniable presence of a metal-rich tail, 
exhibiting colder, disc-like kinematics warrants special attention and is discussed in the subsequent sections of this manuscript.
\begin{figure}
\begin{center}
\includegraphics[width=0.5\textwidth]{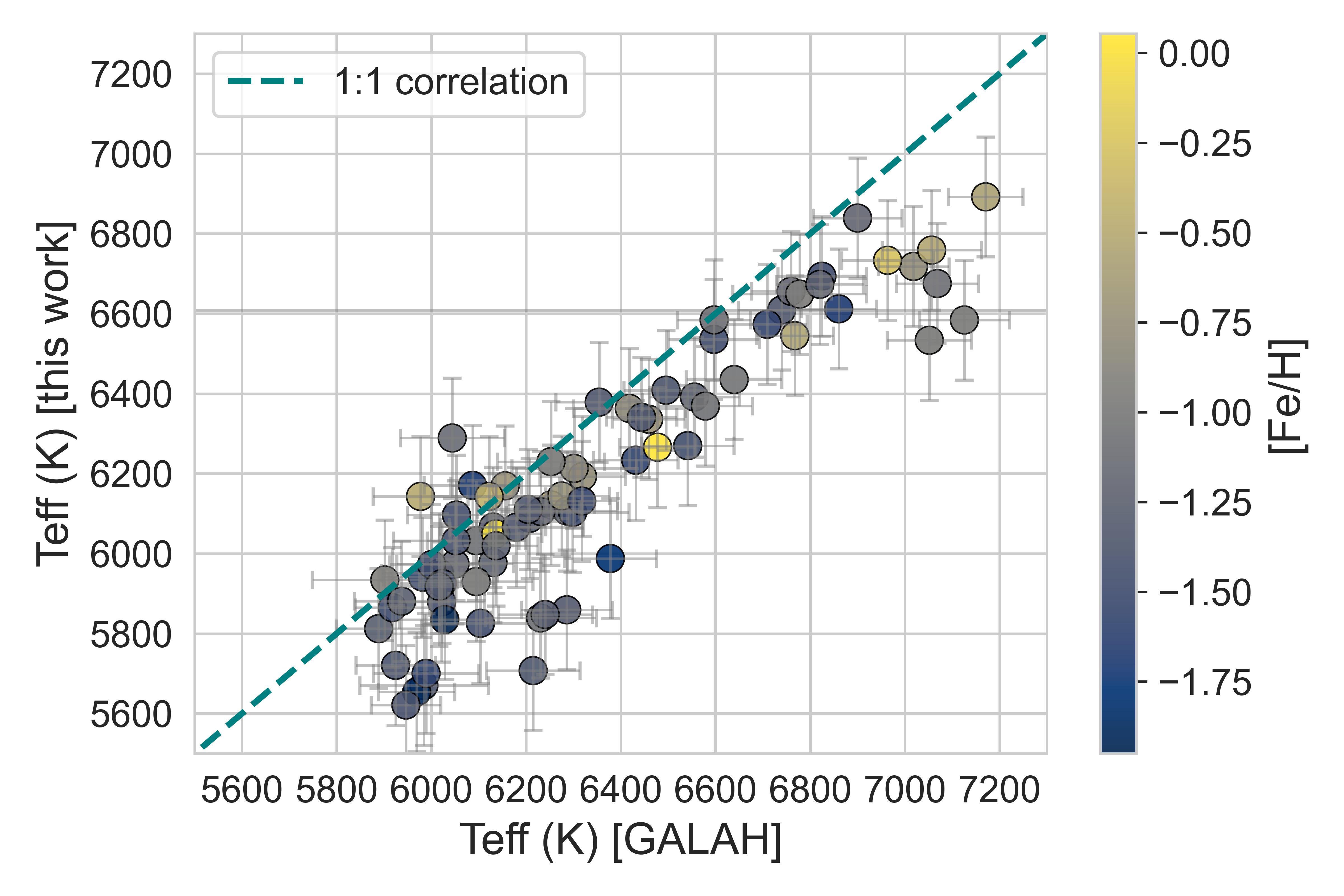}
\caption{Comparison of this study with \teff~ estimates from the GALAH DR3.}\label{fig:compTeff}
\end{center}
\end{figure}
\begin{figure}
\begin{center}
\includegraphics[width=0.4\textwidth]{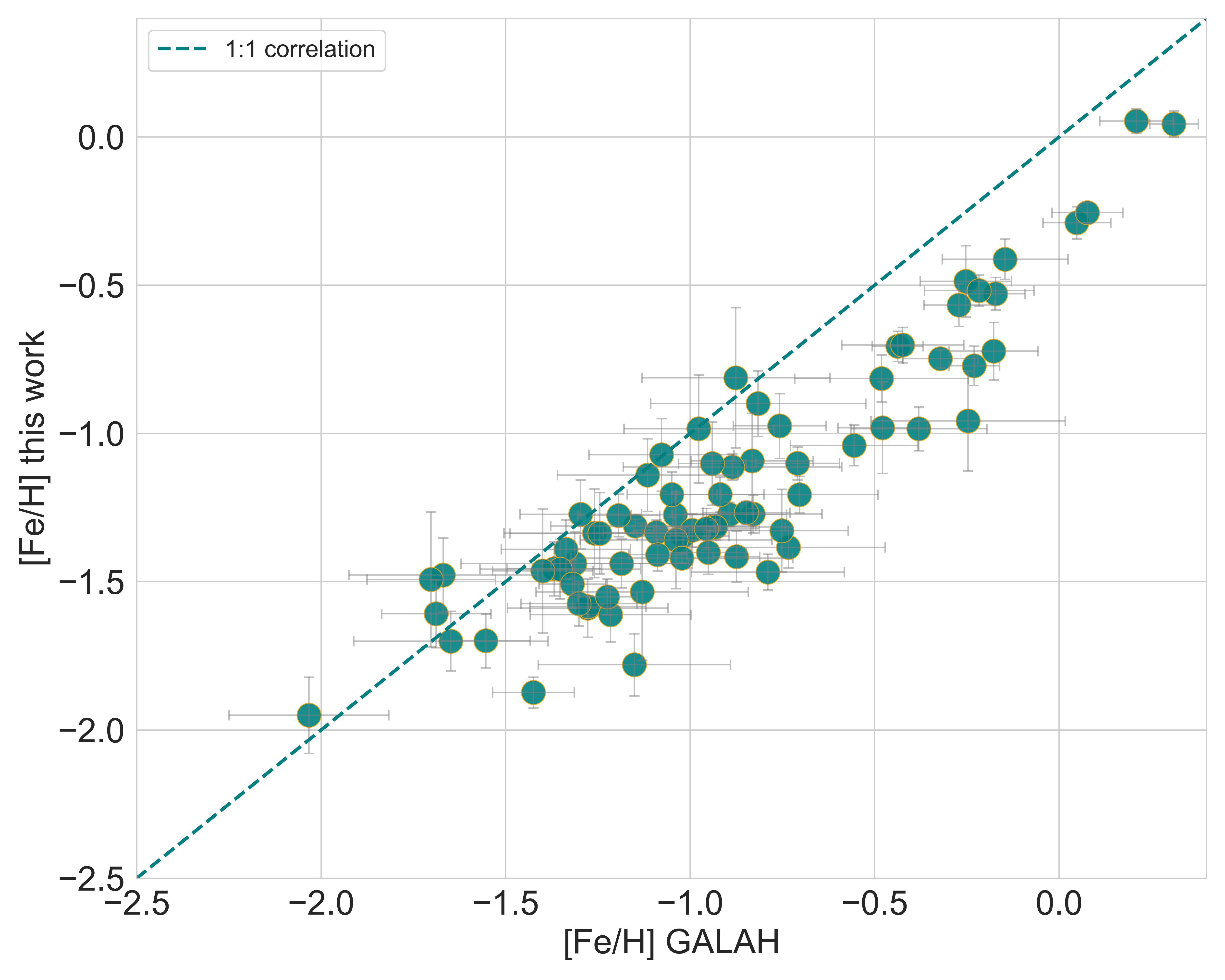}
\caption{Comparison of our [Fe/H] values with those from the GALAH DR3.}\label{fig:compFe}
\end{center}
\end{figure}
We have compared our derived parameters, including [Fe/H], with those published in the official GALAH DR3. Notably, the adoption of NLTE in parameter and abundance determination distinguishes GALAH and is absent in surveys like Gaia-ESO or APOGEE (e.g., \citealt{randich2022}; \citealt{abdurrouf2022}). Consequently, we will confine our comparison for [Fe/H] and atmospheric parameters strictly to GALAH. The difference in \teff~values between our estimates and those of GALAH is $\Delta$\teff=$-152.7 \pm 16$ K (see Figure \ref{fig:compTeff}). This discrepancy can serve as a conservative gauge for the systematic uncertainties influencing our \teff~and abundance scale. The higher \teff~observed in GALAH can be partially attributed to increased gravities, which were not optimised spectroscopically in the GALAH survey and typically measure +0.4$-$0.5 dex more than ours. Additionally, as noted earlier, the GALAH survey reports lower ${\xi_t}$ values ranging between 1.5 $-$ 2 km s$^{-1}$. Cumulatively, these factors lead to a higher metallicity reported by the GALAH survey, being $\Delta$[Fe/H]=$+0.28 \pm 0.02$ dex with respect to our inferred values. The comparison between our estimates and GALAH DR3 metallicity values is reported in Figure \ref{fig:compFe}.

We identified 4 stars in common with the high-resolution spectroscopic analysis by \cite{crestani2021b}, namely CM Ori, V413 CrA, BN Aqr, and AR Oct. Except for CM Ori (for which the Crestani's [Fe/H] value is $-$2.05$\pm$0.15 to be compared with our $-$1.46$\pm$0.09), the two studies agree fairly well, once observational uncertainties are taken into account: $\Delta$ [Fe/H]=+0.11, +0.14, $-$0.13 dex, respectively (in the sense ours minus Crestani et al.). 
To better understand the discrepancies for Cm Ori, we compared our findings with those reported by Crestani et al. (2021). They employed a medium-resolution (R = 37,000) spectrum from the HRS at SALT with a signal-to-noise ratio (SNR) of 38. In contrast, our analysis used a GALAH spectrum with an SNR of 79 in the red band. Both spectra were obtained at a similar phase ($\phi=0.3$), showing comparable effective temperatures (Teff of 5879 K in our study versus 5820 K by Crestani et al.). However, the difference in logg values is notable, with our measurement at 2.81 dex compared to 1.22 by Crestani et al. Upon re-analyzing the spectrum used by Crestani et al., we determined a Teff of 5860$\pm$60 K, logg of 2.75$\pm$ 0.20 dex, microturbulent velocity (Vt) of 3.00$\pm$0.25 km/s, and metallicity ([Fe/H]) of -1.58$\pm$0.15 dex. These findings are consistent with our measurements from the GALAH spectrum. It is crucial to highlight that the lower SNR of their spectrum limited the number of FeI and FeII lines measured. Additionally, the omission of non-local thermodynamic equilibrium (NLTE) effects in Crestani et al.'s study further amplifies the discrepancies, especially since their gravity measurements do not match photometric values. In our forthcoming paper (Pipwala et al., 2024, in preparation), we aim to re-analyze all stars in the Crestani sample using spectral synthesis calculations that include NLTE effects, similar to the methods used in this study.

\begin{figure*}
\begin{center}
\includegraphics[width=0.9\textwidth]{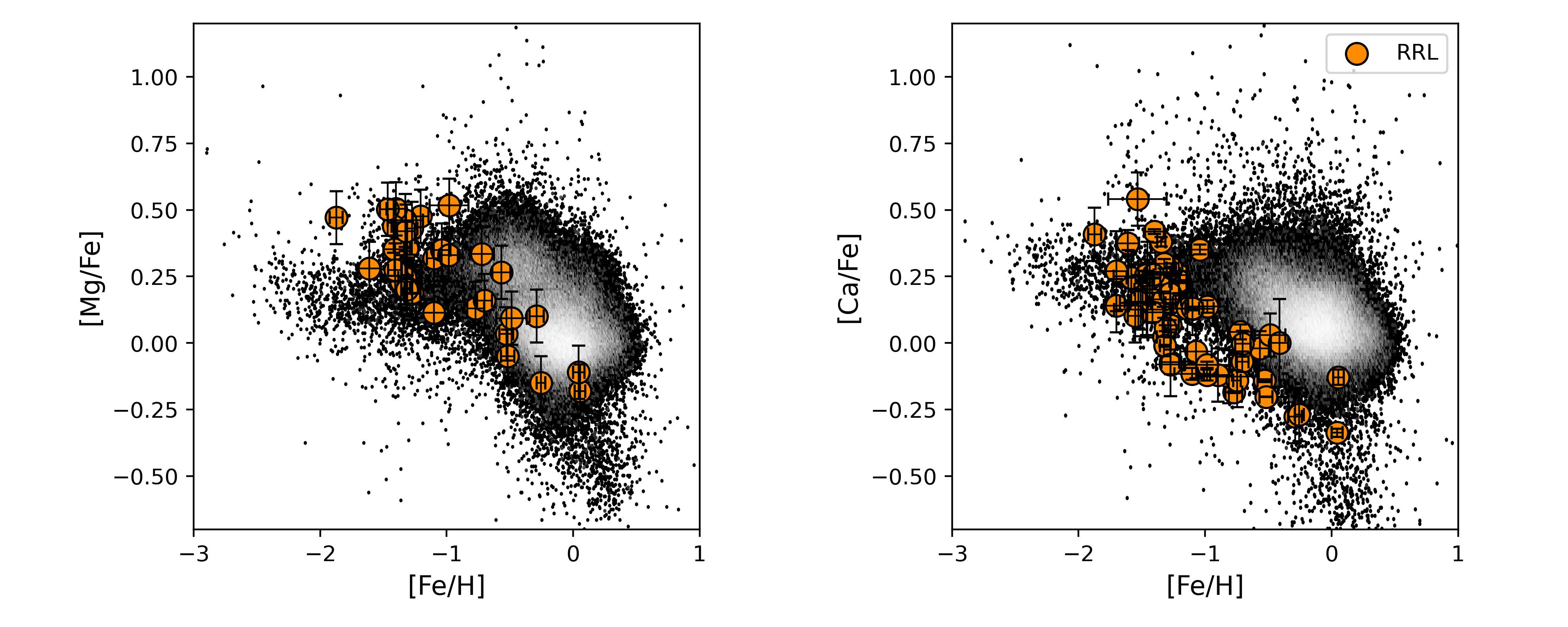}
\caption{[X/Fe] ratios of $\alpha$ elements Mg, Ca for our RRLs compared to stars in the GALAH survey.}
\label{fig:alpha}
\end{center}
\end{figure*}

Our sample star, DT Hya, is analysed in both \cite{marsakov2019} and \cite{preston2019}, with reported [Fe/H] values of $-1.23$ and $-1.22$, respectively. In comparison, our derived value is [Fe/H]=$-1.21 \pm 0.05$. Notably, these estimates are somewhat higher than the $-1.43$ reported by \citealt{chadid2017}, a value that aligns with the earlier result from \cite{for2011}. Considering the systematic uncertainties inherent to abundance analysis of warm metal-poor stars, an offset of 0.2 dex can be deemed reasonably acceptable.
Stellar parameters and abundances for our 78 RRLs are reported in Table \ref{tab:chemistry}, where we list atmospheric parameters, metallicity [Fe/H], [Mg/Fe], [Si/Fe], [Ca/Fe], [Y/Fe] and [Ba/Fe] along with the internal errors as given by the line-by-line scatter.
\begin{table*}
\centering
\caption{Atmospheric parameters and abundances for our sample RR Lyrae stars. This excerpt is for illustrative purposes; the full dataset can be accessed online via the CDS} \label{tab:chemistry}
\begin{tabular}{ccccrrrrrr}
\hline
\hline
GALAH$_{ID}$ & \teff & $\log g$ & $\xi_t$ & [Fe/H] & [Mg/Fe] & [Si/Fe] & [Ca/Fe] & [Y/Fe] & [Ba/Fe]\\
        & (K)  & (dex)  & (km s$^{-1}$) &   &  &  & &  & \\
\hline
140118002001313 & 5879$\pm$70 & 2.80$\pm$0.20 & 2.96$\pm$0.09 & $-1.45\pm0.09$ &  --- & --- & --- & --- & $-0.35\pm0.10$ \\
140312004501064 & 6030$\pm$50 & 2.75$\pm$0.18 & 2.77$\pm$0.19 & $-1.18\pm0.08$ & 0.43$\pm$0.10 & 0.44$\pm$0.09 & 0.30$\pm$0.07 & --- & --- \\
    ...     &   ...   &  ... &  ... &  ... & ...  & ...  & ...  &   ... & ...\\
\hline
\hline
\end{tabular}
\end{table*}
In Figure \ref{fig:alpha}, we compare abundance patterns for [Mg/Fe] and [Ca/Fe] for RRLs to stars in the GALAH survey. We adhered to the quality criteria outlined in \cite{buder2021} and conducted a selective filtering, ensuring that: \texttt{flag\_sp==0}, \texttt{flag\_fe\_h==0}, \texttt{flag\_alpha\_fe==0}, \texttt{snr\_c3\_iraf > 50}, \texttt{e\_fe\_h < 0.10}, and \texttt{e\_alpha\_fe < 0.10}. From an initial catalogue of 588,571 sources, we retained 161,952 stars (28\%). 
In our analysis, the derivation of Mg abundances was limited to a subset of 36 out of 78 stars. This limitation arose from the combination of the intrinsic faintness of the 5711 \AA\ line in the parameter range of our stars (\teff~and metallicity) and the relatively low SNR of the spectra. Conversely, Ca abundances were determined for 62 RR Lyrae stars, representing 78\% of our sample. Notably, the standard deviation from different lines for [Ca/Fe] ratios was relatively small, typically less than 0.1 dex, leading us to conclude that Ca is the most reliable tracer of $\alpha$-element abundances for the RR Lyrae stars discussed in this manuscript. 

\begin{figure}
\begin{center}
\includegraphics[width=0.4\textwidth]{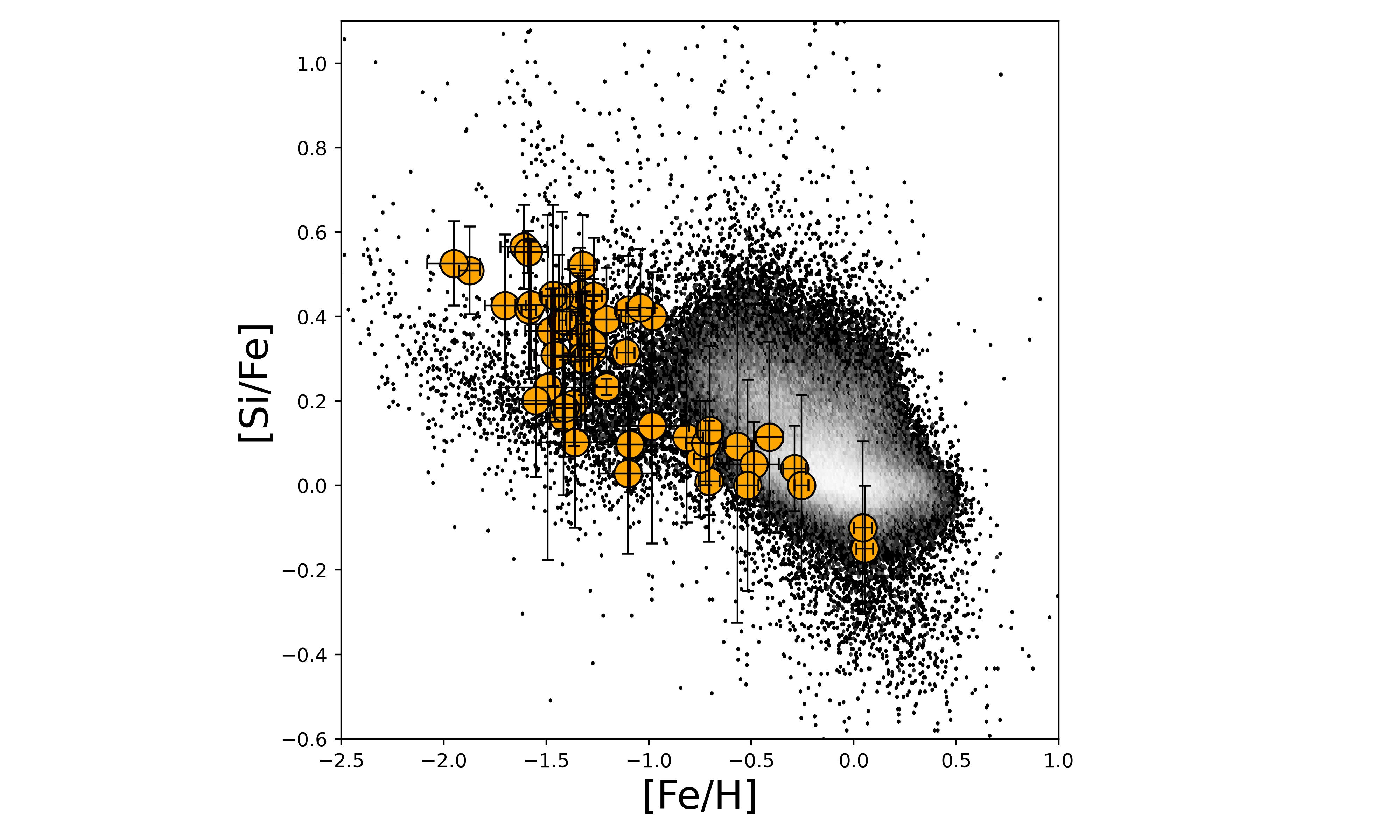}
\caption{[Si/Fe] ratios for our RRLs compared to stars in the GALAH survey.}
\label{fig:si_fe}
\end{center}
\end{figure}

Silicon requires distinct consideration, which is why we have dedicated a separate figure (Figure \ref{fig:si_fe}) to it, distinct from its counterparts Mg and Ca. As extensively discussed by \cite{for2011}, the anomalous behaviour of Si abundances in RR Lyrae stars presents a complex challenge. They noted that the NLTE corrections available at the time, as proposed by \cite{shi2009}, were insufficient to fully explain the observed spread. Specifically, their Figure 33 highlights the significant fluctuations in the [Si/Fe] ratios across different pulsational phases for two RR Lyrae stars.
Our work includes the more recent NLTE corrections from \cite{bergemann2013} and \cite{magg2022}. Despite this, we observed considerable internal scatter in the silicon abundances, reaching up to 1 dex for the most extreme cases. Consequently, we opted to retain only those lines yielding silicon abundances within $\pm$0.2 dex of the [Ca/Fe] values (typically only 1$-$2 lines per star). We advise readers to interpret silicon data cautiously; it should be seen merely as a supplementary confirmation of the general trends of $\alpha$-elements with metallicity, rather than as a standalone reliable indicator. For accurate comparisons, modellers should prioritize calcium over silicon when aligning their predictions to RR Lyrae stars, considering calcium as the more reliable tracer.

Figure \ref{fig:mdf} demonstrates that the bulk of our RR Lyrae stars is predominantly found in regions of the diagram with [Fe/H] below $-$1. These stars are significantly enriched in [Ca/Fe], aligning remarkably well with the distribution patterns of Halo stars as delineated by the GALAH survey. Their chemical content reflects the strong contribution of core-collapse supernovae (CC SNe) in the early evolution of the Milky Way, when nucleosynthesis is dominated by these $\alpha$-element producers (see e.g., \citealt{mashonkina2019} and references therein). 

\begin{figure*}
\begin{center}
\includegraphics[width=0.9\textwidth]{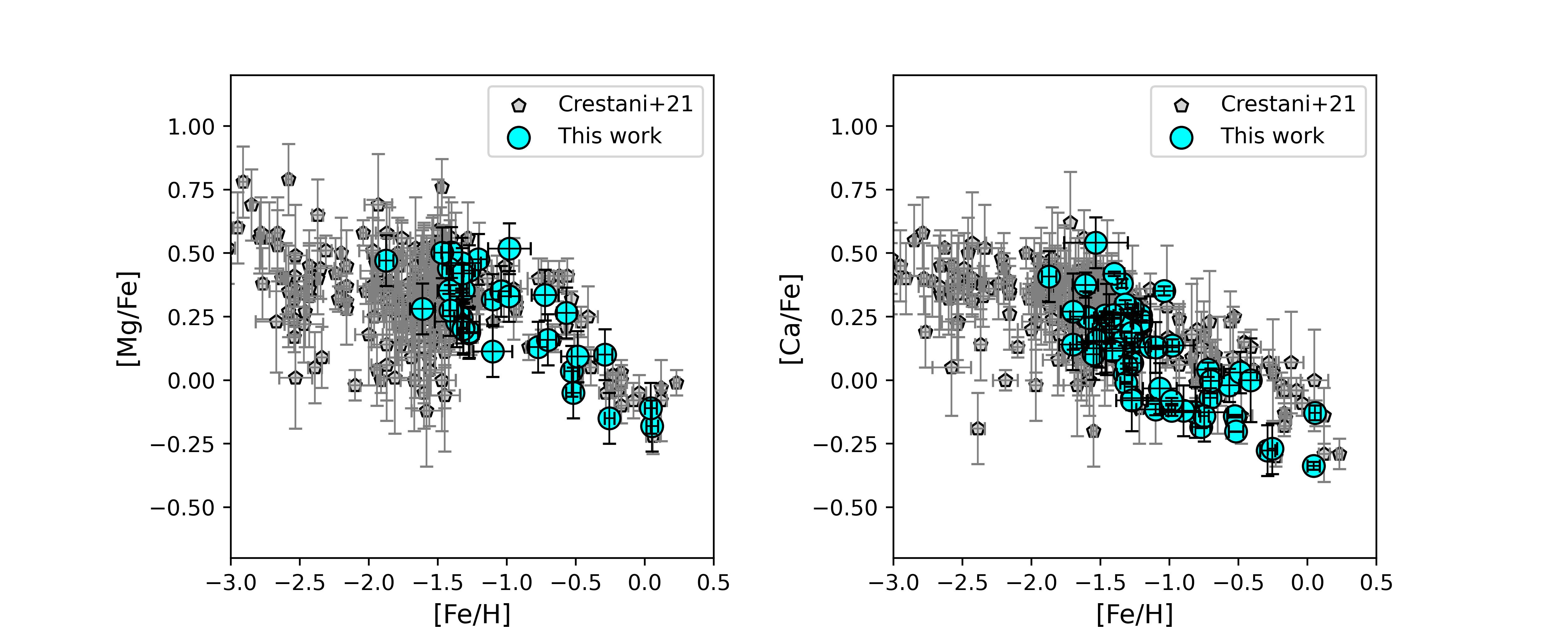}
\caption{Comparison of RRL in the present study with those analysed by \citet{crestani2021b}.}
\label{fig:crestani21}
\end{center}
\end{figure*}

Intriguingly, at a metallicity around $\approx-$1 dex, we observe the emergence of a distinct RRL population, characterised by sub-solar $\alpha$-element compositions. Here, calcium, benefitting from larger statistical representation and greater measurement precision, serves as a proxy for the trio of $\alpha$ elements. This detection of a metal-rich ([Fe/H] $> -$1) and $\alpha$-element poor ([Ca/Fe] between $\approx $0 and $-$0.35) subgroup within RRLs appears to be unique. These stars, exhibiting radial velocities consistent with thin disc kinematics, had been tentatively identified in smaller cohorts by \cite{prudil2020} and more extensively in larger samples by \cite{crestani2021b}. Our findings corroborate the existence of these 'anomalous' RRL stars, a topic currently debated in the field. In a recent study, \cite{iorio2021} posited a new hypothesis: the presence of young, metal-rich RRL stars in the thin disc challenges conventional, single-star evolutionary models, which would require implausibly high rates of mass loss for such metal-rich progenitors. They suggest an alternative scenario in which these stars might be RRL impostors (akin to Binary Evolution Pulsator -BEP), formed through mass transfer in binary systems.
However, our results are not in accord with this hypothesis, as binary evolution alone cannot account for the observed low calcium abundances in these stars. Conversely, these metal-rich RRLs look like a distinct population to the majority of thin-disc stars.
In Figure \ref{fig:crestani21}, we present a comparison of our estimates for [Mg/Fe] and [Ca/Fe] with the findings from \cite{crestani2021b} high-resolution study. It is important to note that Crestani et al. did not adjust their results for NLTE effects. Despite this, there is a general agreement in the trends observed within the RRL population. However, Crestani et al. data show notably more scatter in Mg, particularly for metallicities lower than [Fe/H] = $-$1.5. The cause of this increased scatter, whether it is an inherent feature or a result of methodological choices such as omitting NLTE corrections, will be the subject of future research. 

To further explore this metal-rich population, we determined the abundance of neutron-capture elements, specifically yttrium and barium. Elements heavier than iron are produced via neutron-capture processes: slow (s-process) and rapid (r-process). The dominant mechanism for r-process production, involving neutron star mergers, neutron-star–black hole mergers, and events like hypernovae, collapsars, 
and magnetorotational supernovae, remains debated \citep{chen2024, cowan2021}. The s-process occurs mainly in asymptotic giant branch (AGB) stars, synthesizing elements from strontium to lead through the $^{13}$C($\alpha$,n)$^{16}$O reaction \citep{karakas2014}, and in massive and low-metallicity, fast-rotating stars for elements \citep{pignatari2010, choplin2018, limongi2018}. In these environments, neutron source reactions primarily involve $^{22}$Ne($\alpha$,n)$^{25}$Mg (see \citealt{lugaro2023} and references therein). Studies confirm the s-process as the main source of yttrium and barium in the Solar System, contributing 76\% and 78\% to their formation, respectively, with the r-process accounting for the remainder \citep{busso2021}. Therefore, these elements are commonly classified as s-process elements, although it is important to note that several of the odd Ba isotopes are also produced via the $r$-process. Importantly, at metallicity lower than solar the contribution of the s-process and r-process to the production of Y and Ba might differ. 
For these elements, when the [Fe/H] is less than $\approx -$ 2, their formation is primarily ascribed to the r-process (see e.g., \citealt{tolstoy2008,reichert2020} and references therein), given that low-mass AGB stars evolve on longer timescales ($\approx$ Gyrs). 

\begin{figure}
\begin{center}
\includegraphics[width=0.5\textwidth]{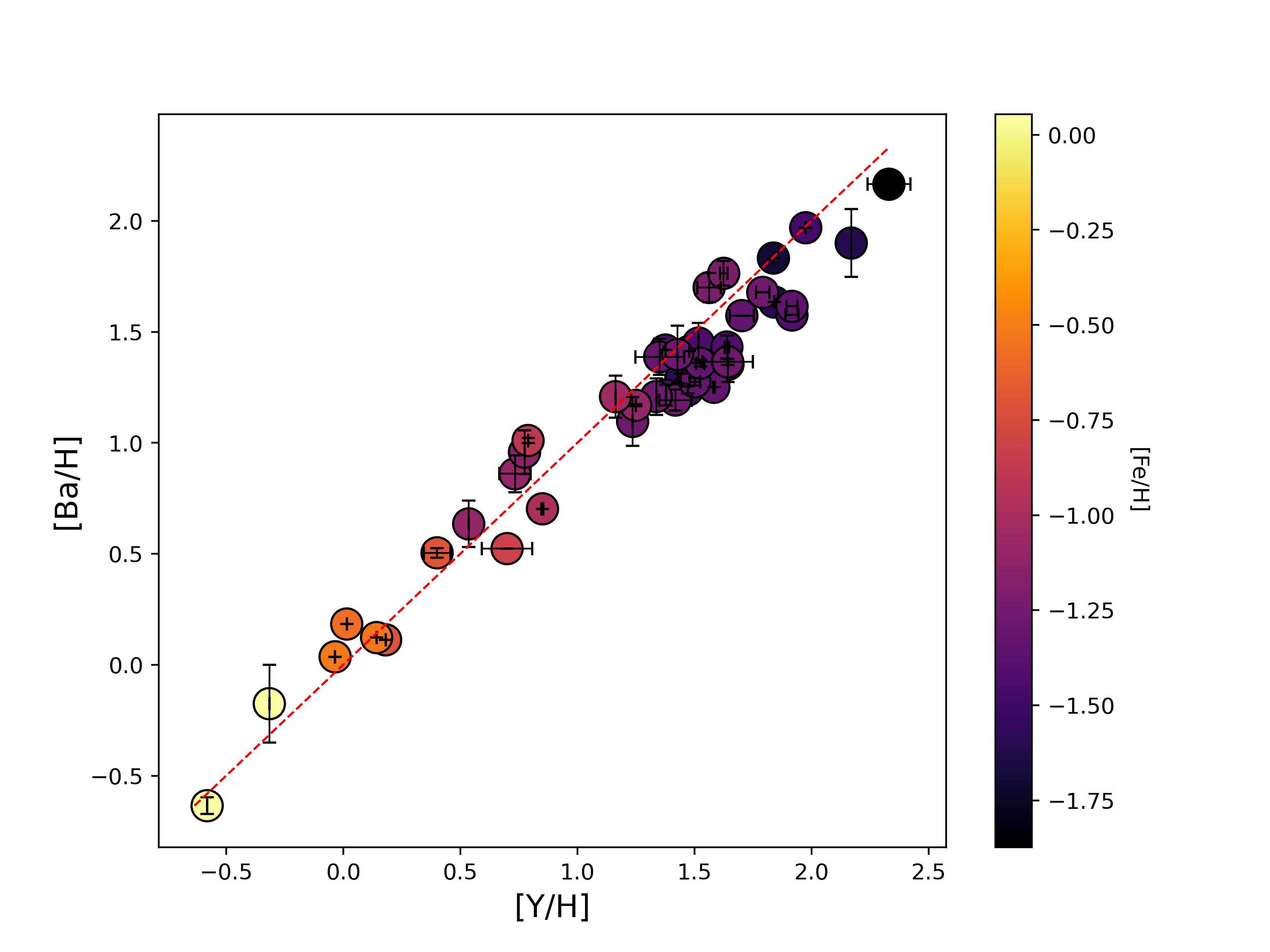}
\caption{Run of [Ba/H] with [Y/H] and the 1:1 correlation for our sample RRLs.}
\label{fig:n-catpure1}
\end{center}
\end{figure}

\begin{figure*}
\begin{center}
\includegraphics[width=0.75\textwidth]{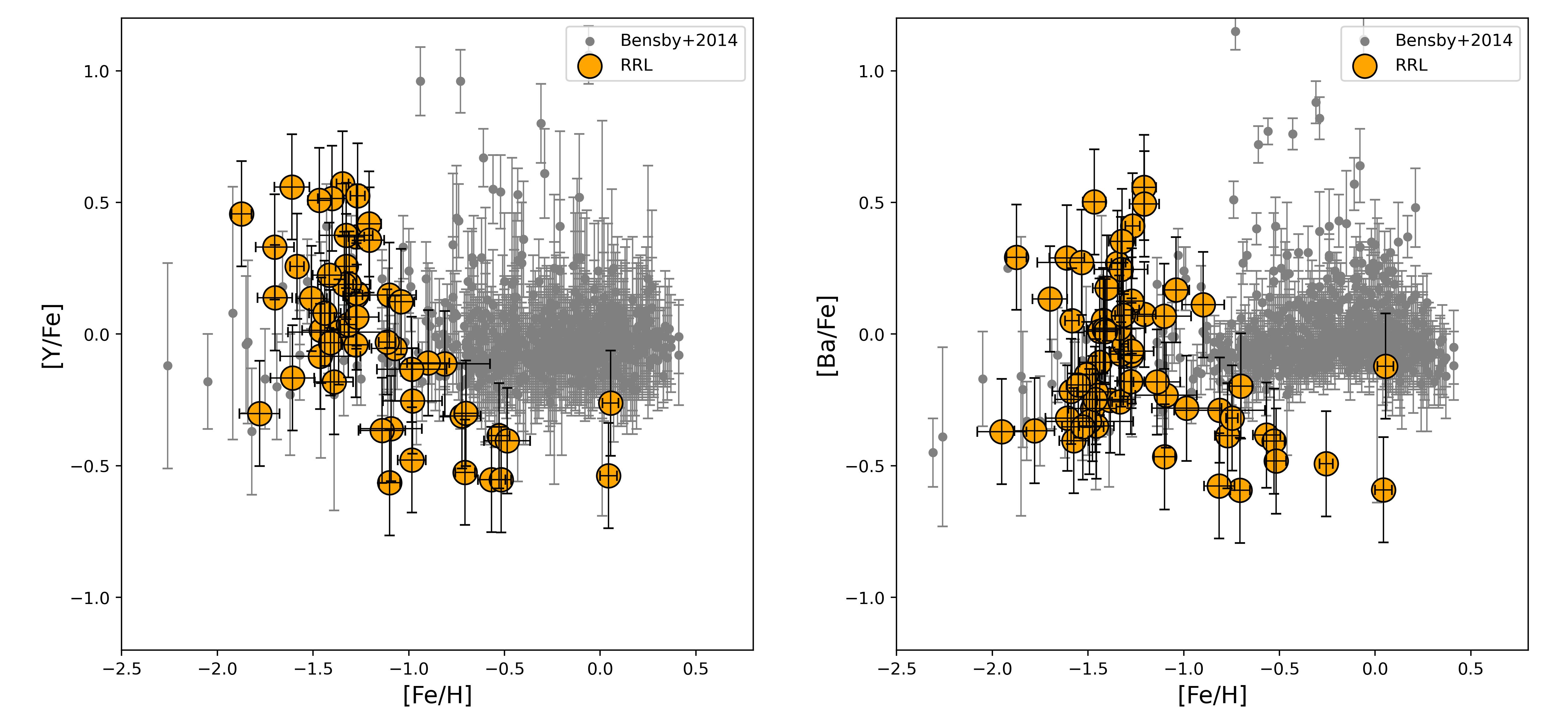}
\caption{[X/Fe] ratios of $n$-capture elements Y, Ba for our RRLs compared to  abundances in disc and halo stars by \citet{bensby2014}.}
\label{fig:n-capture_MW}
\end{center}
\end{figure*}

We derived [Y/Fe] ratios for 52 stars and [Ba/Fe] ratios for 63 stars, with 43 RRLs having measurements for both elements. The 1:1 correlation between [Ba/H] and [Y/H], as shown in Figure \ref{fig:n-catpure1}, lends strong credibility to our determined values. In Figure \ref{fig:n-capture_MW}, we present the trends of [Y/Fe] and [Ba/Fe] with metallicity, comparing our results with those from \cite{bensby2014}. We chose not to compare with GALAH DR3 due to the considerable scatter observed in these heavy-element abundances at each given metallicity bin. However, it is important to note, that the Bensby et al.'s abundances were not corrected for NLTE effects; despite this, the overall trend should remain credible (corrections for Y II and Ba II lines should remain within $\approx$0.2 dex). Our RRL sample exhibits a notable spread in Y or Ba abundances for metallicity below [Fe/H] $\lesssim -$1. 
Our study indicates that metal-rich RRL stars significantly differ from the typical thick/thin disc population, particularly in their lower levels of s-process elements. 

\begin{figure*}
\begin{center}
\includegraphics[width=0.9\textwidth]{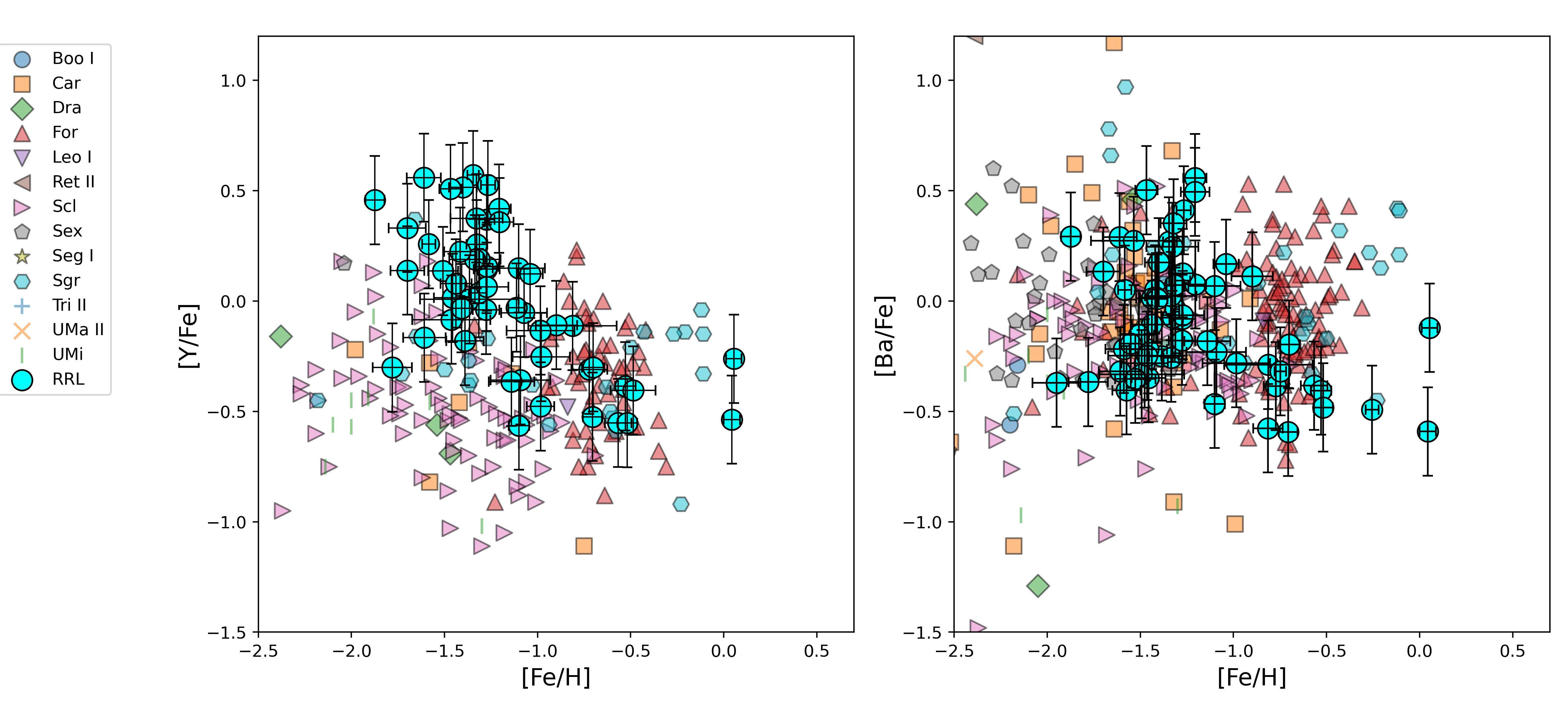}
\caption{[X/Fe] ratios of $n$-capture elements Y, Ba for our RRLs compared to abundances in the sample of dSph stars by  \citet{reichert2020}.}
\label{fig:n-capture_dSph}
\end{center}
\end{figure*}

In a comprehensive study of 380 stars across 13 dwarf spheroidal and ultra-faint galaxies, \cite{reichert2020} analysed the abundance of n-capture elements. Their work revealed that the emergence of the $s$-process (as indicated in the [Eu/Ba] vs [Ba/H] diagram, refer to their Figure 16) appears to start at higher metallicities in galaxies with greater stellar mass. Regardless of this overall pattern, there is significant variation in the levels of Ba and Y at any given metallicity. 
From Figure\ref{fig:n-capture_dSph}, we observe that at higher metallicities ([Fe/H] $\gtrsim -1$), the Ba abundance patterns in the Fornax and Sgr galaxies differ somewhat from those in our RRLs, despite considerable scatter. Conversely, the [Y/Fe] ratios in metal-rich RRLs are comparable to those in Fornax and Sgr galaxies. While our RRLs show solar or sub-solar [X/Fe] ratios for Y and Ba (with [heavy s-process/light s-process, hs/ls] $\approx$ 0), stars in Fornax and Sagittarius exhibit lower Y abundances relative to Ba. This pattern, where the second-peak element Ba is enhanced over the first-peak Y (i.e., a higher [hs/ls] ratio), suggests a significant contribution from low-metallicity AGBs.
\cite{minelli2021} conducted a comprehensive chemical analysis of 30 giant stars in the Large Magellanic Cloud (LMC), 14 in the Sagittarius dwarf galaxy, and 14 in the Milky Way. This study was based on high-resolution spectra obtained using the UVES-FLAMES spectrograph and aimed to homogeneously compare abundance patterns in the three different galaxies. 
The study identified a notable variation in Y abundances in stars from both Sgr and the Large LMC, with the latter exhibiting lower Y levels, similar to the findings in our metal-rich RR Lyrae stars. Contrarily, Zr does not show a deficiency, while barium is found to be significantly abundant. This results in a high [hs/ls] ratio, particularly in stars with a metallicity above -0.5. The study suggests two potential explanations for this trend of increasing [hs/ls] with metallicity: either the presence of comparatively metal-poor AGB stars in the LMC and Sagittarius relative to the Milky Way, or a reduced contribution from the most massive stars (see \citealt{minelli2021} for details).
Interestingly, considering all literature studies from 1995$-$2017, \cite{gozha2021} found that Y abundances in metal-rich RRLs are significantly sub-solar and different from Galactic stars, in line with our findings. In a recent publication, \cite{muccia2023a} reported higher levels of s-process elements like Zr, Ba, and La in the Small Magellanic Cloud (SMC) compared to the Milky Way. A subsequent study by the same team \citep{muccia2023b} analysed the metallicities and s-process elements in three SMC globular clusters: NGC 121 ([Fe/H] = -1.18), NGC 339 ([Fe/H] = -1.24), and NGC 419 ([Fe/H] = -0.58), revealing varied patterns of s-process elements. NGC 121 and NGC 339 showed sub- and super-solar [Y/Fe] and [Ba/Fe] ratios, respectively, while NGC 419, the youngest and most metal-rich, exhibited a mix of ratios, suggesting a potential influence of low-metallicity AGB stars in its chemical evolution. This highlights the distinct chemical characteristics of the SMC's stellar populations with respect to our RRLs. \cite{feuillet2022} analysed APOGEE DR17 data and identified an ancient, metal-rich, accreted component within the ``cool'' Galactic disc of the Milky Way. Alongside red giant stars, they also pinpointed a small group of RR Lyrae variables with disc kinematics that exhibit the same chemical signature as the accreted red giant stars found in the disc. In their study, they identified key indicators for differentiating accreted and in situ stars, specifically focusing on aluminium ([Al/Fe]) and the [Mn/Mg] ratio (as illustrated in their Figure 1). 
For our sample stars, we could not analyse Al and Mn due to limitations of low signal-to-noise ratios and insufficient spectral coverage/resolution (and Mg was inferred only for a small sub-sample because of the weakness of the 5711~\AA~line at the temperature/metallicity of our RRLs). Nonetheless, we examined the trend of cerium, the only s-process element available in APOGEE data. Regrettably, the data on cerium abundance in the accreted red giant population varied widely, showing a mix of both low and high cerium levels, which made it challenging to reach conclusive findings.



\section{DYNAMICS}\label{sec:kinematics}

We explore the dynamical properties of these stars by characterising their orbits. To integrate the orbits we use the code \texttt{AGAMA} \citep{vasiliev19}, and adopted the potential of the Galaxy by \cite{mcmillan17}. We assume the same position of the Sun in the Galaxy adopted by \citet{ceccarelli2024}: height of the Sun above the Galactic plane $z_{\odot} = 20.8$ pc \citep{bennett19}, distance of the Sun from the Galactic centre $R_{\odot} = 8.122$ kpc \citep{gravitycollaboration18}. The solar velocity has been obtained by combining the definition of the local standard of rest by \citet{schonrich2010} with the proper motion of Sgr $\mathrm{A^{\ast}}$ from \cite{reidandbrunthaler04}. In this frame, the 3D components of the Sun's velocity are ($U_{\odot}$, $V_{\odot}$, $W_{\odot}$) = (12.9, 245.6, 7.78) km $\mathrm{s^{-1}}$ \citep{drimmelandpoggio18}. We integrate the orbits for $3$ Gyr, starting from the current position and velocity of the stars. To estimate the uncertainty of the orbital parameters, for each star, we run 100 Monte Carlo simulations of the orbit, by assuming Gaussian errors in distance, proper motion and radial velocity. The final values of the orbital parameters are given by the median of the distributions, whereas the associated uncertainties correspond to their 16th and 84th percentiles.

This allows us to gather information on the areas of the Galaxy where these stars move. In particular, we compute the circularity of the orbit \citep[see e.g.,][]{massari19}, defined as the angular momentum $J_{\rm z}$ along the vertical $z$-axis, normalised by the angular momentum of a circular orbit with the same binding energy $E$:
\begin{equation}
 \lambda_{\rm z} = \frac{J_{\rm z}}{J_{\rm max}(E)} \ .
\end{equation}
Following \citet{Zhu2018} and \citet{Santucci2023}, we separate orbits into four different components: a cold component with near circular orbits (representing the thin disc), a warm component in between (representing the thick disc), a hot component with near radial orbits (representing the halo and/or the bulge),  and a retrograde component. We assign stars to each of these components according to their orbit circularity $\lambda_{\rm z}$: cold orbits are characterised by $\lambda_{\rm z} > 0.8$, warm orbits by $0.25 < \lambda_{\rm z} \leq 0.8$, hot orbits by $-0.25 < \lambda_{\rm z} \leq 0.25$, and retrograde orbits by $\lambda_{\rm z} \leq -0.25$. Implications of the dynamical properties of our sample RRLs are discussed in the next Section.

\section{Discussion}\label{sec:discussion}

\begin{figure*}
\centering
\includegraphics[width=0.45\textwidth]{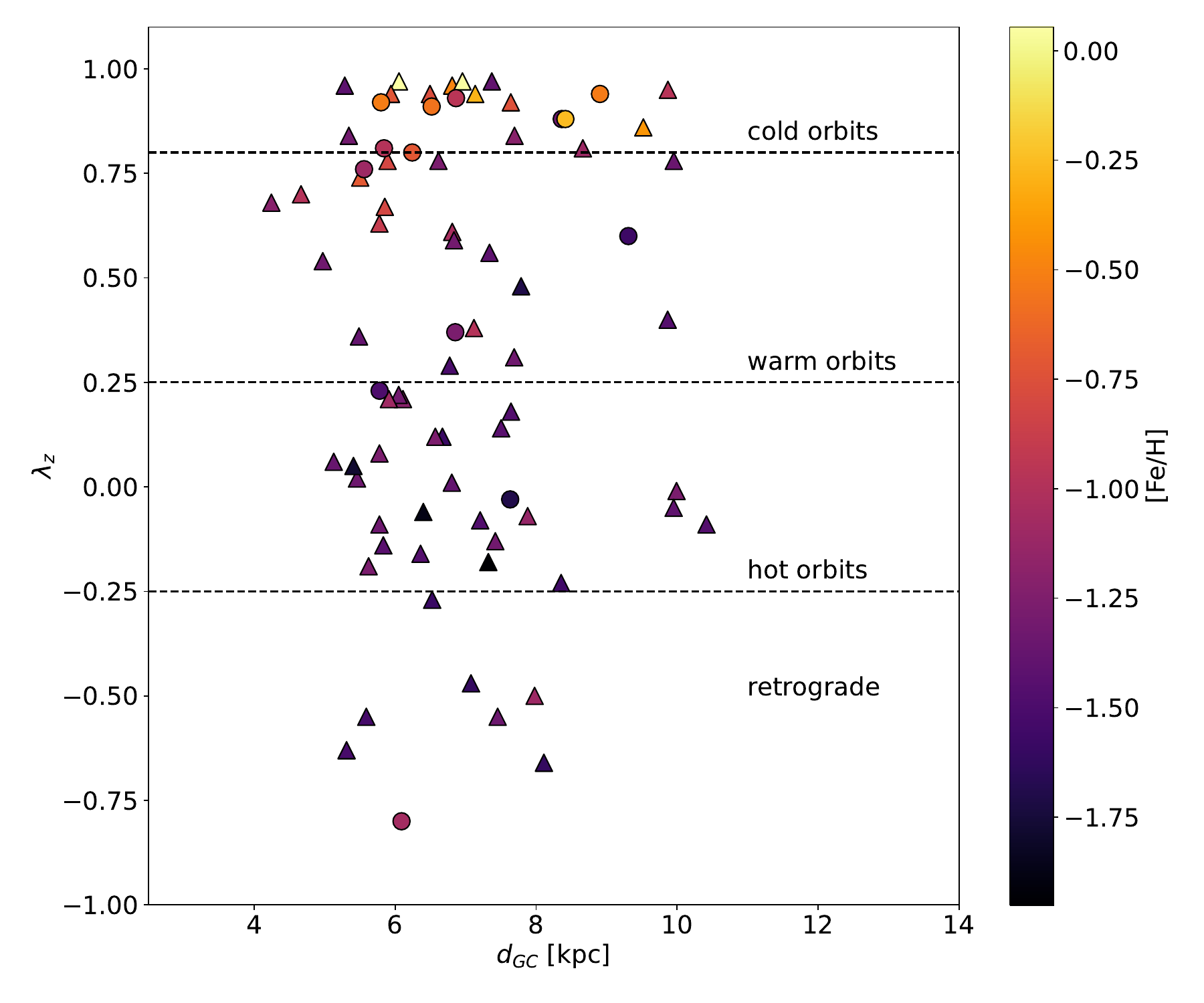}
\includegraphics[width=0.45\textwidth]{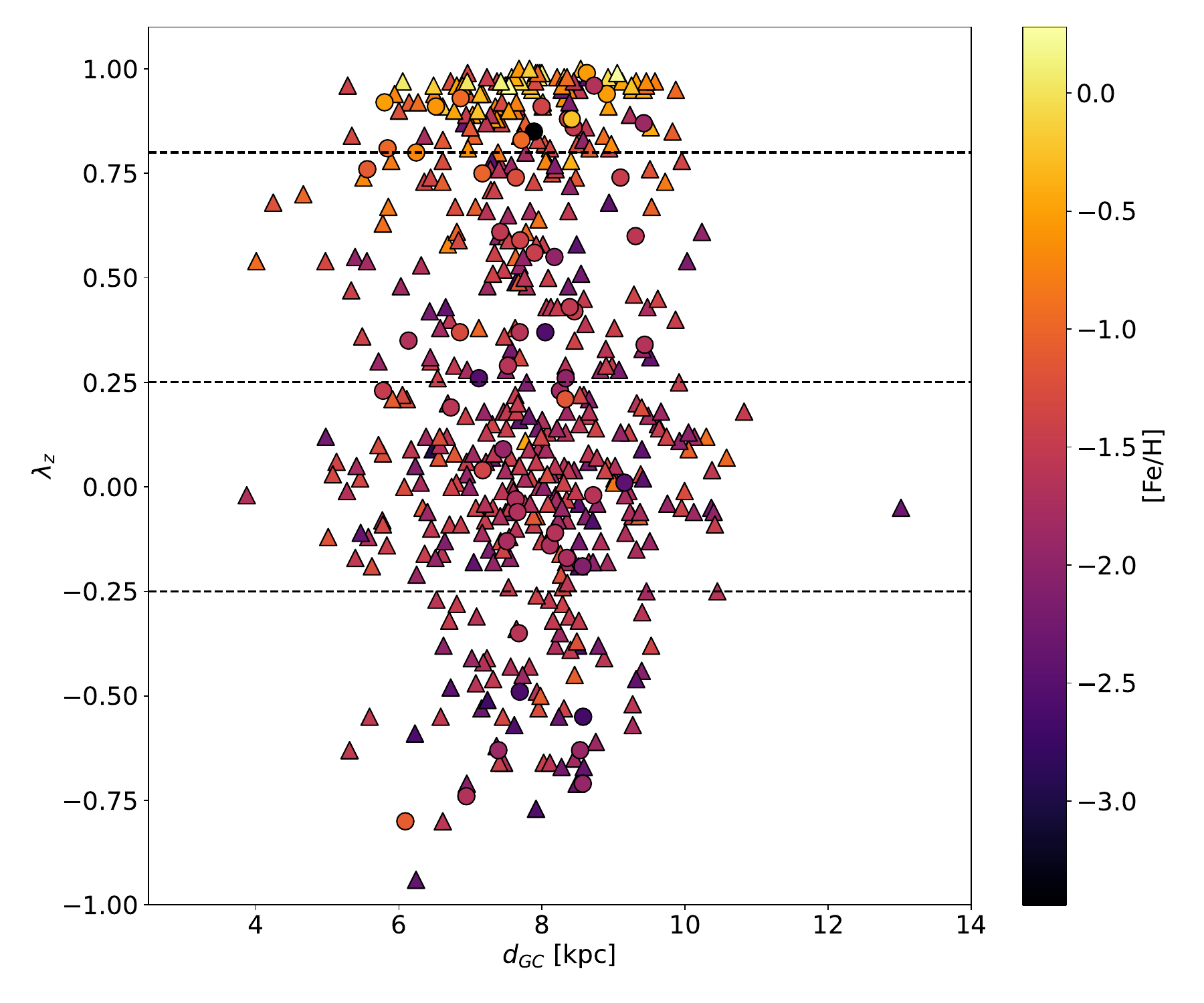}
\caption{Left-hand panel: The circularity $\lambda_{\rm z}$ of the orbits in our GALAH sample as a function of the Galactocentric distance ($d_{\rm GC}$). Filled circles display the position of RRab variables, while the triangles are RRc variables.  The symbols are colour-coded according to the iron abundance (see the colour bar on the right). Horizontal lines divide the sample into the various Galactic components introduced in Sect.~\ref{sec:kinematics}.
Right-hand panel: Circularity of the orbits but for the full sample of 535 RRLs.}
\label{fig:circFe_all}
\end{figure*}

The left-hand panel of Figure~\ref{fig:circFe_all} shows the circularity computed for the orbits of the RRLs in the GALAH sample, as a function of their Galactocentric distance ($d_{\rm GC}$). The horizontal lines separate the orbits in four kinematic bins and the symbols are colour-coded according to their iron abundance (see colour bar on the right). Data reported in this figure display several compelling features that warrant a closer examination and discussion.

{\em Retrograde RRLs} -- We discovered
seven out of the 78 RRLs to display retrograde orbits. 
The key pulsation parameters and chemical abundances are listed in Table~\ref{tab:counter-rot}. Among these, two exhibit metallicities around [Fe/H]$\approx -1$ and have [Ca/Fe] ratios similar to the Sun or slightly lower. The other five stars align closely with typical halo star characteristics, showing low metallicity and enhanced $\alpha$-elements. The neutron-capture elements yttrium and barium in these stars are nearly solar or just below, which does not significantly deviate from the general distribution. However, the large scatter in metallicity below $-$1 for Y and Ba adds complexity to understanding the patterns of these heavy elements.
We cannot conclusively determine whether these stars originated outside our Galaxy, even though their dynamical properties strongly support this interpretation. 
Interestingly, two of the metal-poor retrograde RR Lyrae stars, Gaia DR3 3892831505236595456 and 4737956618117144832, appear to be connected to the Gaia-Sausage-Enceladus structure, based on a match with the catalogue by \citet{helmi2018}.
\begin{table*}
\centering
\scriptsize
\caption{Properties of the retrograde RRLs.}
\label{tab:counter-rot}
\begin{tabular}{lcccccr}
\hline
\hline
RRL & P & Amp & [Fe/H] & [Ca/Fe] & [Y/Fe] & [Ba/Fe] \\
    & (days) & (mag) &  &  & & \\
\hline
    &   &   &   &   &   & \\
Gaia DR3 294072906063827072  & 0.390316 & 0.389 &  -1.90$\pm$  0.20 & \ldots            & \ldots            & \ldots            \\
Gaia DR3 315028326379733760  & 0.563123 & 1.123 &  -1.61$\pm$  0.20 &   0.28$\pm$  0.20 & \ldots            & \ldots            \\
Gaia DR3 593827920716707584  & 0.548512 & 0.869 &  -1.59$\pm$  0.20 & \ldots            & \ldots            & \ldots            \\
Gaia DR3 837077516695165824  & 0.349358 & 0.493 & \ldots            & \ldots            & \ldots            & \ldots            \\
Gaia DR3 1058066262817534336 & 0.660399 & 0.896 &  -1.83$\pm$  0.03 &   0.40$\pm$  0.20 & \ldots            & \ldots            \\
Gaia DR3 1167409941124817664 & 0.340828 & 0.387 &  -1.88$\pm$  0.11 &   0.44$\pm$  0.07 & \ldots            & \ldots            \\
Gaia DR3 1252055779366233344 & 0.609085 & 0.905 &  -1.53$\pm$  0.20 & \ldots            & \ldots            & \ldots            \\
Gaia DR3 1356103870372434560 & 0.630352 & 0.955 &  -2.01$\pm$  0.20 & \ldots            & \ldots            & \ldots            \\
Gaia DR3 1398766021041835648 & 0.469997 & 0.847 &  -1.64$\pm$  0.30 & \ldots            & \ldots            & \ldots            \\
Gaia DR3 1461194435841276288 & 0.648538 & 0.305 &  -1.27$\pm$  0.03 & \ldots            & \ldots            & \ldots            \\
Gaia DR3 1518851996671918208 & 0.558627 & 1.047 &  -1.06$\pm$  0.19 & \ldots            & \ldots            & \ldots            \\
Gaia DR3 1519628565414140160 & 0.668139 & 0.717 &  -2.59$\pm$  0.28 & \ldots            & \ldots            & \ldots            \\
Gaia DR3 1786827307055763968 & 0.547243 & 0.935 &  -1.55$\pm$  0.08 &   0.40$\pm$  0.06 & \ldots            & \ldots            \\
Gaia DR3 2022835523801236864 & 0.594130 & 0.960 &  -1.58$\pm$  0.20 & \ldots            & \ldots            & \ldots            \\
Gaia DR3 2150632997196029824 & 0.602674 & 1.194 &  -1.70$\pm$  0.20 &   0.23$\pm$  0.20 & \ldots            & \ldots            \\
Gaia DR3 2309225008197193856 & 0.784905 & 0.627 &  -2.31$\pm$  0.20 & \ldots            & \ldots            & \ldots            \\
Gaia DR3 2622375506154471680 & 0.469679 & 1.112 &  -1.35$\pm$  0.03 &   0.38$\pm$  0.02 &   0.57$\pm$  0.02 &   0.27$\pm$  0.10 \\
Gaia DR3 2737233514449351680 & 0.496716 & 1.219 &  -1.43$\pm$  0.19 & \ldots            & \ldots            & \ldots            \\
Gaia DR3 2808687029926911360 & 0.554537 & 0.880 &  -1.38$\pm$  0.19 & \ldots            & \ldots            & \ldots            \\
Gaia DR3 2817589255885467520 & 0.607354 & 1.239 &  -1.41$\pm$  0.19 & \ldots            & \ldots            & \ldots            \\
Gaia DR3 2852346261548304896 & 0.306493 & 0.449 &  -1.94$\pm$  0.19 & \ldots            & \ldots            & \ldots            \\
Gaia DR3 2853512332285518592 & 0.566067 & 0.795 &  -1.76$\pm$  0.19 & \ldots            & \ldots            & \ldots            \\
Gaia DR3 2973463347160718976 & 0.581474 & 1.005 &  -1.88$\pm$  0.28 &   0.39$\pm$  0.27 & \ldots            & \ldots            \\
Gaia DR3 3471095334863875456 & 0.778211 & 0.675 &  -1.61$\pm$  0.09 &   0.37$\pm$  0.05 &   0.56$\pm$  0.10 &   0.29$\pm$  0.15 \\
Gaia DR3 3486473757325180032 & 0.650309 & 0.467 &  -2.54$\pm$  0.08 & \ldots            & \ldots            & \ldots            \\
Gaia DR3 3632110703852863616 & 0.651558 & 0.984 &  -2.46$\pm$  0.31 & \ldots            & \ldots            & \ldots            \\
Gaia DR3 3677686044939929728 & 0.525756 & 0.938 &  -1.88$\pm$  0.00 &   0.34$\pm$  0.07 & \ldots            & \ldots            \\
Gaia DR3 3765574712337027456 & 0.537727 & 0.802 &  -1.49$\pm$  0.03 &   0.30$\pm$  0.06 & \ldots            & \ldots            \\
Gaia DR3 3892831505236595456 & 0.767711 & 0.462 &  -1.61$\pm$  0.11 &   0.25$\pm$  0.10 &  -0.17$\pm$  0.10 &  -0.32$\pm$  0.02 \\
Gaia DR3 3915944064285661312 & 0.598638 & 1.040 &  -1.52$\pm$  0.20 & \ldots            & \ldots            & \ldots            \\
Gaia DR3 3917248286939430144 & 0.726806 & 0.524 &  -2.36$\pm$  0.20 & \ldots            & \ldots            & \ldots            \\
Gaia DR3 4179431168210587648 & 0.356331 & 0.432 &  -1.07$\pm$  0.12 &  -0.03$\pm$  0.10 &  -0.05$\pm$  0.05 & \ldots            \\
Gaia DR3 4487364105534330112 & 0.603347 & 0.910 &  -1.58$\pm$  0.19 & \ldots            & \ldots            & \ldots            \\
Gaia DR3 4624119506369712384 & 0.583342 & 0.876 &  -2.57$\pm$  0.20 & \ldots            & \ldots            & \ldots            \\
Gaia DR3 4631934555845355136 & 0.555815 & 0.692 &  -2.07$\pm$  0.20 & \ldots            & \ldots            & \ldots            \\
Gaia DR3 4643606391466365824 & 0.675377 & 0.854 &  -1.72$\pm$  0.20 & \ldots            & \ldots            & \ldots            \\
Gaia DR3 4685757887726594816 & 0.732911 & 0.515 &  -2.70$\pm$  0.20 & \ldots            & \ldots            & \ldots            \\
Gaia DR3 4692528057537147136 & 0.405791 & 0.422 &  -1.60$\pm$  0.20 & \ldots            & \ldots            & \ldots            \\
Gaia DR3 4737725170919494912 & 0.662187 & 0.617 &  -2.39$\pm$  0.20 & \ldots            & \ldots            & \ldots            \\
Gaia DR3 4737956618117144832 & 0.572856 & 0.617 &  -1.10$\pm$  0.14 &  -0.12$\pm$  0.02 &   0.15$\pm$  0.10 &   0.07$\pm$  0.00 \\
Gaia DR3 4760456779256739968 & 0.649671 & 0.387 &  -1.87$\pm$  0.20 & \ldots            & \ldots            & \ldots            \\
Gaia DR3 4788620567737382144 & 0.613053 & 0.500 &  -1.50$\pm$  0.20 & \ldots            & \ldots            & \ldots            \\
Gaia DR3 4818854972838127360 & 0.478862 & 0.795 &  -1.22$\pm$  0.03 &   0.30$\pm$  0.06 & \ldots            & \ldots            \\
Gaia DR3 4854350575436127488 & 0.729404 & 0.724 &  -2.06$\pm$  0.20 & \ldots            & \ldots            & \ldots            \\
Gaia DR3 5011834347435737344 & 0.637068 & 0.763 &  -2.33$\pm$  0.20 & \ldots            & \ldots            & \ldots            \\
Gaia DR3 5094203642556959744 & 0.608816 & 0.945 &  -2.42$\pm$  0.20 & \ldots            & \ldots            & \ldots            \\
Gaia DR3 5281881584407284352 & 0.572398 & 0.862 &  -1.51$\pm$  0.20 & \ldots            & \ldots            & \ldots            \\
Gaia DR3 5555745531172521344 & 0.797322 & 0.171 &  -2.24$\pm$  0.20 & \ldots            & \ldots            & \ldots            \\
Gaia DR3 5638928606644888448 & 0.552546 & 0.590 &  -1.57$\pm$  0.20 & \ldots            & \ldots            & \ldots            \\
Gaia DR3 5696434679682159872 & 0.285668 & 0.220 &  -2.67$\pm$  0.15 &   0.37$\pm$  0.20 & \ldots            & \ldots            \\
Gaia DR3 5743059538967112576 & 0.537228 & 0.751 &  -2.17$\pm$  0.03 &   0.35$\pm$  0.06 & \ldots            & \ldots            \\
Gaia DR3 5909759314361595520 & 0.710373 & 0.801 &  -2.39$\pm$  0.20 & \ldots            & \ldots            & \ldots            \\
Gaia DR3 6120897123486850944 & 0.493990 & 0.789 &  -1.53$\pm$  0.03 &   0.20$\pm$  0.06 & \ldots            & \ldots            \\
Gaia DR3 6341917480568293120 & 0.457998 & 0.825 &  -1.66$\pm$  0.20 & \ldots            & \ldots            & \ldots            \\
Gaia DR3 6345324587928571648 & 0.621858 & 0.899 &  -1.66$\pm$  0.20 & \ldots            & \ldots            & \ldots            \\
Gaia DR3 6380659528686603008 & 0.550081 & 0.995 &  -1.86$\pm$  0.01 &   0.31$\pm$  0.01 & \ldots            & \ldots            \\
Gaia DR3 6428374141448297856 & 0.527141 & 1.050 &  -1.53$\pm$  0.20 & \ldots            & \ldots            & \ldots            \\
Gaia DR3 6434640155133754368 & 0.719611 & 0.879 &  -1.53$\pm$  0.23 &   0.54$\pm$  0.10 & \ldots            &   0.27$\pm$  0.10 \\
Gaia DR3 6502711401043594496 & 0.508403 & 1.004 &  -1.99$\pm$  0.20 & \ldots            & \ldots            & \ldots            \\
Gaia DR3 6519995861275291008 & 0.550309 & 0.928 &  -2.07$\pm$  0.20 & \ldots            & \ldots            & \ldots            \\
Gaia DR3 6541769554459131648 & 0.281110 & 0.238 &  -1.66$\pm$  0.09 &   0.39$\pm$  0.20 & \ldots            & \ldots            \\
Gaia DR3 6625215584995450624 & 0.546740 & 1.048 &  -1.61$\pm$  0.18 &   0.39$\pm$  0.13 & \ldots            & \ldots            \\
Gaia DR3 6662886605712648832 & 0.316907 & 0.436 &  -2.60$\pm$  0.18 &   0.05$\pm$  0.19 & \ldots            & \ldots            \\
Gaia DR3 6673750914465028096 & 0.683173 & 0.793 &  -2.45$\pm$  0.20 & \ldots            & \ldots            & \ldots            \\
Gaia DR3 6749657142798141184 & 0.639409 & 0.718 &  -2.60$\pm$  0.20 & \ldots            & \ldots            & \ldots            \\
Gaia DR3 6795546531894178816 & 0.632184 & 0.595 &  -1.53$\pm$  0.11 &   0.15$\pm$  0.10 & \ldots            &  -0.35$\pm$  0.10 \\
Gaia DR3 6796320308904070016 & 0.544448 & 0.986 &  -1.49$\pm$  0.04 &   0.31$\pm$  0.04 & \ldots            & \ldots            \\
Gaia DR3 6883653108749373568 & 0.447745 & 1.090 &  -1.45$\pm$  0.16 &   0.10$\pm$  0.14 & \ldots            & \ldots            \\
\hline
\hline
\end{tabular}
\end{table*}    

{\em Metal-rich RRLs} -- The bulk of metal-rich RRLs are characterised by cold orbits. Data 
plotted in the left-hand panel in Figure~\ref{fig:circFe_all} shows that almost one-third (23/78) of the sample stars has a $\lambda_{\rm z}$ larger than $\sim$0.8. This preliminary evidence indicates that metal-rich RRLs are far from being rare objects. Their origin should follow an evolutionary channel similar to the RRLs in the metal-intermediate and the metal-poor regime \citep{bono1997,marconi2018}. The absence of these objects in prior spectroscopic \citep{preston1959, walker1991, sneden2018} and photometric \citep{Pietrukowicz2020, mullen2021, mullen2022} studies primarily stems from an observational bias. These investigations were predominantly focused on the Galactic halo, leading to a preference for detecting objects within that region. The present sample, while still confined to a narrow range of Galactocentric distances, includes disc RRLs, thereby providing a broader and more representative sample of the metallicity distribution extremes.
Moreover, there is mounting evidence that metal-rich RRLs are also 
$\alpha$-poor. This evidence was originally brought forward by \cite{magurno2018,magurno2019} 
using a mix of high-resolution spectra for cluster RRLs and similar measurements for 
field RRLs available in the literature.  A similar conclusion was also reached by 
\cite{marsakov2018} by collecting measurements available in the literature. However, 
solid constraints on the $\alpha$-element abundance of metal-rich RRLs were 
only reached with the homogeneous and detailed spectroscopic investigations by \cite{crestani2021a,crestani2021b} based on high-resolution spectra in which metal-rich RRLs approach solar $\alpha$ abundances and in the very metal-rich regime they exhibit sub-solar $\alpha$-elements ratios ([$\alpha$/Fe]$<$~0). 
Arguments grounded in physical principles, including kinematic evidence from cold orbits, iron content consistent with solar iron abundance as reported by \cite{fabrizio2021}, and the ratios of $\alpha$-element to iron abundances ([$\alpha$/Fe] $\approx$ 0, as shown in Figures 7, 8, 9), compellingly indicate that metal-rich RR Lyrae stars share characteristics akin to those of a disc population. Nonetheless, when comparing metal-rich RRLs with non-variable disc stars at similar metallicity levels, we find that RRLs have slightly lower abundances of $\alpha$-elements. Further, based the abundance of the two $s$-process elements Y and Ba are systematically lower than in typical disc stars at the same iron abundance. 
This pattern implies that metal-rich RRLs might not be a 
conventional ``\emph{disc population}''. Instead, they could represent 
either a ``drifting disc population'' or an ``ancient native disc population''. 
The nature of the ``drifting disc population'' is currently unclear, 
being debated as either a purely accreted component, termed ``aliens'' 
by \citet{zinn2020}, or as ``commuters'' as identified by 
\cite{snaith2016,monachesi2019}. These commuters are objects formed 
in situ within the Milky Way’s virial radius but from gas imported by a 
massive satellite.

{\em Metal-intermediate RRLs} -- RRLs characterised by warm orbits roughly 
include 36 \% of the sample (28/78) and display smooth transition when moving 
from RRLs with hot and cold orbits. 

{\em Metal-poor RRLs} -- RRLs with hot orbits are characterised by metal-poor 
iron abundances. The iron abundances of the retrograde component are very 
similar to the RRLs with hot orbits; although, they are considered a typical accreted component their metallicity distribution is similar to Halo RRLs. 


We should note that our sample size is somewhat restricted, limiting our ability to draw definitive conclusions about the evolutionary, pulsation, and kinematic properties of field RRLs. Thus, we have chosen to utilize the RRL sample for which previously determined iron and $\alpha$-element abundances are reliable. These determinations were made using high-resolution, high signal-to-noise ratio spectra and LTE approximations \citep{magurno2019,crestani2021a,crestani2021b}. 

Additionally, we incorporated RRLs with consistent iron abundances and radial velocity (RV) data sourced from existing studies \citep{zinn2020}. This resulted in a spectroscopic catalog of 535 RRLs. The right-hand panel of Figure~\ref{fig:circFe_all} illustrates the circularity of the complete RRL sample in relation to their distance from the Galactic center. The observations presented in this figure corroborate the findings derived from the GALAH sample. Key observations are elaborated on below:

{\em Retrograde RRLs} -- The sample size of field RRLs with retrograde orbits is still of the order of 8\% (44/535) and their iron abundance is either metal-poor or metal-intermediate (below $\approx -1$)

{\em Metal-rich RRLs} -- The metal-rich RRLs include a significant 
fraction (17\% more metal-rich than --1 dex)  of the entire RRL sample. 
The sharp edge for Galactocentric distance smaller than 5~kpc 
and larger than 11~kpc is a consequence of the biases associated with 
the limiting magnitude and the large extinction close to the 
Galactic plane.  

\begin{figure}
\centering
\includegraphics[width=0.45\textwidth]{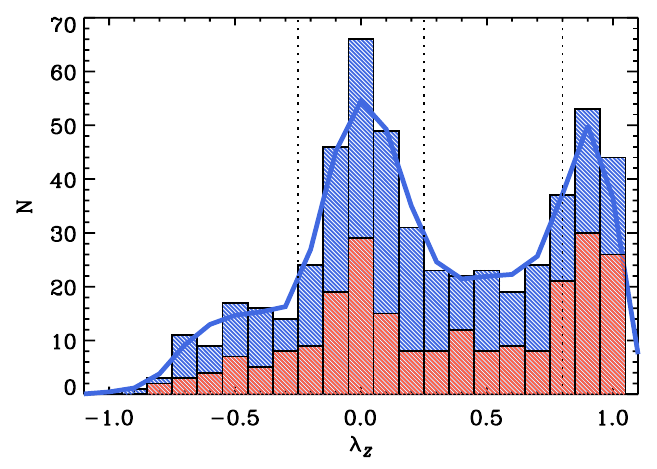}
\caption{Orbit circularity ($\lambda_{\rm z}$) distribution  for the entire RRL sample. 
The red hatched area shows RRLs with iron abundances based on HR spectra, while 
the blue hatched area the cumulative iron distribution function (HR+LR). The blue 
solid line displays the running average over the entire sample, while the vertical dashed lines are the boundaries of the four kinematic groups.}\label{fig:histo_circularity}
\end{figure}

Figure~\ref{fig:histo_circularity} shows the distribution of the orbit circularity 
for the entire RRL sample. The red and the blue hatched areas display the orbit 
circularities based on high-resolution (HR) spectra and on the entire 
(HR+LR [low resolution]) spectroscopic sample. 
The blue solid line shows the smoothed distribution estimated by using a Gaussian kernel with unitary weight and $\sigma$ equal to the error of the individual estimates.
The vertical dashed lines display the four kinematics groups defined in 
Figure~\ref{fig:circFe_all}. Data plotted in this figure display two well-defined 
peaks associated with RRLs with hot ($-0.25 < \lambda_{\rm z} \leq 0.25$) and 
cold ($\lambda_{\rm z} > 0.8$) orbits. 
The current findings suggest that the latter sample is one of the main components of field RRLs and not a minority 
group. This implies a similar evolutionary channel. Moreover, the transition from disc-like to halo/bulge-like RRLs seems to be quite smooth with a well-defined minimum for $\lambda_{\rm z} \sim 0.5$. These are the typical orbit circularities of thick disc stars. Note that 
the current data seem to indicate a smooth transition between thin and thick disc RRLs, thus suggesting that two different disc components formed at similar ages. 
This finding fully supports the results obtained by \citet{beraldoesilva2021}, 
and more recently by \citet{wu2023} using main-sequence turn off field stars.  
Finally, let us mention that RRLs with retrograde orbits 
appear to be an addendum to the long tail of the RRLs with hot orbits. Indeed the smoothed distribution shows a sharp change in the slope for $\lambda_{\rm z}$ between $\sim -0.25$ and $-0.30$. 

\begin{figure*}
\centering
\includegraphics[width=0.75\textwidth]{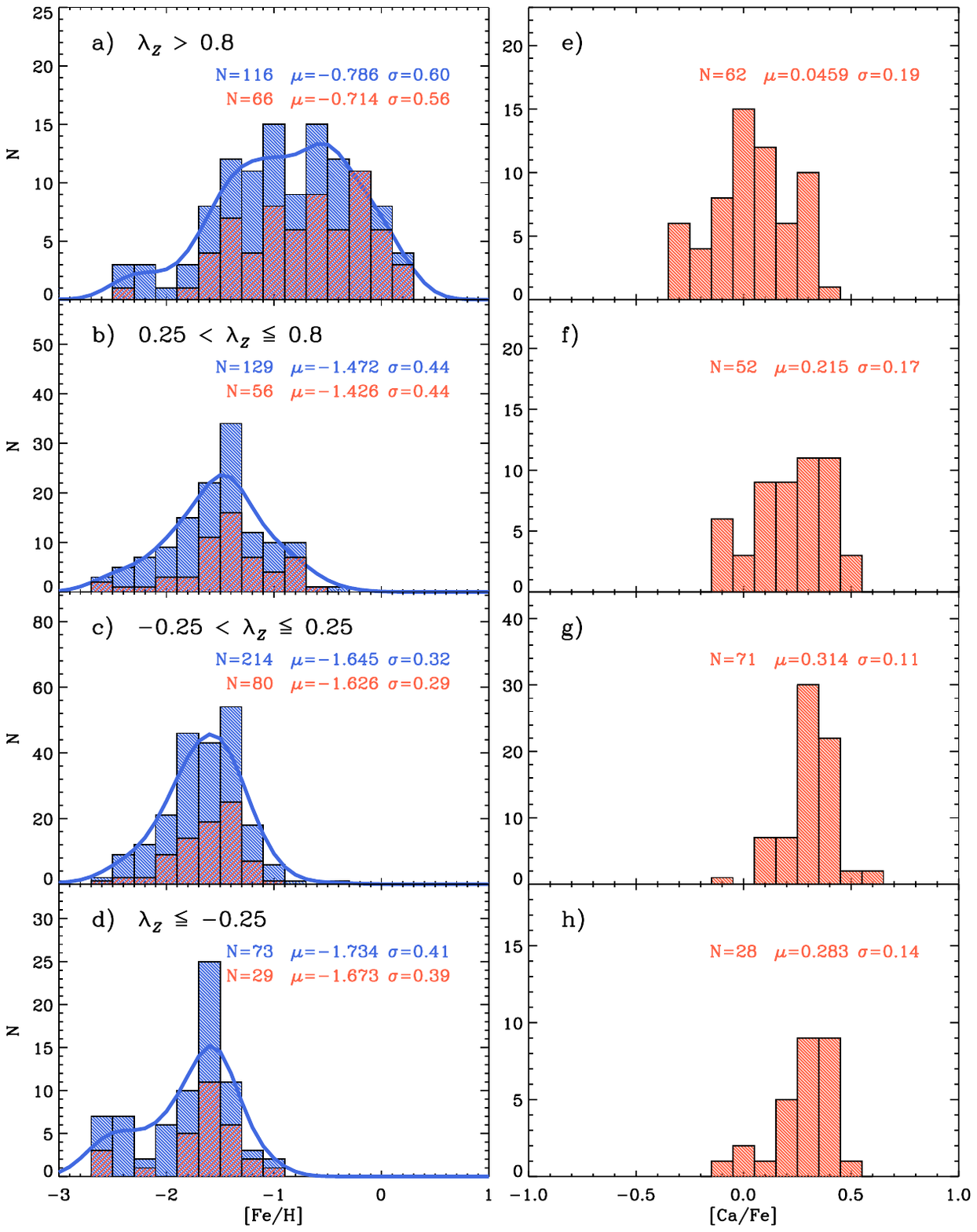}
\caption{Left -- From top to bottom iron distribution function of the entire RRL sample adopted in 
Figure~\ref{fig:circFe_all} for the four kinematic bins defined in Section~\ref{sec:kinematics}. The red hatched 
area shows RRLs with iron abundances based on HR spectra, while the blue hatched area the cumulative iron 
distribution function (HR+LR). The solid blue line displays the running average over the entire sample. The latter was smoothed using a Gaussian kernel with unitary weight and $\sigma$ equal to the error of the individual estimates. 
Right -- Same as the left, but for the [Ca/Fe] distribution function.} \label{fig:Fe_alpha_dist}
\end{figure*}

To investigate on a more quantitative basis the difference in metallicity distribution among the four different kinematic bins, the left panels of Figure~\ref{fig:Fe_alpha_dist} show the iron distribution function based on HR spectra (red hatched area) and entire 
(HR+LR) spectroscopic sample (blue hatched area). The blue solid line shows the smoothed 
iron distribution function. The latter was smoothed using a Gaussian kernel with 
unitary weight and $\sigma$ equal to the error of the individual estimates. The mean and 
standard deviation of the smoothed distributions are also labelled. 
%
%
The iron distribution functions plotted in the left panels show that the 
main peaks agree, within the errors, with the peak in iron abundance for field RRLs found by \citet{fabrizio2021} (Fe/H]=$-$1.26, $\sigma$=0.60 (535 stars) versus [Fe/H]=$-$1.51, $\sigma$=0.41 (9015 stars) using the largest sample of LR spectra and the $\Delta$S method. 

The key difference in the iron distribution function is that the spread of iron abundance for RRLs with cold orbits is systematically larger than for the other kinematic bins. 
This subsample shows an almost flat iron distribution function while the others display an extended metal-poor tail and a sharp decrease in the metal-rich tail as already found by \citet{fabrizio2021}.  Note that this spread in iron abundance is severely 
underestimated. Preliminary evidence based on HR optical spectra indicates that the RRL with disc kinematics recently discovered by \citet{matsunaga2023} is more metal-poor than [Fe/H]$\approx-$3 (D'Orazi et al. 2024, in preparation). The distributions of  [Ca/Fe] for the RRL with HR spectra plotted in the right panels 
of Figure~\ref{fig:Fe_alpha_dist} further strengthen this evidence. The spread in Ca abundance for RRLs with cold orbits is systematically larger when compared with the other 
kinematic components.

\begin{figure*}
\centering
\includegraphics[width=0.48\textwidth]{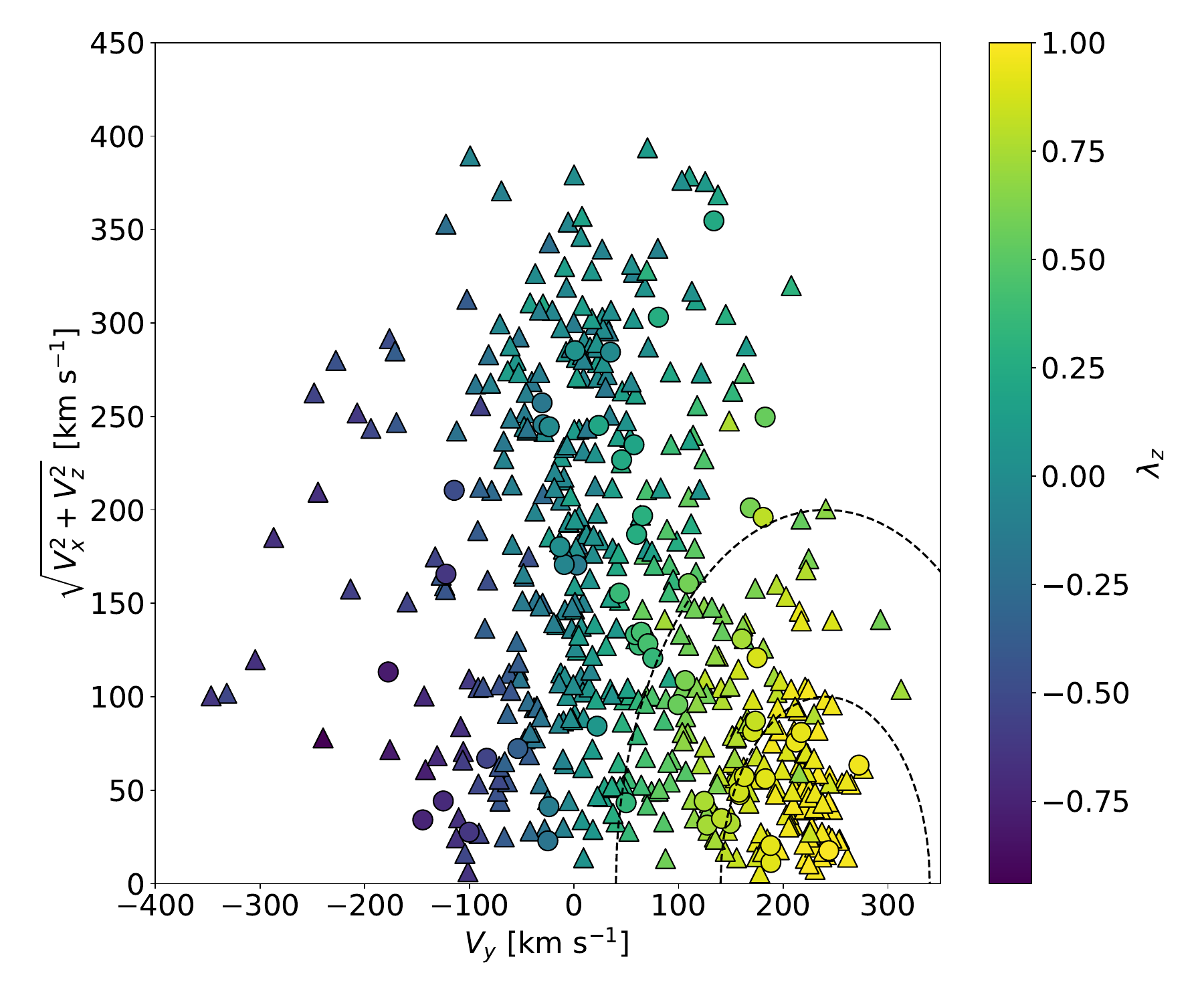}
\includegraphics[width=0.48\textwidth]{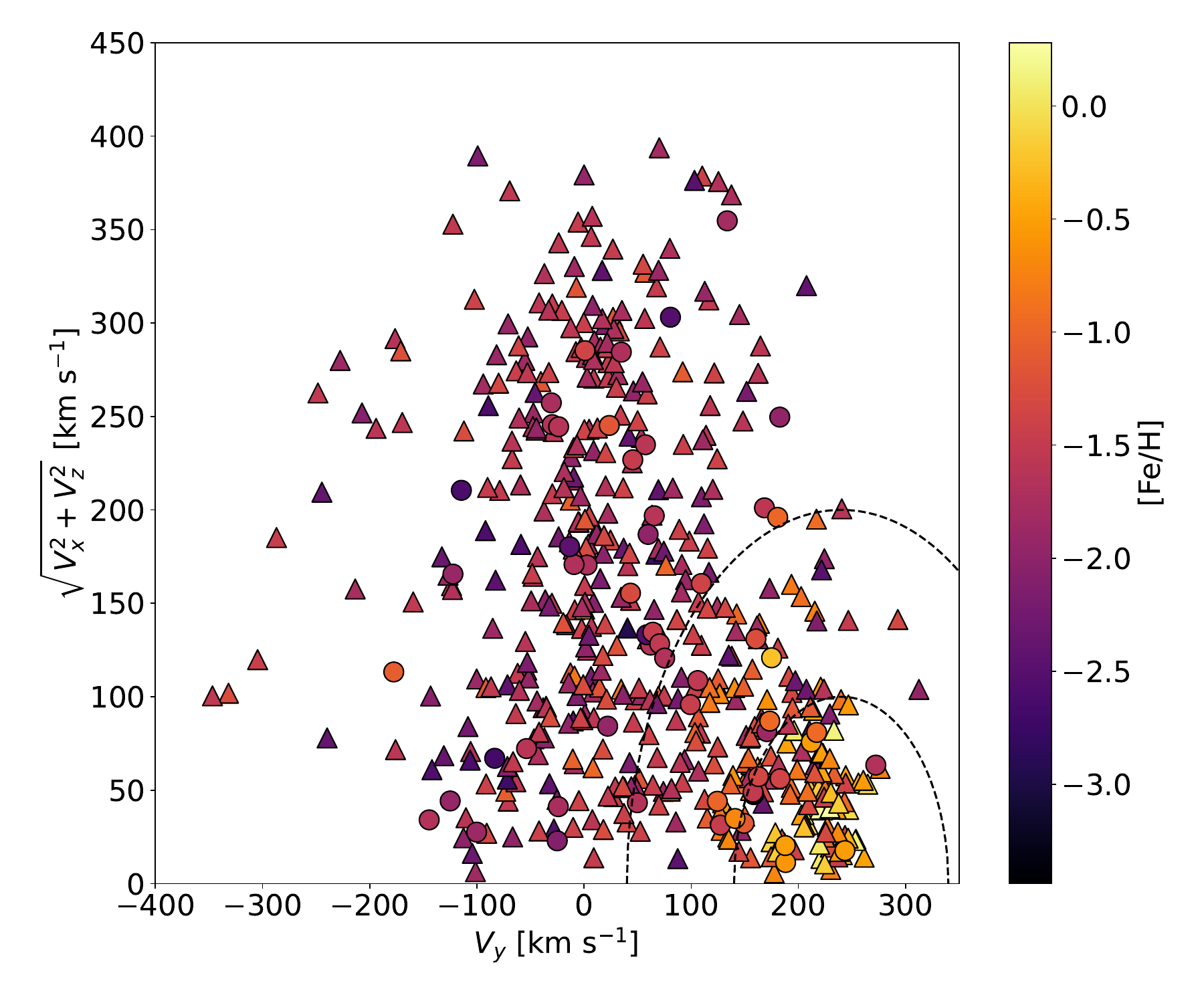}
\caption{Left -- Toomre diagram for the entire RRL sample. The symbols are colour-coded 
according to the circularity of the orbits (see colour bar on the right). 
The dashed semi-circles display the regions of the Toomre diagram of thin 
(V$y$ < 50 km s$^{-1}$) and thick disc (50 < V$y$ < 100 km s$^{-1}$) stars 
according to the definition by \citet{helmi2018}. 
Right: Same as the left, but the symbols are colour-coded according to the 
metallicity [Fe/H].}\label{fig:toomre_circ_iron}
\end{figure*}

Figure~\ref{fig:toomre_circ_iron} shows the Toomre diagram for the RRLs in our sample, where the 
combination of the velocity in the Galactic centre direction ($V_x$), and the velocity component 
perpendicular to the disc plane ($V_z$) is shown as a function of the velocity of stars in the direction of the Galactic rotation, $V_y$. In the left-hand panel, points are colour-coded according to 
the circularity of the corresponding orbit, while in the right-hand panel, the colours indicate iron abundance. Data plotted in this figure soundly support the results based on the circularity of the orbits
plotted in Figure~\ref{fig:circFe_all}. Our sample RRLs shows a clear gradient, becoming progressively more metal-rich and associated with cooler orbits as they transition from the halo through the thick disc to the thin disc regions. For visual clarity in distinguishing the different sub-samples, Figure~\ref{fig:toomre_circ_iron} includes semicircles at V$_{LSR}$=100 and 200 km~$s^{-1}$. These semicircles, as defined by \citet{helmi2018}, demarcate the boundaries between the thin/thick disc and the halo.

\subsection{Kinematic selection of candidate stellar streams}

To identify RRLs belonging to possible stellar streams, we adopted two solid diagnostics: the angular momentum and the total energy. There is vast literature concerning the use of these diagnostics to identify stellar streams in the solar vicinity. Dating back 
to the early identifications of Halo substructures \citep{helmi1999} to more recent 
identifications based on different stellar tracers \citep{refiorentin2005, kinman2007, koppelman2018,zinn2020,malhan2022}.

\begin{figure*}
\centering
\includegraphics[width=0.45\textwidth]{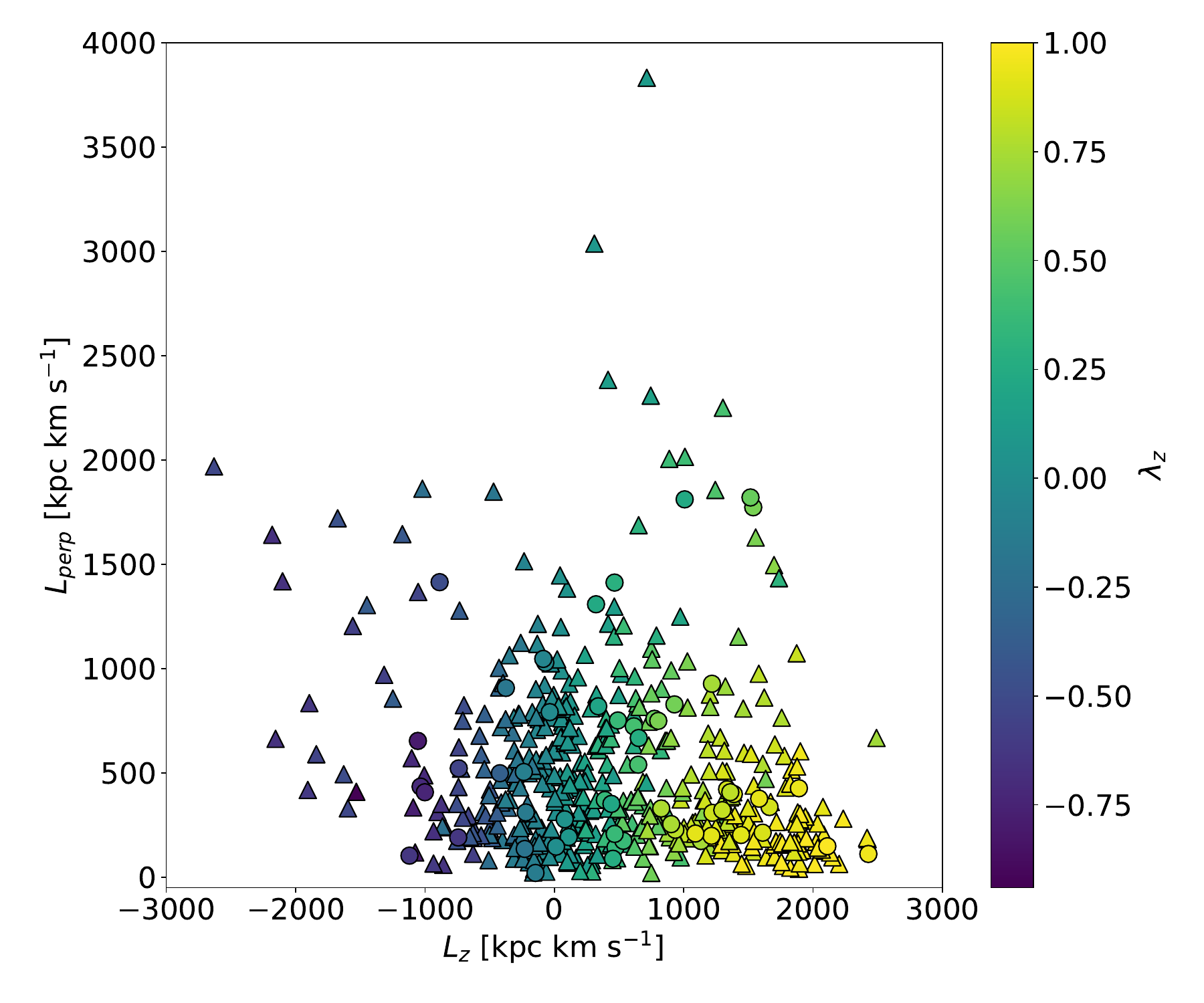}
\includegraphics[width=0.45\textwidth]{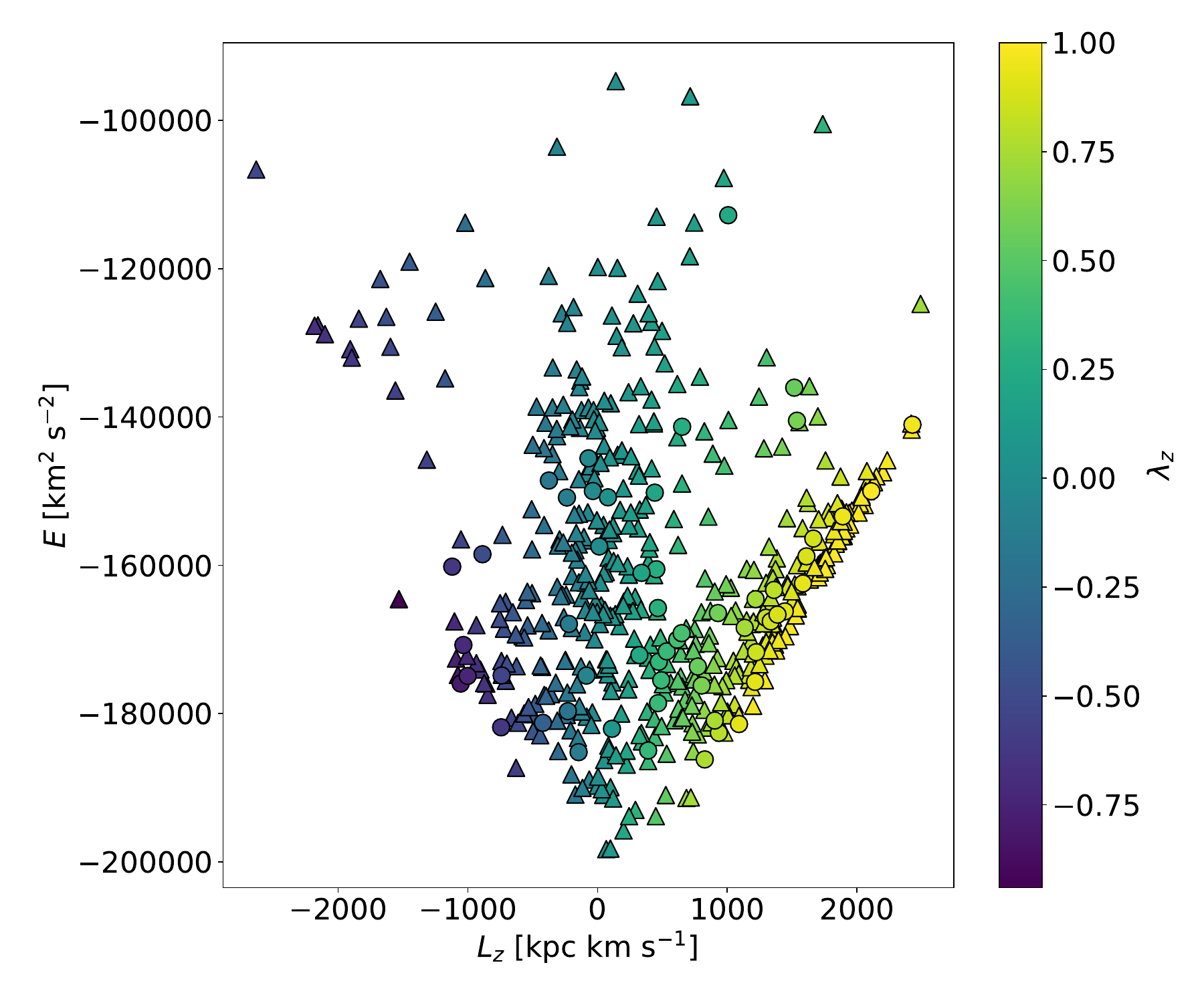}
\caption{Left -- Angular momentum around the Z-axis (L$_{\rm z}$) versus the angular momentum out of the Galactic plane (L$_\perp$) for the entire RRL sample. The symbols are colour-coded according to the circularity of the orbits.
Right: Same as the left, but the angular momentum around the Z-axis is plotted versus the 
total energy.} 
\label{fig:energy_lambda}
\end{figure*}

The left panel of Figure~\ref{fig:energy_lambda} shows the angular momentum around the z-axis (L$_{\rm z}$) versus the angular momentum out of the Galactic plane (L$_\perp$) for the entire RRL sample. The symbols are colour-coded according to the circularity of the orbits ($\lambda_{\rm z}$). As expected the bulk of the cold,  metal-rich RRLs (yellowish symbols) are located at L$_{\rm z}$ values typical of thin disc stars 
(see also Fig.~\ref{fig:toomre_circ_iron}), while warm/thick disc RRLs  
(greenish symbols) attain intermediate  L$_{\rm z}$ values and hot/retrograde 
RRLs (blueish symbols) display a smooth transition toward smaller and 
negative L$_{\rm z}$ values.
A significant fraction of the  RRL sample covers a limited range in 
L$_\perp$--L$_{\rm z}$ values. The exceptions are the 
retrograde RRLs showing two subgroups clustering around 
L$_{\rm z}\sim$-1500 kpc km s$^{-1}$ and L$_\perp\sim$1500 kpc km s$^{-1}$, 
together with a smaller group located at L$_{\rm z}\sim$-2000 kpc km s$^{-1}$ and 
L$_\perp\sim$500 kpc km s$^{-1}$. Similar identifications have been provided by 
\citet{kinman2007, zinn2020} and more recently by \citet{koppelman2018} and by 
\citet{malhan2022} suggesting the association with the so-called Sequoia stream 
\citep{myeong2019}. In the right-hand panel of Figure~\ref{fig:energy_lambda} we plot the angular momentum around the z-axis (L$_{\rm z}$) versus the total energy. As expected, the data display a fan distribution in circularity 
with a smooth transition when moving from RRLs with cold orbits (yellow) to RRLs with retrograde orbits (dark blue). Furthermore, several over-densities show up across the plane associated with warm, hot and retrograde orbits. 

\begin{figure*}
\centering
\includegraphics[width=0.45\textwidth]{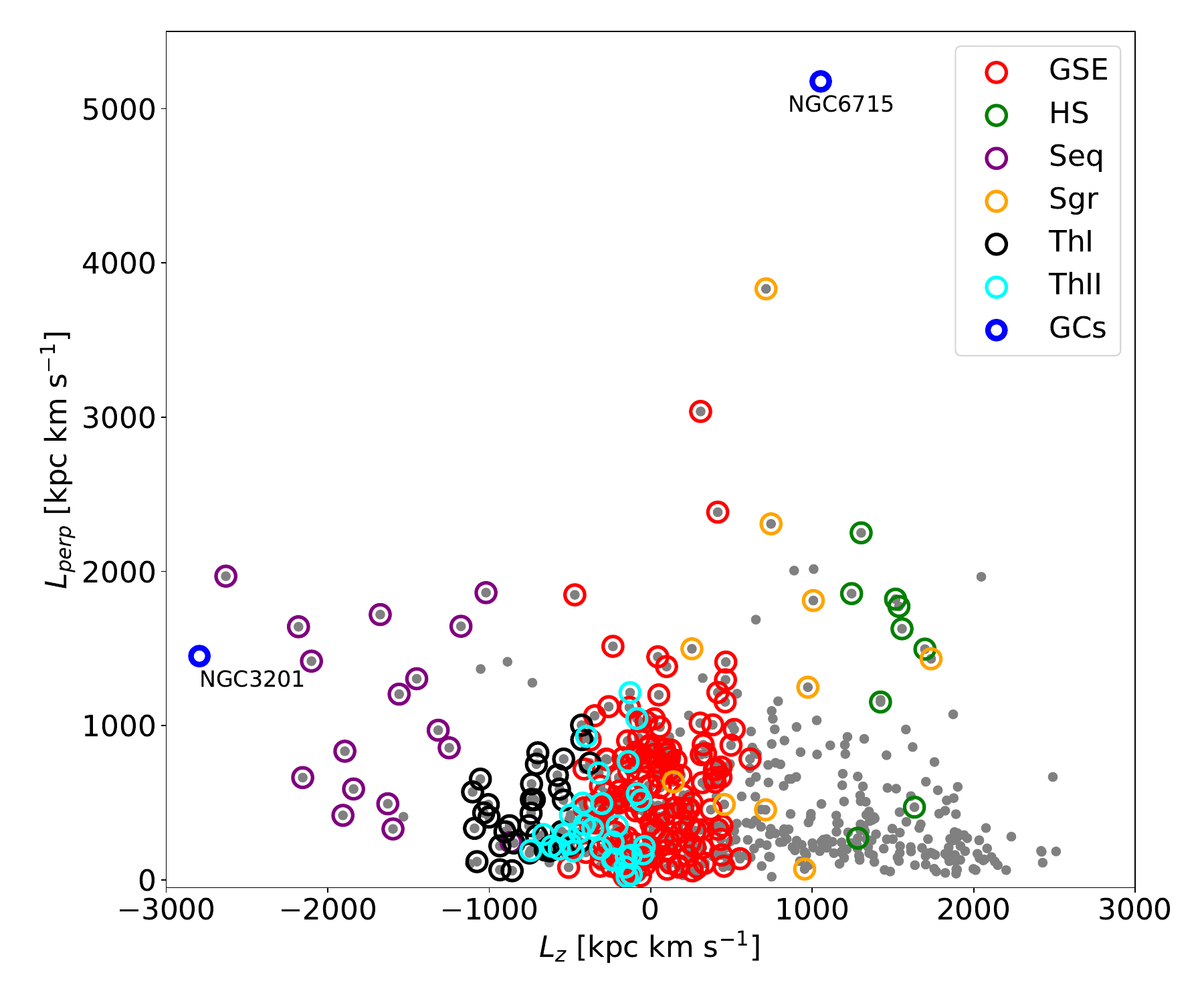}
\includegraphics[width=0.45\textwidth]{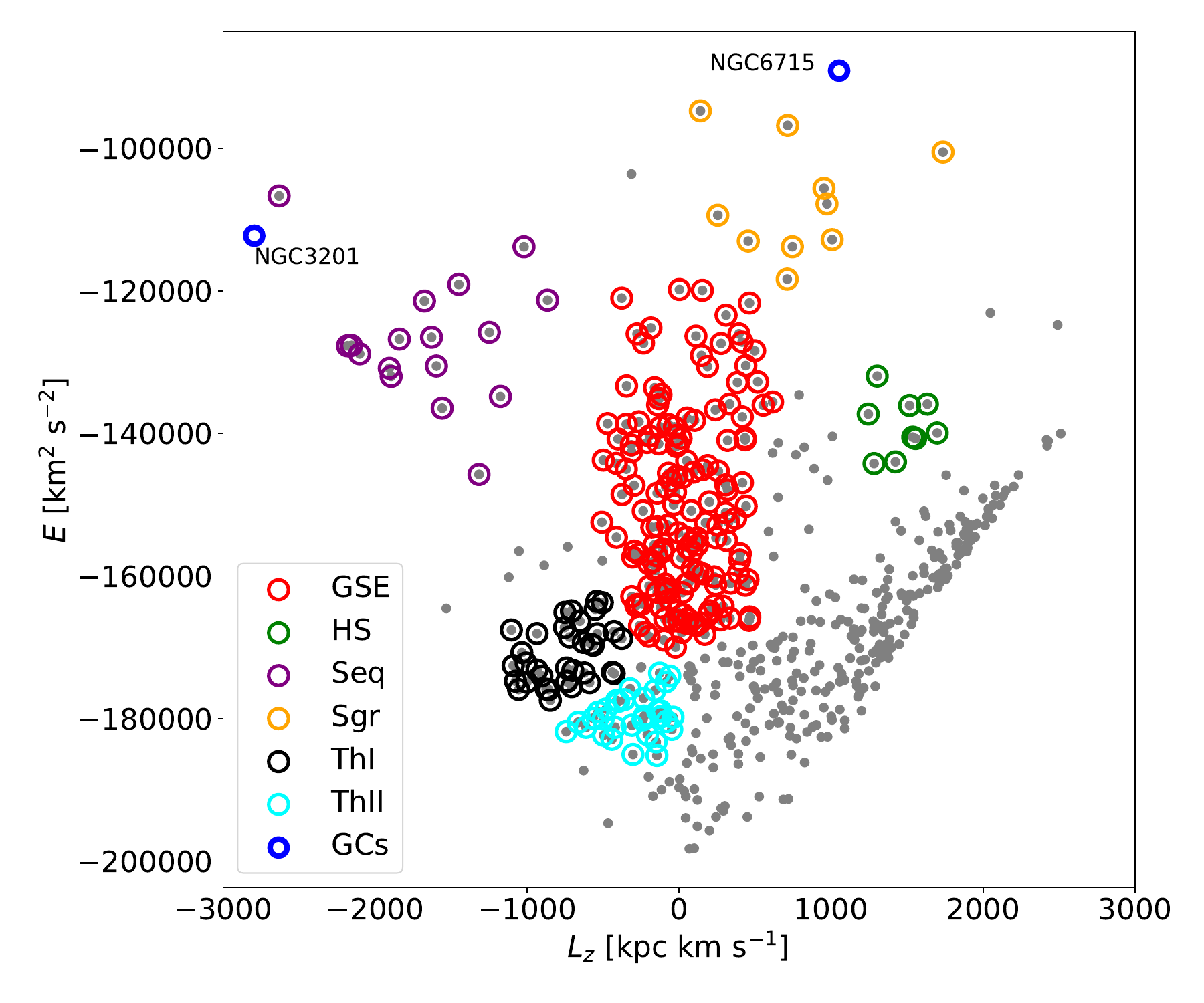}
\caption{Same as Figure~\ref{fig:energy_lambda}, but with RRLs belonging to the 
different streams marked with circles of different colours (see labels): 
Gaia-Sausage-Enceladus (GSE, red), Helmi Stream (HS, green), Sequoia (Seq, brown), 
Sagittarius (Sgr, purple), Thamnos I (ThI, blue) and Thamnos II (ThII, light blue). 
The three thick blue circles mark the location of two globular clusters NGC 6715 and NGC 3201. See text for more details.}\label{fig:energy_streams}
\end{figure*}

To properly identify candidate stellar streams, we took advantage of the stream 
identifications provided by \citet{koppelman2018, zinn2020} and by 
\citet{malhan2022}. The key idea is to trace back in time their origin using 
RRLs. Figure~\ref{fig:energy_streams} shows the same data as in 
Figure~\ref{fig:energy_lambda}, but the different streams are marked with 
different colours. In the following, we discuss in detail the identification of 
the different streams when moving from RRLs with cold/warm orbits to RRLs with 
retrograde orbits. 

{\em Helmi Stream} -- Following \citep{helmi1999,zinn2020} candidate RRLs associated 
with the HS (green empty circles) plotted in Figure~\ref{fig:energy_streams} appear 
relatively concentrated (candidates are listed in Table~\ref{tab:dynamic_rrls}.

{\em Sagittarius Stream} -- Following \citet{malhan2022} candidate RRLs associated 
with the Sgr stream (purple circles) are located in the top right corner of the 
L$_{\rm z}$--E plane, i.e. at high E values and modest L$_{\rm z}$ values. 
The association with the Sgr stream is also supported by the location in the same 
region of M54 \citep[NGC~6715,][]{massari2019}. Note that together with the 
identifications suggested in the literature, we do have several new candidates,
in the high energy regime, that appear to be associated with the Sgr stream. 

{\em Gaia--Sausage--Enceladus Stream} -- This is the major component of the 
candidate stellar streams, located in the L$_{\rm z}$--E plane at small 
 L$_{\rm z}$ and energy values. Solid identifications of  this stellar stream 
 have been provided by \citet{helmi2018, belokurov2018}, but see also 
 \citet{zinn2020}.  Moreover, the current 
 data are indicative of separation from the other candidate stellar streams 
 characterised by lower L$_{\rm z}$ values (see Thamnos streams). 

{\em Sequoia Stream} -- The candidates to the Seq stream can be easily 
identified both in the L$_{\rm z}$--L$_\perp$ plane and the 
L$_{\rm z}$--E plane, since they are characterised by retrograde 
orbits (negative L$_{\rm z}$ values) and small (in absolute sense) energy 
values \citep{myeong2019}. The reader interested in a more detailed 
discussion concerning the occurrence of different sub-groups is referred 
to \citet{zinn2020}. 
\cite{ceccarelli2024} demonstrate that at least two substructures within this area likely originate from the same merger event. It's important to note that due to the small sample size of RRL sampling in this region, we refrain from differentiating between the mentioned sub-groups. In passing we note that the current data fully support 
the association with the globular cluster NGC~3201 
\citep[thick blue circle,][]{massari2019}, which was also preliminary hinted by \citet{kinman2012}. 
Moreover, we also identified a new RRL candidate (see data listed in 
Table~\ref{tab:dynamic_rrls}). 

{\em Thamnos Streams} -- The early identifications of these stellar streams 
were provided by \citet{koppelman2019}. They are quite prominent in the current 
L$_{\rm z}$--E plane, since they show up as separated local overdensities in 
the retrograde regime and the lower boundary of the GSE stream.  
The current data also suggest a few new candidates for the
Thamnos~II stream (see Table~\ref{tab:dynamic_rrls}). Finally, it is worth mentioning that 
there is a small (six) RRL group located between the Seq and the Thamnos~I stream  
with retrograde orbits and L$_{\rm z}$ values similar to the Thamnos streams.

\subsection{Characterisation of candidate stellar streams}

\begin{figure*}
\centering
\includegraphics[width=0.9\textwidth]{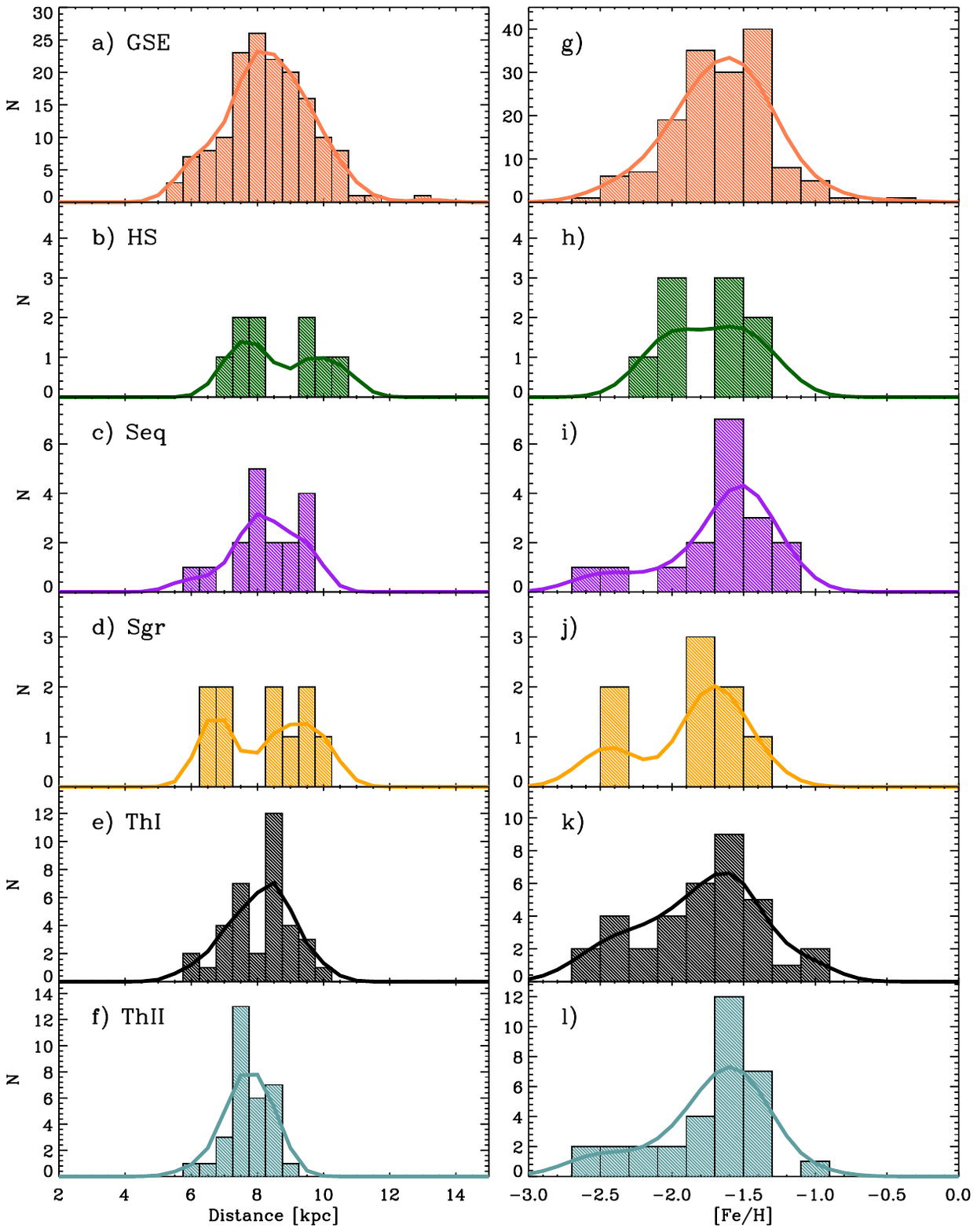}
\caption{Left: From top to bottom Galactocentric distance distribution of RRLs 
identified in the different stellar streams. The colour coding is the same as in 
Fig.~\ref{fig:energy_streams}, and the solid line displays the running average 
over the entire sample. See text and Table~\ref{tab:dist_abund_stream} for more 
details. 
Right: same as the left, but for the iron distribution function.}
\label{fig:histo_streams}
\end{figure*}

The identification of candidate stellar streams using angular momentum 
and total energy is robust, but still affected by possible systematics
and by the natural uncertainties due, for example, to overlapping structures and complex dynamical processes (see e.g., \citealt{belokurov2022, ceccarelli2024}).
The identification of local over-densities is subjective and in some cases 
hampered by limited statistics. The cons of the RRLs as tracers of old stellar 
populations is that they are roughly two orders of magnitudes less abundant 
than old main sequence stars. However, they can be easily recognised and 
their distances are known with an accuracy better than 3\%$-$5\%. 
This means that we can take advantage of this key feature to characterize 
the preliminary identifications of the stellar streams. The idea is to 
use Galactocentric distance distributions and elemental abundances to 
identify the possible presence of a core and to constrain its spatial 
extent.

\begin{table*}
\caption{Mean weighted (mean$\pm$std) Galactocentric distances, iron, calcium and s-process (Y, Ba) 
abundances for the RRLs associated with the different streams. The number of stars is labelled as n.}
\label{tab:dist_abund_stream}
\begin{tabular}{l|rc|rc|rc|rc|rc}
\hline
\hline
Stream & n  & distance  &  n & [Fe/H]  &   n  &    [Ca/Fe]   & n  &  [Y/Fe]  & n &  [Ba/Fe] \\
       &    &  (kpc)    &    &      &      &          &    &      &    &  \\
\hline
GSE & 156 & 8.160$\pm$ 0.856 & 153 &$-$1.646$\pm$0.295 & 47 & 0.314 $\pm$ 0.114  & 12 & $-$0.069 $\pm$ 0.200 & 15 & $-$0.110$\pm$0.196 \\ 
Helmi & 9 & 8.256$\pm$ 0.537 & 9  & $-$1.667$\pm$0.293  & 3 & 0.252 $\pm$ 0.208  & 1  & ~~0.190 $\pm$ 0.100  & 2 & $-$0.221 $\pm$ 0.367  \\ 
Seq & 17 & 8.139$\pm$ 0.788  & 17 & $-$1.601$\pm$0.327  & 5 & 0.199 $\pm$ 0.126  & \dots & \dots  & 2 & $-$0.273 $\pm$ 0.159  \\ 
Sgr & 10 & 8.271$\pm$ 0.689  & 8  &  $-$1.867$\pm$0.399 & \dots & \dots  & \dots & \dots  & \dots & \dots  \\ 
ThI & 36 & 8.275$\pm$ 0.663  & 35  & $-$1.761$\pm$0.429 & 15 & 0.301 $\pm$ 0.173  & 4 & ~~0.122 $\pm$ 0.368  & 4 & ~~0.078 $\pm$ 0.327  \\ 
ThII & 32 & 7.948$\pm$ 0.517 & 32 & $-$1.735$\pm$0.365 & 17 & 0.327 $\pm$ 0.097  & 2 & ~~0.414 $\pm$ 0.313  & 3 & $-$0.017 $\pm$ 0.398  \\ 

\hline
\end{tabular}
\end{table*}

Fig.~\ref{fig:histo_streams} shows the Galactocentric distance 
distributions for the six candidate stellar streams. Panel a) shows the 
distance distribution of RRLs associated with the GSE stream. They are the largest sub-sample (almost one-third) and cover a range in Galactocentric distances similar to the entire RRL sample (4$<$ R$_G$ $<$11 kpc) with a modest dispersion 
(see Table~\ref{tab:dist_abund_stream}). The distribution is skewed toward larger 
distances due to an observational bias, field RRLs located either in the bulge 
or in the inner disc are fainter due to larger reddenings. The core appears to 
be located across the solar circle (R$_G\sim$8 kpc). In passing we also note 
that the iron distribution (panel g) is quite broad with an extended tail in the 
metal-poor regime and a sharp discontinuity in the more metal-rich regime. 
The weighted mean is [Fe/H]=$-$1.646$\pm$0.295, within the uncertainties,  
similar to the mean abundance of the Halo traced by RRL variables 
\citep{crestani2021a, fabrizio2021} and consistent with high-resolution spectroscopic measurement by \cite{ceccarelli2024}. The weighted mean of the $\alpha$-element 
abundance ratio is enhanced ([Ca/Fe]=0.314$\pm$0.114), as expected for 
typical Halo stars. The same outcome applies for 
s-process elements: within the errors, they display a solar-scaled pattern (see weighted mean 
abundances in columns 9--11 of Table~\ref{tab:dist_abund_stream}). 
The current evidence in distance and iron distribution fully supports the 
findings based on kinematical properties that the GSE stream is associated 
with a major merging event with a massive dwarf galaxy. 

The RRLs in the HS display a dichotomic distribution both in Galactocentric 
distance (panel b) and in iron abundance (panel h). They are mainly metal-poor, 
with a moderate enhancement in $\alpha$-element abundances 
(see Table~\ref{tab:dist_abund_stream}). 

The RRLs in the Sequoia (panel c) stream show trends similar to the GSE stream.  Indeed, 
they are skewed in the same direction and the dispersion in Galactocentric distances are 
similar. The mean iron abundance (panel i) is also similar and the Ca abundance ratio 
seems to be slightly less enhanced, but this initial observation is drawn from only 
five RRLs. 
Preliminary evidence suggests that RRLs in the Sgr stream have a dichotomic 
distribution both in Galactocentric distances (panel d) and in iron abundance 
(panel j). Nevertheless, the limited number of RRLs available prevents definitive 
conclusions from being drawn at this stage.

The Galactocentric distance distributions of the RRLs associated with Thamnos~I 
(panel e) and Thamnos~II (panel f) are also concentrated around the solar 
circle but show well-defined peaks at $\sim$9~kpc and $\sim$7~kpc, respectively.
This evidence is suggestive of a possible difference in the 
location of their cores. The iron distribution function of the two streams are quite 
similar with a main peak located at [Fe/H]$\sim -$1.7 and an extended metal-poor 
tail and sharp decrease in the metal-intermediate-regime ([Fe/H]$\sim$-1.3). 
The [Ca/Fe] ratios are enhanced as expected for canonical Halo and thick-disc 
stars. Abundances of the heavy elements Y and Ba exhibit a large scatter with 
super solar and solar-scaled patterns. 

In closing this section two warnings. 
{\em i)}--The RRLs are old stellar tracers, their metallicity distributions
trace the early chemical enrichment and not the current chemical enrichment. 
{\em ii)}--The identification of various streams remains an unresolved issue. A recent study by \cite{donlonnewberg2023} employed a Bayesian Gaussian regression algorithm and proposed that the GSE stream is composed of four distinct components: Virgo, Cronus, Nereus, and Thamnos. This significant finding prompts the need for new selection criteria that rely on observables not affected by uncertainties in distance and reddening.

\section{Summary and final remarks}\label{sec:conclusions}

We performed a detailed abundance analysis of 78 RRLs through high-resolution 
spectra collected within the DR3 of the GALAH survey. We used a fully consistent 
NLTE approach to derive spectroscopic parameters and abundances for Fe, Ca, Mg, Y, and Ba. 
The main results of the current investigation are the following: 

We find evidence of metal-rich RRLs that have either solar or even 
sub-solar $\alpha$-element abundances. This finding fully supports previous measurements based on independent RRL samples provided 
\citet{sneden2018} and more recently by \citet{crestani2021b}. 
However, with respect to previous studies we provide for the first time heavy-element abundances. 
The abundance ratios of neutron-capture elements serve as robust diagnostics for 
determining the origins of metal-rich RRLs. Notably, at a given iron abundance, 
these elements exhibit underabundance relative to the average stellar populations 
of the disc (such as those by \citealt{bensby2014}). This underabundance suggests that 
we may be observing an ancient component of the thin disc, in contrast to 
the comparison stars that cover a broad range in age with typical values 
of $\approx$7$-$8 Gyr (\citealt{kilic2017} and references therein). 
This discrepancy suggests that the s-process nucleosynthesis did not yet contribute substantially to the chemical evolution at the time the metal-rich RRL stars were formed. 


To characterize on a more quantitative basis the nature of metal-rich RRLs 
we investigated the kinematical properties of field RRLs. The GALAH sample 
was complemented with the sample of field RRLs for which our group already 
performed a detailed chemical abundance using high-resolution spectra \citep{for2011, chadid2017, sneden2018, gilligan2021, crestani2021a, crestani2021b}. 
Moreover, we took advantage of a large sample of field RRLs for which 
radial velocities and iron abundance were estimated using low-resolution spectra \citep{zinn2020, fabrizio2021}. To investigate the kinematical properties we adopted the orbit circularity ($\lambda_{\rm z}$). According to this parameter, stars with $\lambda_{\rm z}>0.8$ have cold orbits, while those with $0.25< \lambda_{\rm z} \leq 0.8$ have warm orbits, while those with $-0.25< \lambda_{\rm z} \leq 0.25$ have hot orbits and those with $\lambda_{\rm z} \leq -0.25$ are retrograde.

We found that:

{\em i)}--Approximately 10\% of the GALAH sample is retrograde. 
Two of these stars are metal-intermediate ([Fe/H]$\sim=-$1) and display 
solar or slightly lower [Ca/Fe] abundance ratios. The others are 
metal-poor and $\alpha$-enhanced. Two of them appear to be 
associated with the GSE stream \citep{helmi2018, belokurov2018}. 

{\em ii)}--Interestingly, the bulk of the metal-rich RRLs have cold orbits, 
i.e., a kinematic typical of disc stars. Similar evidence was also 
suggested based on a smaller RRL sample by 
\citet{belokurov2018, iorio2021} and by \citet{prudil2019}. 
Note that this sub-sample includes almost one-third of the sample 
and appears to be a relevant component of field RRLs. Moreover, they 
also have [$\alpha$/Fe] abundance ratios either solar or slightly 
lower. Despite these similarities to the disc stars concerning the iron and the $\alpha$-element 
content, the abundances of neutron-capture elements are, at fixed iron 
abundance, systematically lower than typical disc stars. 

{\em iii)}--Metal-intermediate RRLs include roughly one-third of our sample and 
display a smooth transition from cold to hot orbits. The hot sample 
is characterised by metal-poor iron abundances similar to the 
retrograde sub-sample. 

These findings based on RRL in the GALAH sample were soundly confirmed 
by the entire spectroscopic sample we collected. In particular, we found 
that the distribution of orbit circularity shows two well-defined 
peaks associated with the metal-rich RRLs with disc kinematics 
and metal-poor RRLs with halo/bulge kinematics. Moreover, the 
RRLs with disc kinematics also show a very broad distribution 
in iron abundance ($-$2.5 $<$[Fe/H] $<$0) while the $\alpha$-element 
abundances move from solar to $\alpha$-enhanced. We also 
investigated the possible impact of the topology of the RRL instability 
strip on the chemical and kinematic properties of fundamental and first-overtone RRLs but found no clear difference. This finding brings forward three far-reaching consequences. 

{\em i)}--The evolutionary channel producing metal-rich RRLs with disc kinematics 
is canonical. Their number is too large to be the progeny of either binary 
evolution or stellar collision. This means that they are old, low-mass 
stars in the core-helium burning phase. 

{\em ii)}--The chemical composition of metal-rich RRL stars diverges from that of the typical stars in the thin disc, exhibiting deficiencies in both $\alpha$-elements and neutron-capture elements. Regrettably, the age determinations for thin disc stars, for which uniform measurements of neutron capture elements exist, are exceedingly scarce.

{\em iii)}--The thin disc does not harbour globular clusters, but the current 
preliminary evidence indicates that it includes an old (t$\geq$ 10 Gyr) stellar 
population. Similar conclusions were reached 
by \citet[][and references therein]{wu2023}, who analysed more than 5300 red giant 
stars by combining {\it Kepler} and LAMOST data. The authors found that the oldest 
thin disc stars have an age of 9.5$^{(+0.5,+0.5)}_{(-0.4,-0.3)}$ Gyr (the two 
sources of errors are for random and systematic uncertainties, respectively). 
Crucially, this result is corroborated by our 
recent work on a field  RRL with thin disc kinematics and iron abundance lower 
than $\approx -3$ \citep[][D'Orazi et al., 2024 in preparation]{matsunaga2023}.
The extended iron distribution function, the range covered in 
$\alpha$-element abundances, the neutron capture abundances, the homogeneity 
in orbit circularity and their distribution in the Toomre diagram point toward 
an old stellar population with disc kinematics. The current data do not allow us 
to constrain whether this old stellar component was formed either in situ or 
accreted in the early formation of the Galactic spheroid.  

To investigate the kinematical properties of the entire sample we also performed a 
selection of candidate stellar streams. This idea was motivated by the fact that 
we are dealing with an old stellar tracer for which we do have accurate and 
homogeneous individual distances, radial velocities and elemental abundances. 
Following an approach widely adopted in the literature we took advantage of the 
total energy (E) of the orbits, the angular momentum along the Z-axis (L$_{\rm z}$) 
and of the angular momentum perpendicular (L$_\perp$). We used the 
L$_{\rm Z}$~vs.~L$_\perp$ plane, and in particular, the E~vs.~L$_{\rm z}$ plane 
to identify candidate RRLs belonging to six different stellar streams already 
known in the literature. The key features are the following. 

{\em i)}--GSE stream--Firm identifications of this stellar stream were 
provided by \citet{belokurov2018} and by \citet{helmi2018}. This is the 
major component, and 
indeed, the candidate RRLs are almost one-third of the entire sample and they 
are located in a well defined region of the  E~vs.~L$_{\rm z}$ plane typical 
of stars with warm/hot orbits. 
Moreover and even more importantly, they show a very broad iron distribution 
function and $\alpha$-element abundances are enhanced in the 
metal-poor/metal-intermediate regime and approach solar values or even lower 
in the metal-rich regime. The abundances of s-process elements are around solar 
as expected for typical halo stars. This evidence supports 
recent findings suggesting that the GSE stream is associated with a major 
merging event. 

{\em ii)}--Helmi stream--The RRLs associated with this stream were recently 
complemented with new candidates by \citet{zinn2020}. They show up with 
a well-defined concentration in the E~vs.~L$_{\rm z}$ plane and they also 
display a narrow distribution in Galactocentric distance and in iron 
abundance. More quantitative analyses are hampered by the modest sample size. 

{\em iii)}--Seq stream--After the identification of this stream \citep{myeong2019}
the RRLs associated with this stream have been discussed in detail by \citet{zinn2020}. 
In particular, they suggested the occurrence of different kinematical sub-groups. 
The RRLs associated with this stream display an iron distribution function 
with a main peak similar to RRLs in the GSE and the HS. The same 
outcome applies to the distribution in Galactocentric distances. Moreover, 
we support the association with the Galactic globular cluster NGC~3201 \citep{kinman2012}.

{\em iv)}--Sgr stream--The RRLs associated with this stream display a dichotomic 
distribution both in Galactocentric distances and in iron abundance. This 
preliminary evidence should be treated with care since the sample is 
limited. The current data suggest the association with the Galactic 
globular M54 \citep{massari2019} considered from the very early identifications 
of this stream as the core of the Sgr dwarf galaxy (see \citealt{ibata2009} and references therein). 

{\em v)}--Thamnos streams--The Galactocentric distance distributions of the RRLs 
associated to Thamnos~I and Thamnos~II cluster around the solar circle, but the 
former one shows a peak at 9~kpc, while the latter one at 7~kpc. This evidence 
is suggestive of a possible difference in the location of the core of the two streams. The iron distribution 
functions are similar, moreover, they are $\alpha$-enhanced and the s-process elements 
attain solar values. They appear to be typical halo stars, but their orbit circularity 
and their location in the E~vs.~L$_{\rm z}$ plane suggest that they are retrograde
objects (see however, \citealt{giribaldi2023}, where, based on stellar ages, it is proposed that Thamnos II could be part of an ancient null net rotation structure formed in the Milky Way).

The results of this investigation appear very promising, not only for future 
spectroscopic developments. The ongoing and near future high-resolution spectroscopic 
surveys (e.g., WEAVE, \citealt{jin2023}; 4MOST, \citealt{dejong2019}) 
will allow us to increase the number of field RRLs with 
accurate elemental abundances by order(s) of magnitude. 
Moreover, ongoing (OGLE, \citealt{udalski2004}; ZTF, \citealt{bellm2014}) and upcoming long-term variability surveys (e.g., Vera Rubin Observatory, 
\citealt{dicriscienzo2023}) will provide firm identifications of RRLs across a significant portion of the Galactic thin disc and the Galactic centre. 
The kinematics, the pulsation properties and the chemical abundances of these 
objects will play a key role in assessing on a quantitative basis the early formation 
of the thin disc and the connection with the different components of the Galactic 
spheroid. 

\section*{Acknowledgements}
Parts of this research were supported by the Australian Research Council Centre of Excellence for All Sky Astrophysics in 3 Dimensions (ASTRO 3D), through project number CE170100013. M.M. acknowledges financial support from the Spanish Ministry of Science and Innovation (MICINN) through the Spanish State Research Agency, under
Severo Ochoa Programe 2020-2023. This research has been supported by the Spanish Ministry of Economy and Competitiveness (MINECO) under the grant AYA2020-118778GB-I00. S.W.C. acknowledges funding from the Australian Research Council through Discovery Projects DP190102431 and DP210101299. This research was supported by use of the Nectar Research Cloud, a collaborative Australian research platform supported by the National Collaborative Research Infrastructure Strategy (NCRIS).
M.T. acknowledges support from the Agencia Estatal de Investigación del Ministerio de Ciencia e Innovación (AEI-MCINN) under grant "At the forefront of Galactic Archaeology: evolution of the luminous and dark matter components of the Milky Way and Local Group dwarf galaxies in the Gaia era" with reference PID2020-118778GB-I00/10.13039/501100011033.
V.D. acknowledges the financial contribution from PRIN MUR 2022 (code 2022YP5ACE) funded by the European Union – NextGenerationEU.
Several of us thank the support from Project PRIN MUR 2022 (code 2022ARWP9C) 
"Early Formation and Evolution of Bulge and HalO (EFEBHO)" 
(PI: M. Marconi), funded by the European Union – Next Generation EU and 
by the Large grant INAF 2022 MOVIE (PI: M. Marconi). M.L. acknowledges the support by the Lend\"ulet Program LP2023-10 of the Hungarian Academy of Sciences and the the NKFIH excellence grant TKP2021-NKTA-64.
TZ acknowledges financial support from the Slovenian Research Agency (research core funding No. P1-0188) and the European Space Agency (Prodex Experiment Arrangement No. 4000142234). This work made extensive use of the SIMBAD, Vizier, and NASA ADS databases.
We are grateful to the reviewer for their careful review and constructive feedback, which have substantially improved the quality of our paper.

\section*{Data Availability}
This article's data are available in the DataCentral at \url{https://cloud.datacentral.org.au/teamdata/GALAH/public/GALAHDR3/}. The code \texttt {TSFitPy} can be downloaded at \url{https://github.com/TSFitPy-developers/TSFitPy}.



\bibliographystyle{mnras}
\bibliography{vale} 



\appendix

\section{RR Lyrae sample and properties}



\renewcommand{\arraystretch}{0.6}

\onecolumn

\begin{landscape}
\begin{longtable}{lccccccccr}
\caption{Information on our sample RR Lyrae stars, including GALAH DR3 and Gaia DR3 identifiers, pulsation modes, periods, mean magnitudes, amplitudes and reference epochs. The Source column indicates whether the lightcurve, mean magnitude and amplitude come from ASASSN-V (0), Gaia DR3 (1) or Catalina (2). For 0 and 2, these are provided in the $V$ band, for 1, in the $G$ band.}
\label{tab:photometric} \\
\hline
\hline
Gaia DR3 ID & GALAH ID & myRRLyrSourceId & Type & P & <mag> & Ampl & Source & $T_{max}$ & $T_{ris}$ \\
& & & & d & mag & mag & & & \\
    \hline
    \endfirsthead

    \multicolumn{10}{c}%
    {\tablename\ \thetable{} -- continued from previous page} \\
    \hline
Gaia DR3 ID & GALAH ID & myRRLyrSourceId & Type & P & <mag> & Ampl & Source & $T_{max}$ & $T_{ris}$ \\
& & & & d & mag & mag & & & \\
    \hline
    \endhead

    \hline \multicolumn{10}{r}{Continued on next page} \\
    \endfoot

    \hline
    \hline
    \endlastfoot
 144809087288307584 & 181225002601109 & CSS$\_$J043442.9+214621 & RRab & 0.390592 & 12.799 & 1.21 & 0 & 2456596.3187 & 2456595.8920 \\
 666167814367063296 & 170122003601291 & CSS$\_$J084038.7+231550 & RRc & 0.295777 & 11.653 & 0.50 & 0 & 2456598.0724 & 2456598.0300 \\
2621577222057910528 & 140707003101048 & CSS$\_$J223052.5-074240 & RRab & 0.525980 & 13.398 & 1.07 & 0 & 2456629.9475 & 2456629.8984 \\
2622375506154471680 & 140707003101129 & BN Aqr & RRab & 0.469677 & 12.538 & 1.26 & 0 & 2456629.8657 & 2456629.8233 \\
2683960526815908224 & 170907003601069 & CSS$\_$J220237.0+034216 & RRab & 0.549587 & 13.055 & 0.93 & 0 & 2456231.9541 & 2456231.9022 \\
2697542553435739776 & 171101001201268 & ASAS J215635+0621.3 & RRc & 0.338949 & 12.065 & 0.39 & 0 & 2456597.8348 & 2456597.7711 \\
2710145121353598464 & 160815004301176 & CSS$\_$J222314.6+064802 & RRab & 0.542516 & 13.986 & 0.87 & 0 & 2456593.9129 & 2456593.8602 \\
3181149548774349824 & 181222001801012 & CSS$\_$J044752.9-112047 & RRab & 0.640042 & 12.975 & 0.40 & 0 & 2456631.3672 & 2456631.2903 \\
3201175430791113600 & 171003005101270 & CSS$\_$J043640.7-044014 & RRab & 0.630553 & 12.696 & 0.29 & 0 & 2456594.4686 & 2456593.7368 \\
3230309671428428928 & 151219003101110 & CSS$\_$J043354.9-002532 & RRab & 0.487685 & 12.692 & 1.01 & 0 & 2456626.0455 & 2456625.9987 \\
3323274788307690112 & 140118002001313 & ASAS J060355+0814.5 & RRab & 0.655919 & 12.732 & 1.00 & 0 & 2457008.3687 & 2457008.3043 \\
3406613410300235904 & 170219001601368 & ASAS J045648+1818.3 & RRc & 0.236864 & 11.678 & 0.31 & 0 & 2456001.0366 & 2456000.7479 \\
3471095334863875456 & 170216003301235 & SSS$\_$J123842.8-291327 & RRab & 0.778209 & 13.450 & 0.69 & 0 & 2456793.0069 & 2456792.9248 \\
3479598373678136832 & 160415003101009 & DT Hya & RRab & 0.567980 & 13.015 & 0.98 & 0 & 2456785.1237 & 2456784.5018 \\
3892831505236595456 & 150409002101115 & CSS$\_$J115945.2+025147 & RRab & 0.767693 & 12.068 & 0.43 & 1 & 2456876.4348 & 2456875.5660 \\
4118543955178309376 & 170507010601283 & ASASSN-V J174848.92-215251.6 & RRab & 0.523978 & 11.873 & 0.92 & 0 & 2457070.3386 & 2457069.7651 \\
4179431168210587648 & 151009001601156 & NSVS 16880684 & RRc & 0.356331 & 12.658 & 0.33 & 1 & 2456917.3439 & 2456917.2792 \\
4356714327124981632 & 160514003801381 & PS1 106611 & RRab & 0.571050 & 13.304 & 0.66 & 0 & 2456677.3137 & 2456677.2524 \\
4377598863300947584 & 170711003001180 & CSS$\_$J165135.0-040010 & RRab & 0.445715 & 14.396 & 1.09 & 0 & 2456675.4004 & 2456674.9123 \\
4387211137548084352 & 160421005101223 & NSVS 13688631 & RRab & 0.737684 & 12.247 & 0.42 & 0 & 2456445.6577 & 2456444.8098 \\
4476491413005810176 & 160817002101389 & PS1 137119 & RRab & 0.667544 & 13.869 & 0.28 & 0 & 2456676.6715 & 2456675.8995 \\
4649015060967755520 & 150208002701190 & ASAS J051623-7527.4 & RRab & 0.834004 & 13.279 & 0.50 & 1 & 2456895.7414 & 2456894.8088 \\
4705269305654137728 & 170905003501244 & AG Tuc & RRab & 0.602574 & 12.761 & 1.00 & 1 & 2456899.7921 & 2456899.7367 \\
4717044869029445632 & 171001002901257 & ASAS J013055-5935.2 & RRab & 0.537899 & 13.012 & 1.01 & 0 & 2456809.9257 & 2456809.8759 \\
4737956618117144832 & 161117003501393 & SSS$\_$J023019.4-590805 & RRab & 0.572881 & 12.049 & 0.72 & 1 & 2456869.0362 & 2456868.4045 \\
4894040677457410944 & 150902003701106 & SSS$\_$J044525.5-245536 & RRab & 0.505065 & 13.613 & 1.09 & 0 & 2456848.9581 & 2456848.9108 \\
5377248172118234624 & 160525002701364 & SSS$\_$J111536.5-423619 & RRab & 0.557873 & 13.290 & 0.90 & 0 & 2456784.8975 & 2456784.8417 \\
5378134722086771840 & 170603002101348 & SSS$\_$J115507.0-443135 & RRab & 0.592547 & 13.728 & 0.63 & 0 & 2456793.0803 & 2456792.4260 \\
5380885768898706688 & 170511001101394 & KS Cen & RRab & 0.397422 & 13.119 & 0.99 & 0 & 2456792.6228 & 2456792.5790 \\
5420008488562024064 & 190211002801043 & SSS$\_$J095706.0-391727 & RRab & 0.535917 & 12.455 & 1.21 & 0 & 2456784.7542 & 2456784.7023 \\
5420283950583950464 & 140414002601184 & ASASSN-V J101104.07-400651.4 & RRc & 0.356108 & 13.004 & 0.07 & 3 & 2458165.8780 & 2458165.7950 \\
5469689165544998400 & 160129005201291 & SSS$\_$J104201.2-262653 & RRab & 0.467723 & 13.595 & 1.27 & 0 & 2456792.8896 & 2456792.3786 \\
5500275203911641344 & 171206005101348 & ASAS J061233-5402.6 & RRc & 0.220926 & 12.139 & 0.34 & 0 & 2456785.5643 & 2456785.5151 \\
5583138458927986816 & 161117005201008 & SSS$\_$J064733.9-324722 & RRab & 0.586934 & 13.475 & 0.62 & 0 & 2456785.8740 & 2456785.8110 \\
5721724473609473024 & 150210003201109 & ASAS J081624-1513.4 & RRab & 0.370581 & 13.231 & 1.07 & 0 & 2457010.0936 & 2457010.0579 \\
5790862043943782784 & 170511001601385 & ASASSN-V J131525.38-752744.2 & RRab & 0.544882 & 14.061 & 0.49 & 1 & 2456898.3322 & 2456897.7242 \\
5811341925478455680 & 160813002101375 & SSS$\_$J173422.2-690828 & RRab & 0.588727 & 13.629 & 0.60 & 2 & 2453587.3840 & 2453587.3276 \\
5820212372985755648 & 140312004501064 & ASAS J153830-6906.4 & RRab & 0.622437 & 13.072 & 0.52 & 0 & 2457455.9435 & 2457455.8820 \\
5821920567383108224 & 160327006101183 & UCAC4 121-145252 & RRab & 0.405418 & 13.663 & 0.84 & 1 & 2456902.4877 & 2456902.4482 \\
6009536215017666560 & 160326001601102 & SSS$\_$J154651.6-380040 & RRab & 0.576051 & 12.958 & 1.01 & 0 & 2457456.8897 & 2457456.8340 \\
6025101210857181056 & 170516003101297 & ASASSN-V J162036.29-322435.7 & RRab & 0.502336 & 14.254 & 1.13 & 0 & 2457457.3335 & 2457456.7838 \\
6081475783346889600 & 170508003301387 & ASASSN-V J130646.56-501617.8 & RRab & 0.508149 & 13.795 & 0.75 & 0 & 2457423.1197 & 2457422.5578 \\
6096455220523450112 & 190206007201270 & SSS$\_$J142022.9-443200 & RRab & 0.607117 & 11.948 & 0.57 & 0 & 2457458.0988 & 2457458.0274 \\
6104846246589810304 & 160418004101394 & ASAS J143814-4025.6 & RRc & 0.374718 & 13.262 & 0.45 & 0 & 2457457.9696 & 2457457.8959 \\
6109120799902812928 & 170413004101159 & ASASSN-V J135041.67-421434.7 & RRab & 0.332638 & 12.776 & 0.77 & 1 & 2456899.2747 & 2456899.2287 \\
6120342346853804160 & 160328004201274 & SSS$\_$J142437.2-332942 & RRc & 0.184775 & 12.578 & 0.49 & 0 & 2456790.7022 & 2456790.4943 \\
6121588956818597760 & 160424004201158 & SSS$\_$J135658.5-364111 & RRab & 0.703782 & 13.187 & 0.70 & 0 & 2456794.0784 & 2456793.3001 \\
6137933300244848384 & 160519003601266 & SSS$\_$J131629.2-410407 & RRab & 0.836563 & 13.194 & 0.24 & 1 & 2456897.6068 & 2456897.4721 \\
6162709690968204288 & 150607002601302 & SSS$\_$J132745.4-374659 & RRab & 0.548049 & 13.008 & 0.90 & 0 & 2456793.8831 & 2456793.8289 \\
6180274359158473728 & 160415003601070 & ASASSN-V J130615.31-323202.9 & RRc & 0.330064 & 13.502 & 0.06 & 3 & 2457457.9144 & 2457458.1721 \\
6182771968540082176 & 170515003101037 & SSS$\_$J131301.2-292531 & RRab & 0.700974 & 13.040 & 0.53 & 1 & 2456897.6910 & 2456896.9053 \\
6225380999055363456 & 170531003301371 & SSS$\_$J145702.0-264238 & RRab & 0.586276 & 13.380 & 0.49 & 1 & 2456904.0275 & 2456903.9300 \\
6226585956422527616 & 160513002601168 & XX Lib & RRab & 0.698501 & 12.375 & 0.75 & 1 & 2456904.8504 & 2456904.0831 \\
6242022966539339136 & 140414004601373 & SSS$\_$J160733.6-240311 & RRab & 0.576510 & 12.840 & 0.62 & 1 & 2456908.1175 & 2456907.4852 \\
6273401001165886336 & 170415004101377 & SSS$\_$J142624.1-225311 & RRab & 0.600104 & 13.645 & 0.39 & 2 & 2453478.8643 & 2453478.7865 \\
6367478755094103808 & 170905002101379 & ASAS J192639-7438.6 & RRc & 0.316734 & 12.432 & 0.39 & 0 & 2456794.9731 & 2456794.5781 \\
6378877082899249664 & 150830004601021 & AR Oct & RRab & 0.394029 & 12.769 & 1.23 & 0 & 2456792.0435 & 2456792.0064 \\
6409071321466282752 & 161008003001285 & SSS$\_$J215601.0-612912 & RRab & 0.616088 & 12.752 & 1.10 & 0 & 2456790.0647 & 2456790.0068 \\
6409095201484462208 & 150828003701073 & ASAS J215855-6109.1 & RRab & 0.802121 & 12.665 & 0.16 & 0 & 2456790.1041 & 2456789.9391 \\
6434640155133754368 & 170516004101029 & ASASSN-V J185553.56-664412.0 & RRab & 0.719612 & 13.363 & 0.94 & 0 & 2456793.3233 & 2456792.5323 \\
6454841894587413376 & 160816003201145 & SSS$\_$J203305.8-611320 & RRab & 0.577596 & 13.945 & 0.80 & 0 & 2456791.1779 & 2456791.1070 \\
6463159287733222016 & 150603004301337 & SSS$\_$J211324.5-552407 & RRab & 0.646437 & 13.784 & 0.31 & 0 & 2456793.2197 & 2456792.4753 \\
6473684637667905280 & 170614004101002 & SSS$\_$J201425.0-525542 & RRab & 0.575418 & 13.237 & 0.41 & 0 & 2456790.1162 & 2456790.0458 \\
6562247309988278144 & 140808003201227 & SSS$\_$J213829.6-490054 & RRab & 0.477468 & 13.686 & 0.91 & 0 & 2456793.0378 & 2456792.5084 \\
6564092943335079808 & 140811004501392 & SSS$\_$J214753.4-471332 & RRab & 0.632508 & 13.799 & 0.77 & 0 & 2456790.2502 & 2456790.1856 \\
6637089279786058880 & 140611004001172 & SSS$\_$J190236.4-563640 & RRab & 0.638228 & 13.160 & 0.34 & 0 & 2456785.3688 & 2456785.2804 \\
6638021596927943808 & 180628003301156 & SSS$\_$J190001.9-551546 & RRab & 0.460542 & 13.608 & 0.97 & 0 & 2456795.2034 & 2456794.6989 \\
6640222746188019200 & 150603003801282 & SSS$\_$J192907.7-554517 & RRc & 0.343012 & 13.078 & 0.42 & 0 & 2456784.9368 & 2456784.8515 \\
6650047986394047104 & 170805003101128 & ASAS J184324-5458.9 & RRab & 0.359691 & 13.196 & 1.29 & 0 & 2456784.9482 & 2456784.9147 \\
6664639188589394688 & 170713004101098 & SSS$\_$J191941.6-443402 & RRc & 0.247099 & 13.408 & 0.26 & 0 & 2456814.9134 & 2456814.8569 \\
6665135721170434816 & 170713004101347 & SSS$\_$J192157.1-432821 & RRab & 0.593109 & 12.143 & 0.49 & 0 & 2456790.1657 & 2456789.5019 \\
6674570703462167680 & 170905002601269 & ASAS J203616-4628.1 & RRab & 0.497461 & 13.566 & 0.81 & 0 & 2456790.2918 & 2456789.7442 \\
6679308288613240064 & 160916001801011 & SSS$\_$J202812.2-423609 & RRab & 0.479459 & 13.719 & 1.03 & 0 & 2456792.9185 & 2456792.3870 \\
6681635649786400384 & 150705005401263 & SSS$\_$J203652.4-391206 & RRab & 0.654473 & 14.172 & 0.48 & 0 & 2456793.1912 & 2456793.0987 \\
6681944303315818624 & 150705005401344 & SSS$\_$J204110.0-390818 & RRab & 0.522306 & 13.403 & 1.07 & 0 & 2456792.9871 & 2456792.9386 \\
6694638955332768512 & 160530005501015 & ASAS J202817-3806.6 & RRc & 0.244849 & 13.215 & 0.40 & 0 & 2456789.9191 & 2456789.8588 \\
6698926702788332416 & 170507011701145 & ASAS J195927-3400.1 & RRc & 0.379658 & 12.372 & 0.60 & 0 & 2456790.0545 & 2456789.9786 \\
6730211038418525056 & 170710002201295 & V0413 CrA & RRab & 0.589343 & 10.591 & 0.60 & 0 & 2456797.3124 & 2456796.6565 \\
6768220266332023040 & 151008001601166 & ASAS J193559-2418.6 & RRab & 0.628529 & 13.100 & 0.35 & 0 & 2456790.1843 & 2456789.4531 \\
6795546531894178816 & 171003002101302 & SSS$\_$J205254.6-285212 & RRab & 0.632154 & 13.913 & 0.66 & 0 & 2456788.9960 & 2456788.9308 \\
\hline											  		  		    
\end{longtable}
\end{landscape}

\renewcommand{\arraystretch}{0.8}

\begin{landscape}
\begin{longtable}{lccccccc}
    \caption{Kinematics properties of our RRLs}
    \label{tab:kinematics} \\
    \hline
\hline
Gaia DR3 ID & HJD(begin) & HJD(end) & <$\phi_{Tris}$> & <$\phi_{Tmax}$> & distance & RV & v$_{\gamma}$ \\
& & & & & pc & km/s & km/s \\
\hline
\endfirsthead

    \multicolumn{7}{c}%
    {\tablename\ \thetable{} -- continued from previous page} \\
\hline
Gaia DR3 ID & HJD(begin) & HJD(end) & <$\phi_{Tris}$> & <$\phi_{Tmax}$> & distance & RV & v$_{\gamma}$ \\
& & & & & pc & km/s & km/s \\
\hline
\endhead

    \hline \multicolumn{7}{r}{Continued on next page} \\
    \endfoot

    \hline
    \hline
    \endlastfoot
 144809087288307584 & 2458477.9938 & 2458478.0355 & 0.6396 & 0.5470 & 1579 $\pm$  55 &   85.14 $\pm$  0.51 &   70.24 $\pm$  2.45 \\
 666167814367063296 & 2457776.0683 & 2457776.1100 & 0.9251 & 0.7815 & 1535 $\pm$  24 & -117.35 $\pm$  2.71 & -124.36 $\pm$  3.01 \\
2621577222057910528 & 2456846.2095 & 2456846.2512 & 0.2930 & 0.1996 & 2981 $\pm$ 169 &  -60.68 $\pm$  1.48 &  -45.37 $\pm$  2.73 \\
2622375506154471680 & 2456846.2096 & 2456846.2513 & 0.7574 & 0.6672 & 2334 $\pm$ 140 & -128.73 $\pm$  1.82 & -151.05 $\pm$  3.05 \\
2683960526815908224 & 2458004.0651 & 2458004.1068 & 0.5731 & 0.4786 & 3147 $\pm$  49 & -212.13 $\pm$  1.97 & -221.77 $\pm$  2.93 \\
2697542553435739776 & 2458058.9482 & 2458058.9899 & 0.9682 & 0.7801 & 1598 $\pm$  48 & -299.83 $\pm$  3.09 & -302.98 $\pm$  3.26 \\
2710145121353598464 & 2457616.1448 & 2457616.1864 & 0.3766 & 0.2795 & 2604 $\pm$ 165 & -140.15 $\pm$  0.91 & -133.49 $\pm$  2.31 \\
3181149548774349824 & 2458474.9484 & 2458474.9901 & 0.5610 & 0.4407 & 2390 $\pm$  95 &  222.88 $\pm$  0.87 &  215.46 $\pm$  1.90 \\
3201175430791113600 & 2458030.2157 & 2458030.2574 & 0.1586 & 0.9980 & 2284 $\pm$  74 &  -61.70 $\pm$  2.56 &  -39.93 $\pm$  3.02 \\
3230309671428428928 & 2457376.0569 & 2457376.0986 & 0.0410 & 0.9451 & 1852 $\pm$  70 &   32.21 $\pm$  3.51 &   61.76 $\pm$  4.16 \\
3323274788307690112 & 2456676.0137 & 2456676.0553 & 0.4287 & 0.3304 & 2087 $\pm$  85 &   28.84 $\pm$  2.13 &   31.47 $\pm$  3.04 \\
3406613410300235904 & 2457803.9068 & 2457803.9485 & 0.7278 & 0.5092 &  811 $\pm$  12 &   90.75 $\pm$  0.89 &   84.10 $\pm$  1.23 \\
3471095334863875456 & 2457801.1824 & 2457801.2241 & 0.6400 & 0.5346 & 4384 $\pm$ 356 &  214.82 $\pm$  1.22 &  202.92 $\pm$  1.89 \\
3479598373678136832 & 2457494.0181 & 2457494.0597 & 0.2301 & 0.1351 & 3121 $\pm$  50 &   59.80 $\pm$  1.67 &   81.86 $\pm$  2.72 \\
3892831505236595456 & 2457121.9123 & 2457121.9540 & 0.9187 & 0.7870 & 2146 $\pm$ 131 &  204.26 $\pm$  2.89 &  180.21 $\pm$  3.16 \\
4118543955178309376 & 2457881.2209 & 2457881.2625 & 0.6847 & 0.5902 &  875 $\pm$  14 &   16.39 $\pm$  1.93 &    0.60 $\pm$  2.90 \\
4179431168210587648 & 2457304.8828 & 2457304.9244 & 0.8199 & 0.6383 & 2239 $\pm$ 102 & -149.51 $\pm$  0.82 & -157.26 $\pm$  1.21 \\
4356714327124981632 & 2457523.0586 & 2457523.1003 & 0.1780 & 0.0707 & 2186 $\pm$  89 &   -7.78 $\pm$  1.86 &   16.31 $\pm$  2.65 \\
4377598863300947584 & 2457945.9675 & 2457946.0092 & 0.7680 & 0.6728 & 3573 $\pm$ 296 &  -57.80 $\pm$  0.85 &  -79.32 $\pm$  2.46 \\
4387211137548084352 & 2457500.2115 & 2457500.2532 & 0.7235 & 0.5742 & 1624 $\pm$  38 & -184.85 $\pm$  1.99 & -198.11 $\pm$  2.36 \\
4476491413005810176 & 2457617.9238 & 2457617.9655 & 0.2101 & 0.0537 & 3240 $\pm$ 164 &   30.25 $\pm$  0.89 &   48.02 $\pm$  1.82 \\
4649015060967755520 & 2457061.9095 & 2457061.9511 & 0.3845 & 0.2663 & 3417 $\pm$ 124 &   98.50 $\pm$  2.19 &  105.42 $\pm$  2.56 \\
4705269305654137728 & 2458002.1540 & 2458002.1957 & 0.5479 & 0.4559 & 3232 $\pm$ 150 &  187.17 $\pm$  1.57 &  178.74 $\pm$  2.67 \\
4717044869029445632 & 2458028.1148 & 2458028.1565 & 0.8470 & 0.7544 & 3218 $\pm$  51 &  135.46 $\pm$  0.63 &  113.08 $\pm$  2.32 \\
4737956618117144832 & 2457710.0254 & 2457710.0670 & 0.1392 & 0.0365 & 1903 $\pm$  50 &   92.35 $\pm$  5.42 &  120.56 $\pm$  5.76 \\
4894040677457410944 & 2457268.2982 & 2457268.3399 & 0.4046 & 0.3108 & 4232 $\pm$  63 &  293.64 $\pm$  1.04 &  298.16 $\pm$  2.53 \\
5377248172118234624 & 2457533.8594 & 2457533.9011 & 0.6689 & 0.5688 & 3344 $\pm$  54 &  183.06 $\pm$  1.03 &  166.84 $\pm$  2.32 \\
5378134722086771840 & 2457907.8635 & 2457907.9052 & 0.4818 & 0.3776 & 3291 $\pm$ 181 &  189.60 $\pm$  1.36 &  187.41 $\pm$  2.31 \\
5380885768898706688 & 2457884.9340 & 2457884.9756 & 0.6516 & 0.5414 & 2380 $\pm$  76 &  -19.38 $\pm$  0.45 &  -33.83 $\pm$  2.27 \\
5420008488562024064 & 2458526.0867 & 2458526.1284 & 0.3908 & 0.2940 & 2325 $\pm$  37 &  257.27 $\pm$  1.81 &  263.38 $\pm$  3.01 \\
5420283950583950464 & 2456761.9448 & 2456761.9865 & 0.8541 & 0.6208 & 1474 $\pm$  31 &   67.70 $\pm$  1.52 &   65.52 $\pm$  1.54 \\
5469689165544998400 & 2457417.1770 & 2457417.2186 & 0.8741 & 0.7815 & 4063 $\pm$  62 &  136.58 $\pm$  0.76 &  111.41 $\pm$  2.57 \\
5500275203911641344 & 2458094.2102 & 2458094.2518 & 0.7845 & 0.5616 & 1493 $\pm$  29 &  101.36 $\pm$  0.46 &   93.56 $\pm$  1.03 \\
5583138458927986816 & 2457710.2116 & 2457710.2533 & 0.9996 & 0.8922 & 3327 $\pm$ 127 &  150.50 $\pm$  1.42 &  148.58 $\pm$  2.34 \\
5721724473609473024 & 2457064.0265 & 2457064.0681 & 0.6885 & 0.5921 & 3139 $\pm$  50 &   30.85 $\pm$  0.60 &   13.91 $\pm$  2.37 \\
5790862043943782784 & 2457884.9844 & 2457885.0261 & 0.9180 & 0.8020 & 3333 $\pm$ 165 &   17.32 $\pm$  1.18 &   -2.54 $\pm$  2.16 \\
5811341925478455680 & 2457613.8864 & 2457613.9280 & 0.4636 & 0.3677 & 3525 $\pm$ 180 &  140.87 $\pm$  1.10 &  140.19 $\pm$  2.15 \\
5820212372985755648 & 2456729.2493 & 2456729.2909 & 0.6346 & 0.5357 & 2344 $\pm$  83 &  -17.85 $\pm$  0.82 &  -30.09 $\pm$  1.96 \\
5821920567383108224 & 2457475.2462 & 2457475.2879 & 0.9093 & 0.8120 & 2872 $\pm$ 133 &  -11.50 $\pm$  0.64 &  -34.15 $\pm$  2.20 \\
6009536215017666560 & 2457474.2372 & 2457474.2789 & 0.2475 & 0.1508 & 1938 $\pm$  72 & -100.57 $\pm$  1.43 &  -79.92 $\pm$  2.60 \\
6025101210857181056 & 2457890.1071 & 2457890.1487 & 0.6581 & 0.5637 & 2890 $\pm$ 181 &  -22.59 $\pm$  0.74 &  -38.13 $\pm$  2.45 \\
6081475783346889600 & 2457882.0293 & 2457882.0710 & 0.2477 & 0.1420 & 2637 $\pm$ 119 &   92.68 $\pm$  0.90 &  110.40 $\pm$  2.21 \\
6096455220523450112 & 2458521.2590 & 2458521.3007 & 0.3129 & 0.1953 & 1456 $\pm$  35 &  131.55 $\pm$  1.04 &  143.65 $\pm$  2.09 \\
6104846246589810304 & 2457497.1301 & 2457497.1718 & 0.7590 & 0.5623 & 3485 $\pm$  52 & -124.21 $\pm$  0.81 & -134.02 $\pm$  1.44 \\
6109120799902812928 & 2457857.1145 & 2457857.1562 & 0.7281 & 0.5896 & 2056 $\pm$  83 &    9.58 $\pm$  0.30 &   -7.61 $\pm$  2.06 \\
6120342346853804160 & 2457476.1962 & 2457476.2378 & 0.1228 & 0.9978 & 2252 $\pm$  85 &   -3.44 $\pm$  1.67 &   12.60 $\pm$  2.10 \\
6121588956818597760 & 2457503.0730 & 2457503.1147 & 0.5426 & 0.4367 & 3498 $\pm$  63 &  271.08 $\pm$  2.11 &  265.14 $\pm$  2.56 \\
6137933300244848384 & 2457527.9897 & 2457528.0313 & 0.7253 & 0.5643 & 3460 $\pm$ 190 &  196.12 $\pm$  0.98 &  183.93 $\pm$  1.53 \\
6162709690968204288 & 2457180.9302 & 2457180.9718 & 0.3641 & 0.2651 & 2498 $\pm$ 110 &  179.40 $\pm$  1.93 &  187.29 $\pm$  2.89 \\
6180274359158473728 & 2457494.0569 & 2457494.0986 & 0.7836 & 0.5644 & 1700 $\pm$  55 &  -22.20 $\pm$  1.28 &  -24.19 $\pm$  1.30 \\
6182771968540082176 & 2457888.9837 & 2457889.0254 & 0.3157 & 0.1948 & 3512 $\pm$  64 &   11.02 $\pm$  1.18 &   23.35 $\pm$  1.79 \\
6225380999055363456 & 2457904.9774 & 2457905.0191 & 0.5025 & 0.3363 & 2839 $\pm$ 189 &  -60.34 $\pm$  1.52 &  -63.94 $\pm$  2.32 \\
6226585956422527616 & 2457522.0118 & 2457522.0535 & 0.6790 & 0.5805 & 2273 $\pm$  80 &  104.09 $\pm$  4.88 &   88.35 $\pm$  5.26 \\
6242022966539339136 & 2456762.1564 & 2456762.1980 & 0.9524 & 0.8556 & 2277 $\pm$  93 &   87.10 $\pm$  1.32 &   65.18 $\pm$  2.28 \\
6273401001165886336 & 2457859.1199 & 2457859.1616 & 0.3313 & 0.2016 & 3727 $\pm$ 310 &  -59.76 $\pm$  1.18 &  -50.02 $\pm$  2.05 \\
6367478755094103808 & 2458001.9410 & 2458001.9827 & 0.9815 & 0.7346 & 2217 $\pm$  82 &    4.38 $\pm$  2.28 &    2.25 $\pm$  2.51 \\
6378877082899249664 & 2457265.0970 & 2457265.1386 & 0.7024 & 0.6083 & 2051 $\pm$  82 &   -7.48 $\pm$  0.46 &  -26.24 $\pm$  2.46 \\
6409071321466282752 & 2457669.9739 & 2457670.0155 & 0.3488 & 0.2548 & 3071 $\pm$  52 &  171.30 $\pm$  2.45 &  182.48 $\pm$  3.32 \\
6409095201484462208 & 2457263.0445 & 2457263.0862 & 0.8439 & 0.6382 & 2355 $\pm$  64 &   64.47 $\pm$  4.16 &   48.60 $\pm$  4.31 \\
6434640155133754368 & 2457890.2094 & 2457890.2510 & 0.4014 & 0.3023 & 4142 $\pm$  78 &  184.54 $\pm$  1.02 &  191.17 $\pm$  1.89 \\
6454841894587413376 & 2457617.0217 & 2457617.0634 & 0.9534 & 0.8305 & 4528 $\pm$ 414 &  -84.16 $\pm$  0.79 & -107.67 $\pm$  2.15 \\
6463159287733222016 & 2457177.2618 & 2457177.3034 & 0.2738 & 0.1224 & 3435 $\pm$ 176 &   76.04 $\pm$  1.24 &   89.55 $\pm$  2.03 \\
6473684637667905280 & 2457919.1806 & 2457919.2223 & 0.3233 & 0.2010 & 3103 $\pm$ 173 &  -73.40 $\pm$  0.53 &  -62.96 $\pm$  1.77 \\
6562247309988278144 & 2456878.1021 & 2456878.1437 & 0.3094 & 0.2008 & 3291 $\pm$ 160 &   -1.21 $\pm$  1.22 &   11.65 $\pm$  2.48 \\
6564092943335079808 & 2456881.1618 & 2456881.2034 & 0.8668 & 0.7647 & 3753 $\pm$ 227 &  -78.15 $\pm$  0.58 &  -98.70 $\pm$  2.06 \\
6637089279786058880 & 2456820.1916 & 2456820.2332 & 0.7328 & 0.5942 & 2278 $\pm$ 116 &  340.00 $\pm$  0.90 &  325.55 $\pm$  1.87 \\
6638021596927943808 & 2458298.1093 & 2458298.1510 & 0.4818 & 0.3863 & 3597 $\pm$  55 &   48.94 $\pm$  0.93 &   46.20 $\pm$  2.39 \\
6640222746188019200 & 2457177.2098 & 2457177.2515 & 0.9225 & 0.6737 & 3316 $\pm$ 182 & -107.25 $\pm$  1.74 & -113.43 $\pm$  2.07 \\
6650047986394047104 & 2457970.9907 & 2457971.0324 & 0.5474 & 0.4544 & 3140 $\pm$  47 &  -55.55 $\pm$  0.46 &  -64.54 $\pm$  2.51 \\
6664639188589394688 & 2457948.0900 & 2457948.1316 & 0.2341 & 0.0055 & 2639 $\pm$ 106 &  -26.87 $\pm$  0.77 &  -19.91 $\pm$  1.06 \\
6665135721170434816 & 2457948.0900 & 2457948.1317 & 0.4507 & 0.3316 & 1819 $\pm$  46 &  -40.44 $\pm$  1.20 &  -40.07 $\pm$  2.13 \\
6674570703462167680 & 2458001.9963 & 2458002.0379 & 0.9207 & 0.8198 & 3289 $\pm$ 159 &   47.83 $\pm$  1.53 &   24.93 $\pm$  2.58 \\
6679308288613240064 & 2457647.9346 & 2457647.9762 & 0.4469 & 0.3382 & 4425 $\pm$ 387 & -104.88 $\pm$  2.20 & -104.47 $\pm$  3.15 \\
6681635649786400384 & 2457209.2180 & 2457209.2596 & 0.8397 & 0.6984 & 3928 $\pm$ 341 &  -45.60 $\pm$  0.68 &  -62.26 $\pm$  1.88 \\
6681944303315818624 & 2457209.2179 & 2457209.2596 & 0.0427 & 0.9499 & 3899 $\pm$  62 &  -57.32 $\pm$  1.45 &  -26.67 $\pm$  2.71 \\
6694638955332768512 & 2457539.2330 & 2457539.2747 & 0.6412 & 0.3951 & 3167 $\pm$ 204 &   19.11 $\pm$  0.92 &   12.32 $\pm$  1.41 \\
6698926702788332416 & 2457881.2933 & 2457881.3350 & 0.5224 & 0.3225 & 1678 $\pm$  47 &  -43.14 $\pm$  0.63 &  -48.34 $\pm$  1.68 \\
6730211038418525056 & 2457945.0908 & 2457945.1325 & 0.7048 & 0.5919 &  841 $\pm$  13 &  -76.44 $\pm$  1.89 &  -92.12 $\pm$  2.64 \\
6768220266332023040 & 2457303.8817 & 2457303.9233 & 0.4975 & 0.3341 & 2477 $\pm$ 139 & -195.92 $\pm$  1.02 & -198.94 $\pm$  1.93 \\
6795546531894178816 & 2458029.9037 & 2458029.9454 & 0.1195 & 0.0163 & 3321 $\pm$ 212 & -210.85 $\pm$  2.50 & -182.10 $\pm$  3.13 \\
\hline
\end{longtable}
\end{landscape}

\begin{table}
\caption{Dynamic properties of the total sample of RRLs. This excerpt is shown for guidance, the entire table is made available through CDS.}\label{tab:dynamic_rrls}
\begin{tabular}{rrrrrrrrl}
\hline
Gaia DR3 ID & $V_x$ & $V_y$ & $V_z$ & $L_z$ & $L_{perp}$ & E & $\lambda_{\rm z}$ & Stream \\
  & \multicolumn{3}{c}{[km/s]} & \multicolumn{2}{c}{[kpc km/s]} & [km$^2$/s$^2$] &  &  \\
\hline
\hline
  15489408711727488 &      26.11 &     -28.13 &     -13.22 &    -237 &     130 & -180178 & -0.19 & ThII \\
  80556926295542528 &      79.69 &     170.59 &     -16.25 &    1663 &     336 & -156398 &  0.87 & \ldots \\
 144809087288307584 &     -61.04 &     162.08 &     -12.61 &    1537 &     122 & -159975 &  0.86 & \ldots \\
 182142003881848832 &     -39.07 &     246.99 &       7.12 &    2201 &      63 & -147465 &  0.98 & \ldots \\
 234108363683247616 &       5.83 &     240.99 &     -14.35 &    2059 &     126 & -151882 &  1.00 & \ldots \\
 289662047665943552 &       0.79 &     243.46 &      17.66 &    2110 &     148 & -150013 &  0.99 & \ldots \\
 294072906063827072 &    -165.64 &    -122.33 &       2.47 &   -1120 &     103 & -160181 & -0.63 & \ldots \\
 305829816397138688 &     -26.91 &     127.94 &      -9.26 &    1198 &     142 & -164522 &  0.73 & \ldots \\
 315028326379733760 &    -149.65 &     -30.00 &      11.23 &    -505 &     224 & -163751 & -0.30 & ThI \\
 317254635562605184 &     -95.21 &      34.67 &     -41.98 &     202 &     548 & -164905 &  0.12 & GSE \\
\hline
\end{tabular}
\end{table}

\begin{table}
\caption{Average abundances of RRLs in bins of $\lambda_{\rm z}$.}
\label{tab:abundance_vs_lambdaz}
\begin{tabular}{l|rr|rr|rr|rr|rr}
\hline
\hline
$\lambda_{\rm z}$ & \multicolumn{2}{c}{[Fe/H]$_{HR}$} & \multicolumn{2}{c}{[Fe/H]$_{LR}$} & \multicolumn{2}{c}{[Ca/Fe]} & \multicolumn{2}{c}{[Y/Fe]} & \multicolumn{2}{c}{[Ba/Fe]} \\
& n & mean$\pm$std& n & mean$\pm$std& n & mean$\pm$std& n & mean$\pm$std& n & mean$\pm$std \\
\hline
$\lambda_{\rm z}$>0.8           & 66 & --0.714$\pm$0.561 & 116 & --0.786$\pm$0.605 & 62 &  0.046$\pm$0.195 & 11 & --0.249$\pm$0.321 & 12 & --0.234$\pm$0.268 \\
0.25<$\lambda_{\rm z} \leq$0.8  & 56 & --1.426$\pm$0.442 & 129 & --1.472$\pm$0.440 & 52 &  0.215$\pm$0.170 & 17 & --0.004$\pm$0.315 & 19 & --0.149$\pm$0.290 \\
-0.25<$\lambda_{\rm z} \leq$0.25& 80 & --1.626$\pm$0.288 & 214 & --1.645$\pm$0.321 & 71 &  0.314$\pm$0.112 & 18 &   0.085$\pm$0.283 & 25 & --0.027$\pm$0.268 \\
$\lambda_{\rm z}\leq$-0.25      & 29 & --1.673$\pm$0.393 &  73 & --1.734$\pm$0.408 & 28 &  0.283$\pm$0.144 &  6 &   0.219$\pm$0.306 &  7 &   0.040$\pm$0.275 \\
\hline
\end{tabular}
\end{table}

\renewcommand{\arraystretch}{1.0}

\begin{table}
\caption{Line list for atomic transitions employed in the present study.}
\label{tab:line-list}
\begin{tabular}{l|c|r|r}
\hline
\hline
Wavelength & Species & E.P. & $\log gf$ \\
(\AA)    &         & (eV) & (dex) \\
\hline
            &       &       & \\
5711.088 & Mg 1&    4.35 & $-$1.742\\
5690.425 & Si 1&    4.93 & $-$1.802 \\
5708.410 & Si 1&    4.95 & $-$1.370 \\
5721.021 & Si 1&    5.86 & $-$0.852   \\
5772.146 & Si 1&    5.08 & $-$1.643 \\
5793.073 & Si 1&    4.93 & $-$1.894 \\
6555.462 & Si 1&    5.98 & $-$0.886 \\
5857.451 & Ca 1&    2.93 &  0.240\\
6493.781 & Ca 1&    2.52 & $-$0.109\\
4727.413 & Fe 1&   3.69 &  $-$1.083 	\\ 
4736.774 & Fe 1&   3.21 &  $-$0.674 	\\ 
4798.265 & Fe 1&   4.19 &  $-$1.174 	\\ 
4799.845 & Fe 1&   4.39 &  $-$2.322 	\\ 
4878.210 & Fe 1&   2.89 &  $-$0.887 	\\ 
4891.490 & Fe 1&   2.85 &  $-$0.111 	\\ 
4892.861 & Fe 1&   4.22 &  $-$1.290  	\\ 
5655.459 & Fe 1&   5.03 &  $-$0.436 	\\ 
5662.532 & Fe 1&   4.18 & $-$0.447 	\\ 
5679.025 & Fe 1&   4.60 & $-$0.820  	\\ 
5686.531 & Fe 1&   4.55 & $-$0.445 	\\ 
5701.537 & Fe 1&   2.56 & $-$2.193 	\\ 
5706.004 & Fe 1&   4.61 & $-$0.460  	\\ 
5715.083 & Fe 1&   4.28 & $-$0.970  	\\ 
5717.836 & Fe 1&   4.28 & $-$0.990  	\\ 
5753.123 & Fe 1&   4.26 & $-$0.623 	\\ 
5762.989 & Fe 1&   4.21 & $-$0.360  	\\ 
5775.080 & Fe 1&   4.22 & $-$1.126 	\\ 
5780.566 & Fe 1&   3.24 & $-$2.540  	\\ 
5816.366 & Fe 1&   4.55 & $-$0.601 	\\ 
6494.979 & Fe 1&   2.40 & $-$1.268 	\\ 
6592.912 & Fe 1&   2.73 & $-$1.473 	\\ 
6593.868 & Fe 1&   2.43 & $-$2.420  	\\ 
6609.107 & Fe 1&   2.56 & $-$2.691 	\\ 
6633.751 & Fe 1&   4.56 & $-$0.799 	\\ 
6663.441 & Fe 1&   2.42 & $-$2.473 	\\ 
6677.982 & Fe 1&   2.69  & $-$1.418 	\\ 
4893.818 & Fe 2&   2.83  & $-$4.267 	\\ 
5651.523 & Fe 2&   10.63 & $-$0.615 	\\ 
5806.822 & Fe 2&   10.68 & $-$1.324 	\\ 
6495.212 & Fe 2&   11.09 &  ~~0.031    \\ 
6516.077 & Fe 2&    2.89 &  $-$3.31  	\\ 
4883.682 & Y 2&    1.08  &  0.190     \\
4900.119 & Y 2&    1.03 &  0.030     \\
5853.668 & Ba 2&   0.60 & $-$0.907     \\
6496.897 & Ba 2&   0.60 & $-$0.407     \\
\hline
\hline
\end{tabular}
\end{table}

\clearpage

\noindent \textbf{List of institutions}:
\\
$^{1}$Department of Physics, University of Rome Tor Vergata, via della Ricerca Scientifica 1, 00133 Rome, Italy\\
$^{2}$ INAF - Osservatorio Astronomico di Padova, vicolo dell' Osservatorio 5, 35122 Padova, Italy\\
$^{3}$ Department of Physics and Astronomy, Monash University, Clayton, VIC 3800, Australia\\
$^{4}$ ARC Centre of Excellence for All Sky Astrophysics in 3 Dimensions (ASTRO 3D), Australia\\
$^{5}$ Max Planck Institute for Astronomy,  K\"onigstuhl 17, 69117 Heidelberg, Germany\\
$^{6}$ INAF - Osservatorio Astronomico di Roma, via Frascati 33, Monte Porzio Catone, Italy\\
$^{7}$ Department of Astrophysics,  Türkenschanzstraße 17 (Sternwarte), 1180 Wien,
 Austria\\
$^{8}$ Space Science Data Center - ASI, via del Politecnico s.n.c., I-00133, Rome, Italy \\
$^{9}$ Department of Astronomy and McDonald Observatory, The University of Texas, Austin, TX 78712, USA\\
$^{10}$ INAF - Osservatorio di Astrofisica e Scienza dello Spazio di Bologna, via P. Gobetti 93/3, 40129 Bologna, Italy \\
$^{11}$ Institut d’Astronomie et d’Astrophysique, Université libre de Bruxelles, CP 226, Boulevard du Triomphe, 1050 Brussels, Belgium.\\
$^{12}$ Research School of Astronomy and Astrophysics, The Australian National University, Canberra ACT2611, Australia\\
$^{13}$ School of Mathematical and Physical Sciences, Macquarie University, Balaclava Road, Sydney, NSW 2109, Australia\\
$^{14}$ Astrophysics and Space Technologies Research Centre, Macquarie University, Balaclava Road, Sydney, NSW 2109, Australia\\
$^{15}$ International Space Science Institute--Beijing, 1 Nanertiao, Zhongguancun, Beijing 100190, China
$^{16}$ Konkoly Observatory, HUN-REN Research Centre for Astronomy and Earth Sciences, Konkoly Thege Mikl\'os út 15-17., H-1121, Hungary\\
$^{17}$CSFK, MTA Centre of Excellence, Budapest, Konkoly Thege Mikl\'os út 15-17., H-1121, Hungary\\
$^{18}$ELTE E\"{o}tv\"{o}s Lor\'and University, Institute of Physics and Astronomy, Budapest 1117, P\'azm\'any P\'eter s\'et\'any 1/A, Hungary\\
$^{19}$ Lund Observatory, Department of Geology, S\"olvegatan 12, SE-22362 Lund, Sweden \\
$^{20}$ Department of Physics and Astronomy, Dartmouth College, Hanover, NH03755, USA\\
$^{21}$ INAF - Osservatorio Astronomico di Capodimonte, salita Moiariello 16, 80131, Naples, Italy \\
$^{22}$ Department of Physics, Florida State University, 77 Chieftain Way, Tallahassee, FL32306, USA\\
$^{23}$ Gemini Observatory/NSF’s NOIRLab, 670 N.A’ohokuPlace, Hilo, HI96720,USA\\
$^{24}$ Department of Astronomy, School of Science, The University of Tokyo,7-3-1 Hongo, Bunkyo-ku, Tokyo 113-0033, Japan\\
$^{25}$ Laboratory of Infrared High-resolution spectroscopy, Koyama Astronomical Observatory, Kyoto Sangyo University, Kyoto 603-8555, Japan\\
$^{26}$ IAC- Instituto de Astrof\'isica de Canarias, Calle V\'ia Lactea s/n,
E-38205 La Laguna, Tenerife, Spain\\
$^{27}$ Departmento de Astrof\'isica, Universidad de La Laguna, E-38206 La
Laguna, Tenerife, Spain\\
$^{28}$ Department of Physics and Astronomy, Iowa State University, Ames, IA50011, USA \\
$^{29}$ Center for Astrophysical Sciences and Department of Physics and Astronomy, The Johns Hopkins University, Baltimore, MD 21218, USA\\
$^{30}$ Universit'e de Nice Sophia-Antipolis, CNRS, Observatoire de la Cote d’Azur, Laboratoire Lagrange, BP4229, F-06304 Nice, France\\
$^{31}$ INAF - Istituto di Radioastronomia, via Gobetti 101, 40129 Bologna, Italy \\
$^{32}$ Institute for Astronomy, University of Hawaii at Manoa, 2680 Woodlawn Dr., Honolulu, HI 96822, USA\\
$^{33}$ Universit\"ats-Sternwarte, Fakult\"at f\"ur Physik, Ludwig-Maximilians Universit\"at M\"unchen, Scheinerstraße 1, D-81679 M\"nchen, Germany\\
$^{34}$Sydney Institute for Astronomy, School of Physics, A28, The University of Sydney, NSW 2006, Australia\\
$^{35}$ Faculty of Mathematics and Physics, University of Ljubljana, Jadranska 19, 1000 Ljubljana, Slovenia\\
$^{36}$ Department of Astronomy, Stockholm University, AlbaNova Research Centre, SE-10691, Stockholm, Sweden\\
$^{37}$ School of Physics, UNSW, Sydney, NSW 2052, Australia\\

\bsp	
\label{lastpage}
\end{document}